# First-principles calculation of the stabilities of lithium garnet compositions against hydration

## James R. Rustad[*]

## Corning Incorporated, Corning NY 14830
(21 May 2016)

**Abstract-** A series of density functional electronic structure calculations were carried out to better understand the crystallographic factors governing the stability of $Li_nA_3B_2O_{12}$ lithium garnet phases against hydration. The reaction studied is $H_2O + Li_nA_3B_2O_{12} = LiOH + H_nA_3B_2O_{12}$. Most of the compositions are stable against pure water; the main driving force for instability in the atmosphere is the reaction of lithium hydroxide with $CO_2$ to make lithium carbonate. The calculated hydration resistance scales with the Pauling bond valence on the oxygen atom contributed by the coordinating A and B ions. In the unexchanged Li-garnets, this bond valence must be balanced by lithium, so there is also a good overall correlation of hydration stability with lithium stoichiometry ($n$): hydration resistance increases in the order $Li_8$-garnet < $Li_7$-garnet < $Li_6$-garnet < $Li_5$-garnet < $Li_3$-garnet. Only $Li_3A_3B_2O_{12}$ garnets have proton exchange energies sufficiently positive to overcome the decomposition energy of lithium hydroxide into lithium carbonate + water; the $n$=3 garnets are predicted to be stable to hydration under atmospheric conditions, in agreement with observations (*Galven et al. Chem. Mater.* **2012** *24*, 3335-3345). At a given lithium ion stoichiometry, hydration resistance is greater for A, B ions having smaller ionic radii.

## Introduction

$Li_nA_3B_2O_{12}$ garnets have been proposed for use as solid electrolyte membranes in lithium-based batteries[1]. Materials such as $Li_7La_3Zr_2O_{12}$ (LLZO) are stable against reduction and can withstand contact against metallic lithium. It has been increasingly appreciated, however, that the lithium garnets are prone to exchange of $Li^+$ for $H^+$. When exposed to $H_2O$, hydrolytic breakdown can occur by reaction of $nH_2O + Li_n$-garnet = $nLiOH + H_n$-garnet. In the presence of $CO_2$, breakdown of LiOH into $Li_2CO_3$ +$H_2O$ is strongly exothermic[2,3,4,5]. Recent work[4] has shown that a wide variety of Li-garnet compositions are sensitive to humidity/$CO_2$ in the atmosphere, the exception being $Li_3Nd_3Te_2O_{12}$. It was noted in Ref. 4 that once the number of lithium atoms in the formula unit exceeds three, the garnet becomes more susceptible to proton exchange under atmospheric conditions. In this paper, first-principles calculations are carried out on a series of Li-garnet compositions to assess their relative resistance to hydration. The calculations should yield further insight into the reactivity trends observed in Ref. 4.

---

[*] present address University of California, Davis, One Shields Avenue, Davis, CA 95616 (jrrustad@ucdavis.edu)

**Methods**

The hydration reaction is written as a lithium-hydrogen exchange reaction:

$$nH_2O + Li_nA_3B_2O_{12} = nLiOH + H_nA_3B_2O_{12} \qquad (1)$$

where, for referential purposes, we take $H_2O$ as proton-ordered ice. Additional reactions may be of interest, for example, the breakdown of lithium hydroxide: LiOH + ½$CO_2$ = ½$Li_2CO_3$ + ½$H_2O$ (-64.6 kJ/mol according the Materials Genome Project reaction calculator). Whatever other auxiliary reactions might be of interest, Reaction 1 is convenient for comparing the relative hydration stabilities of variety of $Li_nA_3B_2O_{12}$ compounds. The calculations here were done with VASP 5.2[6,7,8,9] using the PBE exchange-correlation functional[10,11] and the recommended Projector Augmented Wave pseudopotentials[12,13] with an energy cutoff of 520 eV. All calculations were done at the gamma point.

The Li-garnets investigated in this study are given in Table 1. Each structure has eight formula units in the unit cell. For LNdTeO, LNdWO, as well as t-LLZO (tetragonal LLZO) and LLSnO (isostructural with t-LLZO) compounds, the lithium atoms occupy definite sites. No ordered structures are available for the rest of the Li-bearing compounds, nor for any of the hydrogarnet phases. For the hydrogarnets, the problem of configurational disorder may be simplified to some extent by considering only structures with complete replacement of $Li^+$ by $H^+$ according to Reaction (1). Although this would rarely, if ever, occur in any real material, the energetics of the complete hydration reaction should serve as a guide to the overall susceptibility of a given lithium garnet composition to replacement of $Li^+$ with $H^+$.

For the disordered Li-bearing garnets, the *24d* and *48g* sites in the Ia-3d space group were semi-randomly chosen for occupation (as described presently) with starting lattice parameters as given in Ref. 4. Before describing the procedure for generating the configurations, a couple of remarks on Li-garnet crystal chemistry are relevant. First, one often hears of tetrahedral and octahedral proton sites in lithium garnet ion conductors. This is potentially confusing. In the Ia-3d space group, the tetrahedral *lithium* sites are the *24d* sites and the octahedral *lithium* sites are the *48g* sites (in addition, occupying essentially the same *48g* octahedron, are the *96h* tetrahedral lithium sites. These are displaced around 0.4 Å from the *48g* sites, in a direction roughly toward the 24d sites).

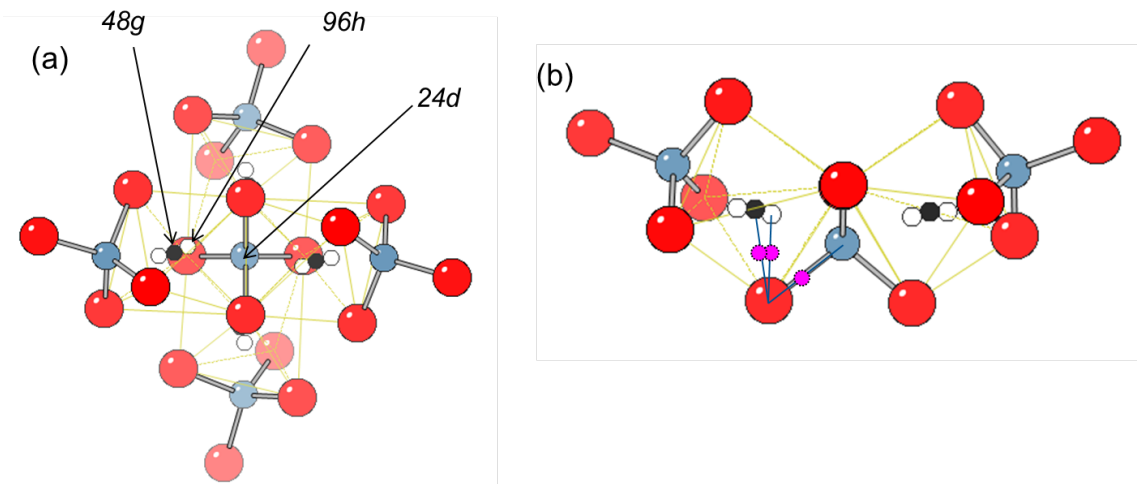

**Figure 1.** Distribution if lithium sites in LinA$_3$B$_2$O$_{12}$ lithium garnet superionic conductors. red: oxygen, teal: 24d (tetrahedral) lithium sites; black 48g (octahedral) lithium sites; white 96h (tetrahedral) lithium sites. Small magenta circles represent how protons can "occupy" 24d, 96h, and 48g sites, while being attached to the same oxygen atom.

As shown in Figure 1, the idea of a *proton* "occupying" a tetrahedral or octahedral site is only a matter of which direction the proton points; a proton sitting at one particular oxygen site can change from tetrahedral to octahedral "occupation" without ever physically hopping to another oxygen atom, so it isn't really useful to talk about tetrahedral or octahedral protons.

Figure 1 also clarifies that the oxygen atoms around the *24d* sites are the same oxygen atoms that surround the *48g/96h* sites. Given this, one can conveniently group the 96 oxygen atoms in the unit cell into 24 groups of four, each of the groups associated to the particular *24d* lithium ion to which the oxygen atom would have been coordinated if the lithium site were occupied. As a screening process for generating good proton configurations, it makes sense to distribute the protons as evenly as possible in the *24d* groups. For example it wouldn't likely be a good idea to put four protons in one of the *24d* sites (i.e. one on each of the oxygen atoms which would have been coordinated to the lithium atom in that particular *24d* site) and no protons at all in another *24d* site. To avoid this type of situation (which could occur if the protonation sites were chosen randomly) the following protocol is used: if the unit cell needs 24 protons to balance charge (i.e. Li$_3$Nd$_3$Te$_2$O$_{12}$) then each of the *24d* groups would get one proton (which particular one of these four oxygen atoms does receive the proton *is* chosen randomly). If the unit cell needs 40 protons to balance the charge (i.e. Li$_5$La$_3$Nb$_2$O$_{12}$) then 16 of the *24d* groups would get two protons and 8 would get one proton (the 16 sites that get two protons *are* chosen randomly). If 56 protons were needed to balance the charge, then 24 groups would have two protons and eight randomly chosen sites would have three protons. One could get more prescriptive about this process; in the last example one could ensure that the 8 sites having three protons are not neighboring sites, but this is not done here. For the disordered lithium garnet phases, lithium positions were chosen randomly on the *24d* and *48g* sites.

**Table 1.** Garnet compositions investigated in this study. All structures started with a cubic unit cell except for t-LYZO; t-LLZO; and LLSnO.

| composition | abbreviation | ΔE (kJ/mol) | protons exchanged | Li-garnet energy (eV) per formula unit[a] | H-garnet energy (eV) per formula unit |
|---|---|---|---|---|---|
| $Li_3Nd_3Te_2O_{12}$ | LNdTeO[b] | 89.3 | 3 | -131.9959 | -129.1833 |
| $Li_3Nd_3W_2O_{12}$ | LNdWO[b] | 67.3 | 3 | -159.9114 | -157.7843 |
| $Li_4Nd_3SbTeO_{12}$ | LNdSbTeO | 48.1 | 4 | -137.7801 | -135.7375 |
| $Li_5Nd_3Sb_2O_{12}$ | LNdSbO | 26.7 | 5 | -143.7860 | -142.3412 |
| $Li_5La_3Ta_2O_{12}$ | LLTO | 21.2 | 5 | -170.8652 | -169.7071 |
| $Li_5La_3Nb_2O_{12}$ | LLNO | 17.7 | 5 | -166.1852 | -165.2096 |
| $Li_6Ca_3W_2O_{12}$ | LCaWO | 6.1 | 6 | -159.2178 | -158.7670 |
| $Li_6BaLa_2Nb_2O_{12}$ | LBLNO | 2.3 | 6 | -163.9109 | -163.6956 |
| $Li_6Ba_3W_2O_{12}$ | LBaWO | -2.1 | 6 | -156.4501 | -156.5088 |
| t-$Li_7Y_3Zr_2O_{12}$ | t-LYZO | 4.6 | 7 | -176.3555 | -175.9402 |
| c-$Li_7Y_3Zr_2O_{12}$ | c-LYZO | 2.9 | 7 | -176.2344 | -175.9402 |
| t-$Li_7La_3Zr_2O_{12}$ | t-LLZO[c] | 4.1 | 7 | -171.8274 | -171.4464 |
| c-$Li_7La_3Zr_2O_{12}$ | c-LLZO | 1.6 | 7 | -171.6441 | -171.4464 |
| $Li_7La_3Sn_2O_{12}$ | LLSnO[d] | 0.8 | 7 | -152.5767 | -152.4393 |
| $Li_8Ba_3Nb_2O_{12}$ | LBNO | -31.5 | 8 | -158.9567 | -161.4754 |

[a]Li-ordered structures are taken for LNdTeO,LNdWO,t-LLZO,LLSnO; [b]24 Li placed in 24d sites; [c]Structure taken from Ref.14; [d]Starting structure taken from t-LLZO.

One hundred proton/lithium configurations were generated for each composition according to the procedures just outlined for the garnet compositions listed in Table 1. All 100 configurations were given to VASP for optimization. While the overall goal is to find the lowest energy configuration and use that as the best representative, here the entire range between the most favorable configuration and the least favorable configuration is recorded, and this range is used to get a rough idea of the uncertainty in the calculation due to the incomplete exploration of proton configurations. Note also that while the starting configurations for the disordered Li-garnet and hydrogarnet compositions have equal *a*, *b*, and *c* lattice parameters characteristic of the disordered structure with cubic symmetry, optimization takes place in $P_1$ symmetry, so that the final configuration may be significantly distorted from the original lattice parameters inherited from Ia-3d. VASP CONTCAR files for the lowest-energy structures found for each Li-garnet and hydrogarnet composition are given in the Supporting Information.

### Results

The results of the calculations are given in Figure 2. In Figure 2, the energy obtained in Reaction (1) is divided by $n$ to put the hydration on a per-proton basis. For reference, the energy of the lithium hydroxide breakdown reaction (as taken from the Materials Genome Project Database[15]) is shown with the vertical dashed line. First, note that nearly all compositions are stable against water, at least as modeled in Reaction (1) where the source of water is proton-ordered ice. In terms of atmospheric stability, the real concern is $CO_2$, which creates a problem due to the highly favorable reaction of LiOH with $CO_2$ to make $Li_2CO_3$. Overall, the order of stability against hydration increases in the series $Li_8A_3B_2O_{12} < Li_7A_3B_2O_{12} < Li_6A_3B_2O_{12} < Li_5A_3B_2O_{12} < Li_3A_3B_2O_{12}$. The finding that the tellurium composition is most stable against hydration is in qualitative agreement with the findings Ref. 4. It appears that when the (positive) energy of the hydration reaction exceeds the (negative) energy of the reaction of LiOH to make $Li_2CO_3$, the reaction will become unfavorable. Of course, in detail, this finding assumes the ultimate reaction product is $Li_2CO_3$, and, furthermore, proton-ordered ice is used as the pool from which the water molecules are drawn. In a more careful calculation corrections would need to be made to calibrate the reactivity to water vapor or liquid water. Aside from such complications, the calculations would predict, as observed in Ref. 4 that the only compound definitely stable against $H^+/Li^+$ exchange is $Li_3Nd_3Te_2O_{12}$. The calculations also predict that $Li_3Nd_3W_2O_{12}$ (which was not studied in Ref. 4) should be marginally stable.

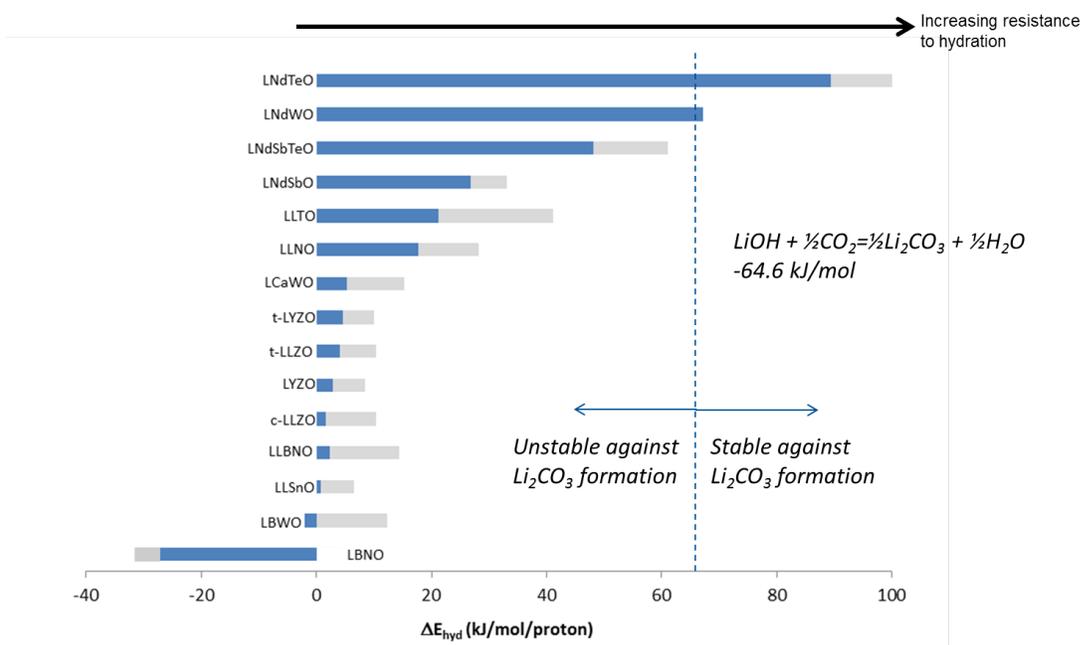

**Figure 2.** Energy of hydration reaction $nH_2O + Li_nA_3B_2O_{12} = nLiOH + H_nA_3B_2O_{12}$. Energies are given on a "per proton" basis (i.e. the energy of reaction 1 is divided by $n$). The total range over the 100 generated proton configurations are given by the grey part of the bar. Vertical dashed line gives the (exothermic) energy of the LiOH breakdown reaction. Hydration reactions falling below this line are unstable, those above are stable.

In Ref. 4 it was stated that all $Li_nA_3B_2O_{12}$ compounds with $n > 3$ were susceptible to proton exchange in a humid atmosphere. The density functional electronic structure calculations indicate, more generally, that the hydration resistance scales inversely with $n$ in these compounds. The correlation is given in Figure 3. The underlying reason for this reactivity trend is clear. All oxygen atoms in the $Li_nA_3B_2O_{12}$ garnet structure are bound to two "A" ions (the "A" ions are eight-fold coordinated) and one "B" ion (the "B" ions are six fold coordinated). As the charge on either the "A" or "B" metal increases, the Pauling bond valence[16] (PBV) into the oxygen atom increases and it becomes progressively more difficult to protonate the site. For the Te/W compound, the total PBV into the oxygen atom (the sum of the charge divided by the coordination number over all bound cations) is 2(+3/8) + (+6/6) = 1.75, for the Nb/Ta compound the sum is 2(+3/8) + (+5/6) =1.58 and for the Zr/Sn compound it is 2(+3/8) + (+4/6) = 1.42. Of course the average PBV into the oxygen atom scales perfectly with $n$, the number of lithium atoms in the formula unit, as the overall charge must be neutral. As shown in Figure 3, the correlation with $n$ rationalizes the observed reactivity trend. The Li-O bond strength, of course, also becomes less strong as the bond valence increases, but the association between a lithium ion and a particular oxygen atom is not as intimate as the proton: in other words, $Li^+$, being a larger ion than $H^+$, is not as sensitive to the basicity of the $BA_2\equiv O$ "ligands" as is the proton.

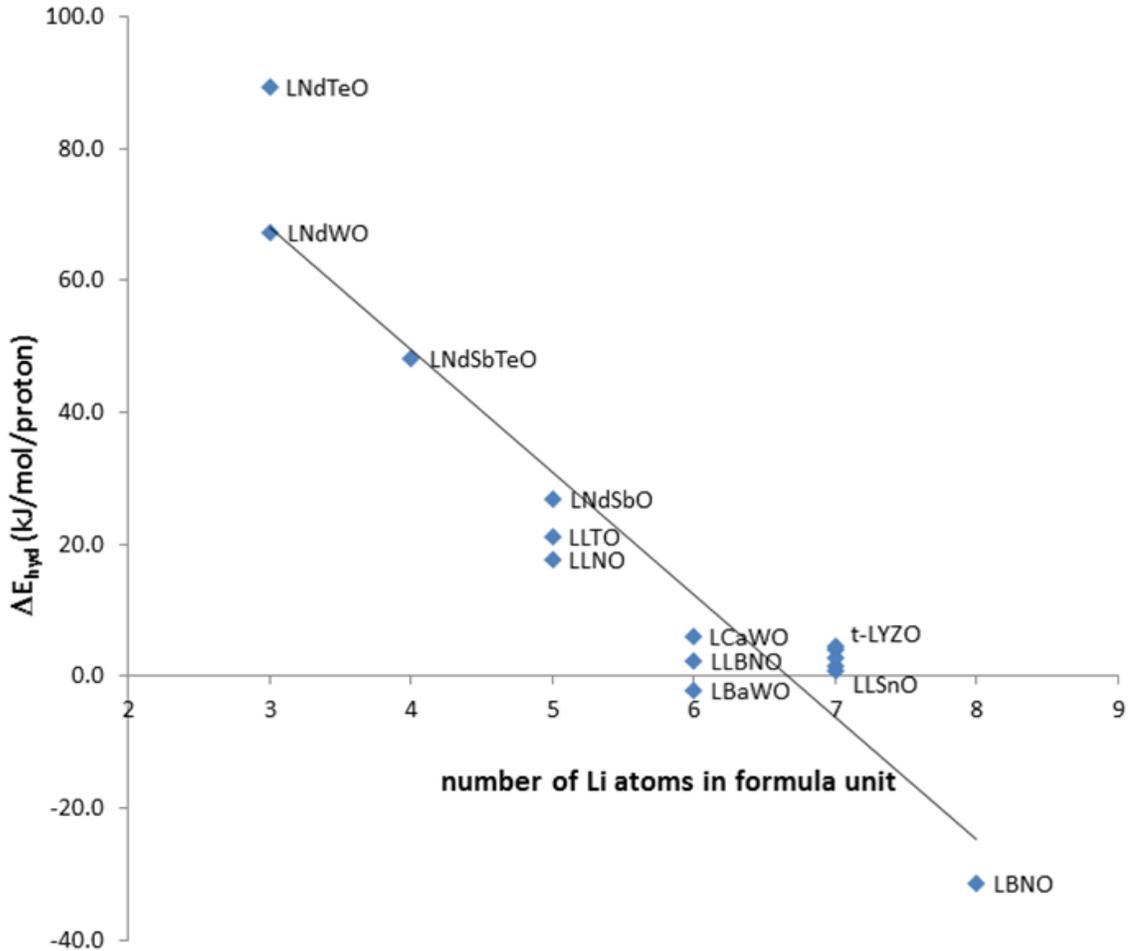

Figure 3. Energy of Reaction (1) as a function of n, the number of lithium ions per formula unit in the $Li_nA_3B_2O_{12}$ compound

The correlation in Figure 3 also suggests that, in terms of the overall correlation, it doesn't matter very much whether one modifies the charge of A or the charge of B, what matters most is simply the overall negative charge on the oxide ion in the Pauling bond valence sense; i.e. (-2 + PBV into the oxide ion). For example, the $Li_6Ba_3W_2O_{12}$ compound, with three divalent A ions and two hexavalent B ions, has nearly the same hydration resistance as the $Li_6La_2BaNb_2O_{12}$ compound, with one trivalent A ion, two divalent A ions and two pentavalent B ions.

Pauling bond valence arguments can clearly differentiate between the +4 ($Li_7$), +5($Li_5$), and +6($Li_3$) compounds. Within a given lithium stoichiometry, one expects that a smaller ionic radius for either the A or B cation would increase the hydration resistance. This is, by and large, true, as the $Te^{6+}$ (ionic radius 70 pm) composition is more resistant that the $W^{6+}$ (ionic radius 74 pm) composition; the LYZO composition ($Y^{3+}$ ionic radius: 104 pm) is slightly more resistant than the LLZO ($La^{3+}$ ionic radius: 117.2); the LCaWO composition ($Ca^{2+}$ radius 114 pm) is more resistant than the LBaWO composition ($Ba^{2+}$ radius 149 pm). While such effects are small relative to changes in bond valence, they could be important in the design of hydration-resistant garnets.

The increase in hydration stability in going from LBNO to LLBNO to LLNO deserves special attention. In an experimental study[17] it was suggested that the extent of $H^+$-$Li^+$ exchange decreased with increasing $x$ for a series of $Li_{5+x}Ba_xLa_{3-x}Nb_2O_{12}$ compounds, in contrast to the correlation shown in Figure 2. The $Li_5La_3Nb_2O_{12}$ compound exchanged nearly 90% of its lithium for $H^+$ in the presence of water while the $Li_6BaLa_2Nb_2O_{12}$ compound exchanged only 20% of its lithium. Given the simplicity of the ideas behind the correlation shown in Figure 3, it is suggested that the experimental data on LBNO-LLNO series be reexamined.  It may be possible, for example that kinetics is playing a role in the measurements reported in Ref. 17.

## Discussion

The calculations indicate that nearly all garnet compositions are stable in water. It is only when the system is exposed to $CO_2$ and LiOH can react to form $Li_2CO_3$, that all the compositions except $Li_3Nd_3Te_2O_{12}$ and $Li_3Nd_3W_2O_{12}$ become unstable in atmospheric conditions. $CO_2$ is more of an issue than water in determining atmospheric stability of these phases.

The calculations also reveal that the total Pauling bond valence at the oxide ion is a good indicator of the hydration resistance: the greater the Pauling bond valence, the greater the hydration resistance.  This also means that the hydration resistance scales inversely with the number of lithium ions in the garnet formula unit, that is, n=3 (e.g. $Li_3Nd_3Te_2O_{12}$) compositions are much more stable against hydration than n=7 (e.g. $Li_7La_3Zr_2O_{12}$) compositions.

Similarly, for a given $n$, the hydration resistance of $Li_nA_mB_2O_{12}$ garnets increases with decreasing ionic radius for the A and B cations.  Thus $Li_3Nd_3Te_2O_{12}$ has greater hydration resistance than $Li_3Nd_3W_2O_{12}$ and $Li_7Y_3Zr_2O_{12}$ has greater hydration resistance than $Li_7La_3Zr_2O_{12}$.

[1] Murugan, R.; Thangadurai, V.; Weppner, W. Fast lithium ion conduction in garnet-type $Li_7La_3Zr_2O_{12}$ *Angew. Chem. Int. Ed.* **2007**, 46, 7778-7781.

[2] Nyman, M.; Alam, T.M.; McIntyre, S.K.; Bleier, G.C.; Ingersoll D. Alternative approach to increasing Li mobility in Li-La-Nb/Ta garnet electrolytes *Chem. Mater.* **2010**, 22, 5401-5410.

[3] Galven, C.; Fourquet, J.L.; Crosnier-Lopez M-P.; Le Berre, F. Instability of the lithium garnet $Li_7La_3Sn_2O_{12}$: $Li^+/H^+$ exchange and structural study *Chem. Mater.* **2011**, 23, 1892-1900.

[4] Galven, C.; Dittmer, J.; Suard, E.; Le Berre, F.; Crosnier-Lopez M-P. Instability of lithium garnets against moisture.  Structural characterization and dynamics of $Li_{7-x}H_xLa_3Sn_2O_{12}$ and $Li_{5-x}H_xLa_3Nb_2O_{12}$ *Chem. Mater.* **2012** 24, 3335-3345.

[5] Wang W.G.; Wang X.P.; Gao Y.X., Yang J.F., Fang Q.F. *Frontiers of Materials Science in China* **2010**, 4, 189-195, 2010.

[6] Kresse, G.; and Hafner, J. Ab initio molecular dynamics for liquid metals. *Phys. Rev. B* **1993,** *47*, 558-561.

[7] Kresse, G.; and Hafner, J. Ab initio molecular-dynamics simulation of the liquid-metal-amorphous-semiconductor transition in germanium. *Phys. Rev. B* **1994,** *49*, 14251-14269.

[8] Kresse, G.; and Furthmueller, J. Efficiency of ab-initio total energy calculations for metals and semiconductors using a plane-wave basis set. *Comp Mater. Sci.* **1996,** *6*, 15-50.

[9] Kresse, G.; and Furthmueller, J. Efficient iterative schemes for ab initio total-energy calculations using a plane-wave basis set" *Phys. Rev. B* **1996**, *54*, 11169-11181.

[10] Perdew, J.P.; Burke K.; Ernzerhof, M. Generalized gradient approximation made simple. *Phys. Rev. Lett.* **1996**, 77, 3865-3868.

[11] Perdew, J.P.; Burke K.; Ernzerhof, M. Erratum: Generalized gradient approximation made simple *Phys. Rev. Lett.* **1997**, *78*, 13956.

[12] Bloechl, P.E.; Projector augmented-wave method. *Phys. Rev. B* **1994**, *50*, 17953-17978.

[13] Kresse, G.; and Jourbert, D. From ultrasoft pseudopotentials to the projector augmented-wave method. *Phys. Rev. B* **1999**, *59*, 1758-1775.

[14] Awaka, J.; Kijima, N.; Hayakawa, H.; Akimoto, J. Synthesis and structure analysis of tetragonal $Li_7La_3Zr_2O_{12}$ with the garnet-related type structure J. of Solid State Chem. **2009**, *182*, 2046-2052.

[15] Jain, A.; Hautier G.; Moore C.J.; Ong S.P.; Fischer C.C.; Mueller T.; Persson K.A.; Ceder G. "A high-throughput infrastructure for density functional theory calculations" Comp Mater. Sci. **2011**, *50*, 2295-2310.

[16] Pauling, L. "The nature of the chemical bond and the structure of molecules and crystals; an introduction to modern structural chemistry" (3rd ed.). Ithaca (NY): Cornell University Press. p. 543-562 ISBN 0-8014-0333-2.

[17] Truong L.; Thangadurai V. "Soft-chemistry of garnet-type $Li_{5+x}Ba_xLa_{3-x}Nb_2O_{12}$ (x=0,0.5,1): reversible $H^+ \leftrightarrow Li^+$ ion-exchange reaction and their X-ray, $^7Li$ MAS, NMR, IR, and AC impedance spectroscopy characterization" Chem. of Mater. 2011, 23, 3970-3977.

Supporting Information: VASP CONTCAR files for the structures reported in Table 1.

c-LLZO
   1.00000000000000
    12.8150559540581472     0.0026051144106524     0.0034338420859114
     0.0026658898138663    13.0771210570781538     0.0023224018967059
     0.0034163921753057     0.0021701040143378    12.8495640433522436
   Zr   O   La   Li
   16   96   24   56
Direct
  0.7517397443229973  0.2520242615756226  0.7503374366495037
  0.2493225408542810  0.7534895225742259  0.2508704168192121
  0.7511425871745987  0.7528761939213026  0.2521188057152878
  0.2508937105595946  0.2513028556222542  0.7486082449204389
  0.2502359706545834  0.7503631110469369  0.7506259622584116
  0.7516286405312449  0.2518347124863028  0.2503671433311067
  0.2502608440386758  0.2520935118062585  0.2494783793780276
  0.7498601133764159  0.7501483403829730  0.7529102211327832
  0.5024135588323392  0.5032369483912481  0.0031807303820865
  0.0001460994766148  0.0009663754335261  0.5011573938802792
  0.5003223074584819  0.0020572348455790  0.5010087458544219
  0.0011932981540119  0.5014956559010555  0.0008648375877814
  0.0009436389805492  0.5020509752339333  0.5014778970925045
  0.5013305338774146  0.0035601022843049  0.0009186062430150
  0.0010162199847431  0.0021004448596588  0.0001018801182945
  0.4995772789134780  0.5009521692358370  0.5018138964340619
  0.5545591392269116  0.3537472285339411  0.0357042821336449
  0.0581022165209487  0.8505658269763499  0.5277370793893126
  0.9455687387810736  0.6529569747744045  0.5278364167121067
  0.4450136475471501  0.1519077717263947  0.0356399248718728
  0.4471689145120170  0.8528365483259351  0.4656526165170185
  0.9441714217433955  0.3516828911241677  0.9750507226341398
  0.0578257744597427  0.1525381517990017  0.9751081334492826
  0.5565848797734686  0.6498296003388061  0.4637899564226254
  0.3457298788723818  0.0294657779695250  0.5551222527328337
  0.8498645921395215  0.5352197481703457  0.0578918706721405
  0.6555075792752597  0.5298165601105778  0.9468365494870501
  0.1514060095705088  0.0350765785401515  0.4442318660205654
  0.8482397671602860  0.4696282202920928  0.4459120924132700
  0.3471879271395665  0.9724002946460449  0.9463086751909288
  0.1532191890140569  0.9688069378742774  0.0554641896820448
  0.6531898731452546  0.4726348214502279  0.5573852897064236
  0.0369243438616616  0.5593091401108533  0.3518438445379393
  0.5360837595058220  0.0587130396059207  0.8487886696452528
  0.5352442863496349  0.9480918586657954  0.6536623834968531
  0.0381431000508920  0.4436886424592755  0.1499085210902533 0.1499085210925330
  0.4637113341457499  0.4542291422270586  0.8516268507174540
  0.9630997881262308  0.9431010498265555  0.3523330052280554
  0.9641921171744789  0.0586105936898221  0.1497576125596538
  0.4637741164429134  0.5623882130087150  0.6504361088172304
  0.9416324698951245  0.1506596680058452  0.4765732397801215
  0.4481012658468834  0.6544797130263801  0.9687834138537285
  0.5576792825990770  0.8552361585875824  0.9683337096172158
  0.0603845290587145  0.3530461935812107  0.4762173287365016
  0.0602739670900643  0.6511669028256356  0.0256935928695435
  0.5539138726078843  0.1513707704609719  0.5365792615524912
  0.4453582420629976  0.3509167177435508  0.5249204609689397
  0.9429198337022767  0.8523793657716704  0.0261538460126436
  0.1538424177363015  0.4673269423667921  0.9450926058504735
  0.6552881656636937  0.9726669545547869  0.4479532109025330
  0.8481906159579831  0.9666586448890028  0.5578356975305589
  0.3464935434900206  0.4722196140016854  0.0553068490395398
  0.6558330663319606  0.0321978960063621  0.0551166687067688
  0.1522526661837476  0.5392081983614377  0.5566261204546458
  0.3448795434938773  0.5341158396933617  0.4473788321316879
  0.8487915621064457  0.0363814283179749  0.9445074922464975
  0.4656530766225412  0.9479984322825569  0.1528583592999195
  0.9643531162933262  0.4445303516756599  0.6502487394710347
  0.9639995523256784  0.5588062664085452  0.8522357994610368
  0.4664109932709169  0.0564064977727854  0.3481302284082897
  0.0373441598111053  0.0579629157746854  0.6497713362059512
  0.5348124970200078  0.5578262140947018  0.1552218796505613
  0.5349068457866074  0.4490312759523651  0.3480091642968527
  0.0373128676446449  0.9436699785263302  0.8515033795979328
  0.3055631269345607  0.7850935054103836  0.0980894161410068
  0.8035381550634656  0.2842481369104338  0.5963131213708246
  0.6951041887369832  0.2849293026964179  0.4028101435075458
  0.1981934588673831  0.7821662146684606  0.9044140743349540
  0.1977982091728094  0.2182279310735574  0.5955034677521256
  0.6951945033165836  0.7195706582742047  0.0998561019227628
  0.8031715534604505  0.7190374149713544  0.9064372894467551
  0.3049142325772093  0.2182220804360723  0.4017487020475112
  0.0980479486127142  0.3046216551934536  0.7862971618292536
  0.6019324068282562  0.8126262442395294  0.2852051603852339
  0.3992515750473457  0.6952091504165677  0.2853910627531090
  0.9043062829078473  0.1978311008658158  0.7861728550279913
  0.6030186375075025  0.1926364458159541  0.2155149644349635
  0.0973076333706329  0.6974416912017204  0.7169190420647010
  0.9026365893164665  0.8039021932233519  0.7163849790753931
  0.4002441196523188  0.3109544893481594  0.2162402825090703
  0.7867493910674894  0.1036157894350194  0.3083305931248032
  0.2802578597060111  0.5987452429246932  0.8031042420759887
  0.2702056358510176  0.4001027996171479  0.6896507782201826
  0.7873973202312211  0.9013368210203778  0.1948503415134244
  0.2154226606013686  0.6045169500464144  0.1937026335942230

```
 0.7260617734312683   0.1012916020827936   0.6966085214492767
 0.7265832825131039   0.9013254312343950   0.8060421535873022
 0.2148093770223531   0.4011125933598664   0.3055859433891256
 0.6958675303335834   0.2209953175335953   0.9038684247291637
 0.1963797732473756   0.7215786318027964   0.4032498775889369
 0.3067163169229734   0.7246288538148810   0.5969470628859126
 0.8052204918252021   0.2206486023574073   0.0971896875841032
 0.8030682720597151   0.7825619389630687   0.4059263406282128
 0.3077005099801920   0.2841314085013087   0.9012928598518340
 0.1985145568908716   0.2828690442085450   0.0955423076437260
 0.6942483032148264   0.7827094786138515   0.5997959031458655
 0.9034626815960453   0.6998719973870896   0.2143715864496785
 0.3993962985641475   0.1897476480825208   0.7160541810796311
 0.6017472882830303   0.3092713447588712   0.7177743439336981
 0.0971318015840618   0.8075631848880552   0.2136796748152240
 0.4003808010918963   0.8066877934403984   0.7866910783996330
 0.9036635703024414   0.3066503678651923   0.2880964640683141
 0.0975665985761060   0.1990200573568272   0.2871269102216827
 0.6001182841540409   0.6907568150874649   0.7869274715537402
 0.2147302433755255   0.9001109556449218   0.6963613869405987
 0.7253312775639708   0.4032208723221427   0.1977726065607140
 0.7259956983958806   0.6009294887923803   0.3030960156996297
 0.2140298413241954   0.1031546158275423   0.8072471123627035
 0.7864364426702509   0.4005706202781179   0.8073890135418795
 0.2751169158862092   0.9036435133319347   0.3054101264400673
 0.2746808404457627   0.1011202413603304   0.1964293260246830
 0.7863469730257783   0.6023243080819243   0.6945050447560049
 0.4998673083453757   0.2532036956668943   0.8735451455734634
 0.0000326766169959   0.7527147514290080   0.3792034981194825
 0.5001402382815331   0.7515805826110217   0.6254166149572313
 0.0009033946280699   0.2517975273886859   0.1220657736433523
 0.2494105564650134   0.8769253264256345   0.4996518071640116
 0.7515299738793551   0.3777675816080985   0.0020760219337683
 0.7505402882873532   0.6258650915908806   0.5004500554116625
 0.2498544129294922   0.1269694651056723   0.0006205067733716
 0.8805713430859030   0.5016814943407438   0.2511783581338943
 0.3688068019173946   0.0015712833734904   0.7507812366745913
 0.6287368188825873   0.4999219900846987   0.7505104878698109
 0.1218379869722883   0.0021362982655548   0.2505228738853768
 0.0013011697529495   0.2515808714676673   0.6206360545200226
 0.5006322357298354   0.7538703839974896   0.1269479093337579
 0.0001003771185482   0.7513721430682141   0.8808667748775263
 0.5004657137430670   0.2515980939062259   0.3749536128283616
 0.2495304057092846   0.6277170482249536   0.0013227696638016
 0.7509290581178196   0.1265410995392358   0.5026965943811962
 0.7512821164090127   0.8779687761839471  -0.0003528003315135
 0.2507685187083754   0.3763570856026879   0.5012703579424347
 0.6326610262231838   0.0028812447887416   0.2509200876133348
 0.1234157561215384   0.5016174661476238   0.7505795725226994
 0.8784060966211849   0.0014002559028004   0.7508170148377399
 0.3687956256029205   0.5029295395192408   0.2515764111240408
 0.3464146861691448   0.6794126035525723   0.4385708898233590
 0.6558738601383394   0.8275651161613996   0.4448272436407657
 0.3087860486390283   0.4306597329868193  -0.0957194767499639
 0.1241794401034941   0.0017784286948000   0.7511824165708971
 0.5800934078054875   0.5882964585546588   0.3077331891484253
 0.8768536404149709   0.5011732216977393   0.7517348834213028
 0.4206378172903277  -0.0832035195967331   0.3042833391462530
 0.5790688836470509   0.0921727471752941   0.6940061997293812
 0.5059355331786768   0.7584440394992130   0.8766279839697838
-0.0516912686291439   0.1700212767002950   0.3342423304795769
 0.8537771602920738   0.3237608418515847   0.4360204620526426
 0.1421708643250084   0.8233549797363530   0.0658804760957336
 0.1464724215339132   0.1809185430890600   0.4350576222081819
 0.3877887656005313   0.4659357734688444   0.7277658903663724
 0.8770482269751516   0.0030815725918653   0.2523483026246462
 0.1244503543198269   0.5039318505177524   0.2499601707748239
-0.0558394815976581   0.6650566608490612   0.6724563791592013
 0.5023530242592673   0.2512715302876518   0.1276396345673074
 0.1878124570352612   0.9261837956957477   0.9079986276022856
 0.6873198612670094   0.4326668358762244   0.3773901131152181
 0.3124660901292417   0.0704455729099429   0.3783562635891520
 0.8133491986063143   0.5752048893051925   0.9096179758991662
 0.8137133708417489   0.4279403677127060   0.5943721764394556
 0.3120419859225159   0.9326952466355640   0.1145137663538902
 0.6886386622036385   0.5716414240029071   0.1294080308089508
 0.1871113001838091   0.0751795616330543   0.5920846071600780
 0.9442835132150998   0.8401520794115087   0.1698530755073560
 0.4955034383804161   0.2597128279732623   0.6284168577635356
 0.0557517233899037   0.3379021538072456   0.3325544336205736
 0.5838659279856243   0.9237342619108301   0.8007651264191668
 0.4280075955338181   0.6101399999863559   0.8192608778595235
 0.0507109600538959   0.6723134348175700   0.1672249090763690
 0.8620722764904916   0.1827561274650378   0.9347873785213009
 0.3490698030742116   0.3261200655352400   0.0616836551166510
 0.8603638356584777   0.8198671874577249   0.5682906298973971
 0.1375714893889723   0.3206445548238057   0.9341607764492386
 0.1249910430330553   0.6858348627822660   0.5670609386105836
 0.6590975872199327   0.1762039015829358   0.0567848618935120
 0.3050105031011122   0.5805252097871709   0.5973893524355591
 0.8144574680517046   0.0765723355099751   0.0920239501886314
 0.1883855323773232   0.4259349543132110   0.0926029735199851
 0.6891186305733310   0.9310602070234867   0.6157990060087040
 0.6897698663245566   0.0740998364502729   0.8850799139800340
 0.1870177800780207   0.5788649705890437   0.4109566562887495
```

```
  0.8125846569232059  0.9262147368719124  0.4116530327393028
  0.5018898712924772  0.7526126736302257  0.3735752547676901
  0.0549168953258515  0.1666404989690396  0.8309941589305794
  0.4197049385171620  0.0862105797903232  0.1965122810227682
  0.0536778957011066  0.8375970862469020  0.6723330217446631
  0.5796967806660831  0.4171648789505315  0.1983406323140808
  0.9478416541629855  0.3359868723433987  0.8312558833737872
  0.6570357406404120  0.3278578729016057  0.5562334202349772
  0.6586562560287677  0.6739074072588361  0.9481526713187259
  0.8542532173243228  0.6814479220278004  0.0666907489760427
  0.3442388668242827  0.1736750672946839  0.5553660587315480
  0.3445791924180083  0.8270810810625635  0.9433427043269922
c-LYZO
   1.00000000000000
     12.5305748819692706     0.0009955001800755     0.0100886891688883
      0.0011525972559614    12.6968921428997135    -0.0031510347403627
      0.0102917395006783    -0.0034534747661083    12.5891676255211475
   Zr   O   Y   Li
    16   96   24   56
Direct
  0.7496329748317957  0.2480188312398433  0.7524575246923315
  0.2501479412687987  0.7539963446856627  0.2509406689394978
  0.7512822190874126  0.7517514871760486  0.2538413535032102
  0.2531613134935922  0.2494584442565278  0.7487205160114594
  0.2493779460821371  0.7474762733055424  0.7496340472079881
  0.7522672229381117  0.2489950234795003  0.2530903995562775
  0.2553963098375677  0.2520672806881105  0.2462665284996847
  0.7471276623564577  0.7499184945801501  0.7542905852407531
  0.5019629418122200  0.5051989745867953  0.0013600248249245
 -0.0021108054847595  0.0003510257032769  0.5004200271622451
  0.4981434939235071  0.0026467277822908  0.4998061640388939
  0.0019362897775369  0.4980807664001978  0.0017460599398131
 -0.0033969020526472  0.4961668832560996  0.5019125152487643
  0.5038039746285069 -0.0000558221533135 -0.0014064477478996
  0.0039585262246771 -0.0011711160369961 -0.0020180773277659
  0.4982641489995129  0.5029303327205492  0.5023823978288917
  0.5536858761747383  0.3483036077763098  0.0327263710881806
  0.0614625037740807  0.8478156227084046  0.5294522848126235
  0.9471197234330382  0.6545719662604002  0.5248784829999488
  0.4368096923866124  0.1499892680084462  0.0380251766695999
  0.4434545695831852  0.8463457910550732  0.4696512671467209
  0.9390117601085218  0.3455992280060648  0.9734460656946200
  0.0561201654199377  0.1569607659630534  0.9756631132645983
  0.5662142332539211  0.6519020173232548  0.4635665253758640
  0.3394096472053861  0.0257655293735822  0.5520098509075954
  0.8503772055973301  0.5324027378921827  0.0653383405584656
  0.6611403448984410  0.5252300846606475  0.9504863035128612
  0.1492877060965575  0.0353314246030946  0.4373224453350774
  0.8384674207744560  0.4726703313495975  0.4499301728849586
  0.3508733299819476  0.9686461534303097  0.9352723800047905
  0.1608041871897464  0.9713467294818512  0.0520878352318418
  0.6485039580915005  0.4703540382696096  0.5698894848911356
  0.0313765858668180  0.5500097885443549  0.3472117806959477
  0.5371864422120762  0.0708892918691528  0.8508092765633493
  0.5278007314797546  0.9482530439920550  0.6549935047088667
  0.0385206730546701  0.4317067716687927  0.1521664746253735
  0.4711656484130807  0.4525385894280108  0.8465790439133992
  0.9641227480504925  0.9323504779969166  0.3507424585327595
  0.9717524807148454  0.0548466966685573  0.1530603952229357
  0.4612228905498494  0.5723982578066199  0.6504870075878029
  0.9311604368547917  0.1521260232927831  0.4820614148674151
  0.4433487555803393  0.6603338891513738  0.9706498626475255
  0.5666400786620193  0.8472265065235407  0.9825192628464372
  0.0585027807580254  0.3416042801984313  0.4703105071280355
  0.0617785644140210  0.6530300146876369  0.0236256341269599
  0.5559456068575037  0.1580130418859167  0.5287793586247692
  0.4365991474266112  0.3491067568112499  0.5184393163082357
  0.9406510587772248  0.8461602830282181  0.0260308137178268
  0.1555969302525546  0.4655475170910480  0.9412038749160975
  0.6555521418888939  0.9700611983043799  0.4486057502903051
  0.8441410058546062  0.9662887820654575  0.5617659186579015
  0.3437751096338541  0.4740942787245393  0.0556338051609956
  0.6611350223413524  0.0303533952699153  0.0567957315549988
  0.1507333822690560  0.5318511993874699  0.5591643010562029
  0.3403573116034317  0.5330666604297282  0.4453155071090210
  0.8501736028692871  0.0353020194599949  0.9417734181629399
  0.4675880542487904  0.9447982089072837  0.1540465503886660
  0.9665389454686839  0.4320714797162792  0.6542768830055430
  0.9669318221253435  0.5569227362851916  0.8490983648758637
  0.4696923511098958  0.0583038251512580  0.3441639152445739
  0.0354600695453711  0.0607112749465677  0.6522921058487551
  0.5317115907944850  0.5614639492393436  0.1561103609648166
  0.5311353912698745  0.4489296623183893  0.3456954325573455
  0.0336667011766927  0.9345669819565940  0.8453187660350814
  0.3100862020446808  0.7838047331496698  0.0972451739249790
  0.7971350194217195  0.2789687036510305  0.5950280712116269
  0.6881372524082200  0.2855105540293171  0.4038079621809496
  0.1976900349524719  0.7774353963654678  0.9070690479039487
  0.1950295489828889  0.2197741595914030  0.5922374776234870
  0.6972965003139069  0.7189982741973882  0.0998774186940325
  0.8079303068331898  0.7177899612876011  0.9087011663548304
  0.3075689336912029  0.2188802027369627  0.4013589490050404
  0.0974453189240940  0.3058999540174277  0.7830821581456566
  0.5975540527803235  0.8155210269646088  0.2824488123993263
  0.4008512845500159  0.6887051379251934  0.2852038533565075
```

```
 0.9061144509322524   0.1933271681159583   0.7798459049712325
 0.6005123185394546   0.1874689911847359   0.2160024475103903
 0.0931992629149718   0.6921870856857438   0.7208379744294154
 0.9026482226293021   0.8060700319349303   0.7181238153987336
 0.4060615383448771   0.3171574460417543   0.2184010583048675
 0.7753050683198891   0.0969508917066980   0.3177115832225287
 0.2798602363060973   0.5911235762994560   0.8056967981766521
 0.2734384931330566   0.4014518616114819   0.6847763662902534
 0.7856082600068495   0.9057935098326171   0.1935956674009212
 0.2171120128668012   0.6030547727852137   0.1856029947222678
 0.7180384731925495   0.0921961407186242   0.6961833562425742
 0.7293482389900532   0.9028723204201855   0.8170822306343553
 0.2198038976282622   0.4073105163326461   0.3075910839395171
 0.6967675576520820   0.2260099861775598   0.9118562032188992
 0.1874975703690825   0.7200693087930159   0.4033500557424978
 0.3021755116332293   0.7259749847075679   0.5913982825233202
 0.8133666418755852   0.2185539036313289   0.0999622508240546
 0.8024197963874690   0.7770233000616906   0.4127758899617560
 0.3188387083808112   0.2806886871574839   0.9005793962015316
 0.2008318478780135   0.2771565590339909   0.0889477458971121
 0.6837661693623743   0.7805525830378257   0.6015505542343255
 0.9087984848513356   0.6992170627776344   0.2220695558973340
 0.4011704230993551   0.1796715946412663   0.7113769069194534
 0.5931370701324260   0.3007640666172982   0.7242501253855489
 0.0976232218948654   0.8150412774167246   0.2138047822398844
 0.4053881495718852   0.7994690808075938   0.7803930071022844
 0.9021084569430966   0.3175377045883092   0.2894656439065254
 0.0966554781563917   0.2035227992676354   0.2810009571642481
 0.5996154751075697   0.6789146287605209   0.7902530763872510
 0.2210947744709524   0.9036136912126617   0.6960354053951312
 0.7231542139164772   0.4029689055821849   0.1914413620502519
 0.7283757635712089   0.5942393420036632   0.3037304978528911
 0.2153564864030086   0.1000336270295707   0.8158502502950936
 0.7801222827632578   0.4039726838215087   0.8061354278924253
 0.2755031867547463   0.9058553839726508   0.3135002445861089
 0.2725382116612833   0.0946697026798389   0.1963353067637447
 0.7868197191810350   0.6006911037119015   0.6858863113513811
 0.5026050191308954   0.2517925958157345   0.8801963323721527
 0.0001558156125519   0.7513486178042208   0.3795678274117543
 0.4983988396849746   0.7537501692269317   0.6249734430959267
-0.0001240359731522   0.2469933606184781   0.1265471194079152
 0.2472666866035131   0.8756975333339292   0.5003223089961721
 0.7522011973425130   0.3747912173573920   0.0045990883572519
 0.7490313362936242   0.6243234756890091   0.5028590208492252
 0.2521058889955839   0.1256151344508351  -0.0009595765366078
 0.8785085550674347   0.4999685128406687   0.2503072549747370
 0.3748059435172975  -0.0019301900362324   0.7505728686259291
 0.6261770363483173   0.4989537015943747   0.7526137331796692
 0.1214325166931572   0.0014313729157566   0.2518367739028565
 0.0015155680945195   0.2480572658258976   0.6199345997733934
 0.5015848273031037   0.7529438468528072   0.1267123302666455
-0.0013372039027071   0.7481774973310836   0.8808723943608274
 0.5003118846820043   0.2517663680948707   0.3745319970434829
 0.2491721824303079   0.6265220893428723   0.0016886675108147
 0.7482452060510163   0.1247588625457366   0.5021766557483791
 0.7529046646296490   0.8745489730903715   0.0018675167074536
 0.2505765929804565   0.3756412674885281   0.5003976052400515
 0.6322715697645996   0.0015132229583302   0.2502753134088428
 0.1241889130636240   0.4975903881328786   0.7503844643834725
 0.8758048953387156  -0.0017021094771099   0.7515272438581657
 0.3693201998281882   0.5040276032286878   0.2521021569603401
 0.3419420747320500   0.6818953340933774   0.4419725421068116
 0.6496615565032751   0.8197560982932652   0.4469554649226943
 0.3063300682864984   0.4311164773699508  -0.0975393393857199
 0.1255214506611288   0.0017425906831549   0.7512220629080476
 0.5781887220344218   0.5899554242913273   0.3049152495308031
 0.8760895439330977   0.5017465746393817   0.7491927470312866
 0.4252294067935730  -0.0866343049695788   0.3027823430999405
 0.5612558364398949   0.1346143616405455   0.6829653026444332
 0.5177787754503703   0.7714097461922208   0.8718414756834013
-0.0222175619378589   0.2181703162419516   0.3658038656460947
 0.8379262010905064   0.3250611129867066   0.4451750575046622
 0.1447795426237971   0.8211334966987326   0.0606741570364662
 0.1544674626823228   0.1848874830849827   0.4367619635259070
 0.3782721686962423   0.4842789059505600   0.7366370548724527
 0.8791244653040516   0.0109389837989790   0.2595389015929964
 0.1260531066295091   0.5074922580793887   0.2429062504932746
-0.0560359806687723   0.6642421216861232   0.6728691972954787
 0.5011065473624141   0.2480503151763645   0.1283565784637619
 0.1968621426185808   0.9269282035364362   0.9025699671757361
 0.6869354602561007   0.4381214364435163   0.3672876550930356
 0.3100096302297162   0.0638258070757905   0.3733549205380944
 0.8110233408704679   0.5700707707221947   0.9116311112022853
 0.8017539671632576   0.4278391766612826   0.6014561960413980
 0.3107131313885423   0.9361043900670365   0.1181236805619093
 0.6921632264120362   0.5650844248348662   0.1286789194375058
 0.1907386981392495   0.0725508118968893   0.5899982906760606
 0.9440999318594627   0.8498185758700356   0.1764665923669057
 0.4727128088554997   0.2887675727396896   0.6424795188213979
 0.0649611808466373   0.3636488865746536   0.3192942750935086
 0.6206988566827273   0.9833535729770985   0.7647725478689077
 0.4355274299469242   0.6278372522233344   0.8164062855654036
 0.0507186982301116   0.6772650105977810   0.1665239921226803
 0.8813313129589452   0.1887937362765260   0.9374564691881832
 0.3539176680621113   0.3230471680753753   0.0554354506163471
```

```
   0.8673885337703156   0.8145968486609333   0.5651249393564464
   0.1233116578710527   0.3119724371349320   0.9374125165932142
   0.1196004010382391   0.6866015798851384   0.5641900779562609
   0.6604294953228990   0.1781780169957738   0.0616141514879491
   0.3005297726340662   0.5774048827715602   0.5949762437268241
   0.8153141301153232   0.0700828146393007   0.0932889559340281
   0.1938585747366829   0.4239741575981766   0.0887888088557718
   0.6925435542850010   0.9304009684063185   0.6003340730086466
   0.7008338392454091   0.0763724925397132   0.9060374304082527
   0.1862688688149022   0.5719596400714981   0.4087069273659165
   0.8068455362544337   0.9256757569522363   0.4103934522559245
   0.5005963892540732   0.7499781321151493   0.3734638237177881
   0.0556830831411339   0.1630703649041889   0.8265967779074538
   0.4222835785619561   0.0857898874726795   0.1950165976499984
   0.0523489696785769   0.8412802615496235   0.6783196867410102
   0.5727205829230047   0.4118785143577813   0.1961074827609844
   0.9456566328729660   0.3424242502544380   0.8228129799215320
   0.6446980802536161   0.3184705714322353   0.5567374550290932
   0.6600813735267010   0.6739584492382950   0.9529531452998238
   0.8636552763468979   0.6852673249130089   0.0671989161696502
   0.3484820749148623   0.1748235997405862   0.5514810150738645
   0.3497428509780088   0.8178126101385708   0.9457928825450225
HBLNO
   1.00000000000000
    12.9675545547608433    -0.0132652679550069     0.0176785561023780
    -0.0131016537779045    12.9650990082923467    -0.0283568569334681
     0.0184951941995765    -0.0281061393168055    12.8914702631027822
   Nb    O    La    Ba    H
   16   96    16     8   48
Direct
   0.7606802534188808   0.2398083917339459   0.7648581896128748
   0.2612730951739911   0.7423046758923323   0.2492428744598347
   0.7594614754692948   0.7705808851427499   0.2530838379943155
   0.2638555930675412   0.2488061264348397   0.7418388478950032
   0.2774848552186316   0.7417844242155197   0.7647910130455592
   0.7587286179022568   0.2614839226314891   0.2586804513187568
   0.2489797001158220   0.2610487076384131   0.2502927349886579
   0.7745168192034541   0.7537321523908159   0.7499606474818703
   0.5090469849450521   0.5058448884936694   0.0024879299117745
  -0.0136841070763782   0.0135805814257225   0.4923261416499753
   0.4996681099132920   0.0086161554135329   0.5218387020200629
   0.0167847467948958   0.5108587723774352  -0.0135039301704182
  -0.0107593075888692   0.5112128612258029   0.5124363483603324
   0.5049836569730348   0.0164345013521243   0.0192629033285488
   0.0102852169863577   0.0077283817910868  -0.0206737662910553
   0.5050800489667999   0.4941801188208207   0.5001171558344739
   0.5509977881035313   0.3681679824186188   0.0410996524936035
   0.0549682142158370   0.8671241785152958   0.5323894714751186
   0.9368601898080543   0.6444584819323322   0.5302318045925516
   0.4372769059311598   0.1435516419620252   0.0260867918845595
   0.4483515748206549   0.8637134633172914   0.4634552837774646
   0.9325416432300305   0.3646896757334537   0.9693243079326791
   0.0640257024784242   0.1374605823170066   0.9650032323801695
   0.5443888834016981   0.6407962861787475   0.4557221440598469
   0.3597141637731310   0.0240350678906094   0.5645115113017064
   0.8590725205319708   0.5362429427733692   0.0547253605358995
   0.6406353813484783   0.5289904319532465   0.9427863262129004
   0.1434570476363018   0.0290538342017870   0.4414974104424092
   0.8528103683586932   0.4799457077299119   0.4408868594847837
   0.3589486106303956   0.9710258053178340   0.9502424011195849
   0.1504033522364618   0.9636061234443052   0.0385268592941614
   0.6375501318018587   0.4914305752532078   0.5586052966960430
   0.0296443742316963   0.5474261074205222   0.3509534385210222
   0.5304113407667460   0.0581477014722074   0.8552951055919342
   0.5362486352170984   0.9550919680810027   0.6511540797021613
   0.0297899930179491   0.4511613669363724   0.1431701282989376
   0.4635346009628578   0.4680649983204280   0.8541126502479519
   0.9704600701999399   0.9572307510469855   0.3454921743702676
   0.9734638792434445   0.0379831783070020   0.1361090836727857
   0.4655737656727796   0.5471035320488790   0.6498980095409873
   0.9481352921948909   0.1451025041730929   0.4625611956397100
   0.4382123519210203   0.6328567926733565   0.9963645764951767
   0.5520225081435063   0.8686520616928207   0.9607897302206138
   0.0475629621403975   0.3595357674103182   0.4558714420243996
   0.0598242742304239   0.6400849754072979   0.0323127504031453
   0.5654287150563900   0.1372999090043428   0.5278512900514960
   0.4385421167361608   0.3577066231933987   0.5421816650378621
   0.9403876202025371   0.8647398422995562   0.0245566738598943
   0.1503289943001231   0.4721252990739118   0.9420758509407360
   0.6494211694882361   0.9691702585501035   0.4506997620000487
   0.8589931152240714   0.9823293376563523   0.5532929228291540
   0.3518800617734823   0.4678056839294154   0.0617385527493362
   0.6472168756456962   0.0306949210313468   0.0556352804318717979
   0.1358275149027589   0.5264089241376066   0.5441834991511342
   0.3500425261883821   0.5340319849294173   0.4576912092053709
   0.8547368462572623   0.0371564889446029   0.9393898516982222
   0.4720325186241702   0.9684761281828040   0.1503256624215583
   0.9666301999972029   0.4591759108242393   0.6467579063165269
   0.9764419485052503   0.5459319606262091   0.8521632504738338
   0.4727539153689941   0.0425088118338890   0.3587484473295796
   0.0332364668043996   0.0529297669306705   0.6410493949433357
   0.5405199710352586   0.5427175040207933   0.1587219560847742
   0.5101308605556046   0.4465779366705578   0.3658180730250641
   0.0216416560298065   0.9539242935916193   0.8470423307607080
   0.2989978685212369   0.7821614321354575   0.1162115928986757
```

```
 0.8004683102623183   0.2841285988076662   0.6088517395831231
 0.7031221524108426   0.3002239207988980   0.4038504100775067
 0.2101070551070071   0.7733524648356925   0.8958217693112183
 0.2094231950059342   0.2025255501988896   0.5969916959078977
 0.7014221842999866   0.7306348420371588   0.0955794350305348
 0.8018569949374450   0.7428902162094391   0.8888539283802898
 0.2890232690752955   0.2238892373021685   0.3874879038102059
 0.1105121724059449   0.2968465383227313   0.7926928369581363
 0.6201059917785350   0.8228977462981850   0.2787580852067402
 0.3914425059943337   0.6902428037423584   0.2918136892451756
 0.8948235736583842   0.1953943796282707   0.7880039901099402
 0.6088345828228598   0.2031680960752121   0.2101686283561647
 0.1239337642510542   0.6867985008965236   0.7111857036957924
 0.8950286361536837   0.8158474376314688   0.7095978105792041
 0.3926057916961576   0.3089778097721956   0.2083894332122265
 0.7566476151701772   0.1224425571789570   0.3197321738976784
 0.2965669035958813   0.6010610558340973   0.8045941462536914
 0.2496290312103267   0.3849448246066871   0.6768426899991850
 0.7919064861304972   0.9029047358287298   0.1903231248700309
 0.2218461171682027   0.6059946098020005   0.2018730040677722
 0.7273055034708923   0.1119756808544945   0.7021873115551314
 0.7150076847805580   0.9001029197374047   0.7875266672833550
 0.2227923081200965   0.3959312701482830   0.2967096178402726
 0.7060574589615353   0.2371098956827922   0.8989460878186958
 0.1992954298422750   0.7088203974890450   0.4087644341659172
 0.3059702335004179   0.7371303713693461   0.6152952295610781
 0.7934714542602439   0.2124967090315029   0.1074825896915510
 0.7981898497987855   0.7833239719126804   0.3925684990744735
 0.2956966989121813   0.2996961317833124   0.8930807336031957
 0.2012972368163293   0.2630822900794081   0.1088679123748195
 0.7108876998126316   0.7821175072796607   0.6007375870973128
 0.8778101932739754   0.6881807248553206   0.2140487079582542
 0.3980167394216669   0.2193094444949469   0.7089049865821418
 0.6169048139182416   0.3143971289355471   0.7147351428019522
 0.1051056924675602   0.8003289804777552   0.2014367579709689
 0.4013947533307709   0.8146124581013304   0.7926364277448439
 0.8908489042928784   0.3073115001336916   0.2832945700021262
 0.1106397224160273   0.1816090878455134   0.2805059860429811
 0.6221577132982100   0.6895528789149258   0.7801727858041332
 0.2185890328659987   0.8986579753535849   0.7116184861620438
 0.7258218049397612   0.3953904055603267   0.1960725717736137
 0.7124068685752538   0.6112218746913376   0.3017659219621923
 0.2424158360145127   0.1146820422483465   0.8060200319093216
 0.7836453310337271   0.4038750075009089   0.8043531342363240
 0.2533124850339291   0.8798358253704540   0.3205086295218343
 0.2851773095051840   0.0951657594557751   0.1922025237663239
 0.7895972198980928   0.6053634525050389   0.6945332835180545
 0.2502035771556705   0.8679148476987344   0.5074380308953418
 0.7424875640769832   0.3854735685296256   0.0096693451142525
 0.7548176434944622   0.6274215931859000   0.4880419297718949
 0.2544903768529605   0.1330872996850890  -0.0051003754331637
 0.8740758693667277   0.5029178688149951   0.2512902216847719
 0.3791961227723722  -0.0007492226810088   0.7537669984911934
 0.6280386124108913   0.5013567898896099   0.7507968868146457
 0.1387903140539148  -0.0138167346939978   0.2406389611399056
 0.2521435477339277   0.6290010330659880   0.0056000089784323
 0.7468367342165466   0.1315773015892554   0.5026523553056774
 0.7457705671388614   0.8830501558202765  -0.0014708593183963
 0.2363072344392662   0.3766172401955817   0.4922883259465702
 0.6326025441920955   0.0038801803131607   0.2508409494518022
 0.1293857075746411   0.5002196396106371   0.7464820100282914
 0.8761379309365555   0.0059522162404944   0.7421418823797726
 0.3692069322982843   0.5086526317132053   0.2539305260263046
 0.5021112807249309   0.2600230782606670   0.8783553345758508
 0.0000889065811565   0.7493812596463726   0.3684629297006572
 0.5061976109312358   0.7543016966042156   0.6261480132252057
-0.0001339620300829   0.2490112093174999   0.1225496148994254
 0.0039255212308319   0.2584838125994219   0.6265662495638655
 0.5001438899569925   0.7604288736967766   0.1345436417124154
 0.0114273488169559   0.7500892165077554   0.8684652074940716
 0.4938355917182615   0.2445834302125361   0.3757510032079421
 0.7133039860227471   0.6652987332182873   0.0596063827400709
 0.2527391573178536   0.1463695593117711   0.5693967493474481
 0.1191053659863680   0.6853606530774060   0.6357466094822024
 0.5788654434545426   0.2378164965067139   0.1499786326957342
 0.1854806814475128   0.6426355043767987   0.4432110474680388
 0.5964250556370829   0.6991404747082365   0.8507232567109915
 0.0757391936683532   0.1867475113227762   0.3477939707098130
 0.6423023878414570   0.5869461024566984   0.2893387088554926
 0.7451525606054418   0.1589083153442988   0.0834441336120221
 0.8573111671905803   0.5701339409403973   0.6913917998042699
 0.3533083551320121   0.0615597396283974   0.1902953920227352
 0.9040340744947009   0.9648525627034198   0.3099179737709669
 0.4009549959030780   0.5068830200970316   0.8322505254751857
 0.9395885870715860   0.3277902663420578   0.9033576033350382
 0.4763086558116338   0.8405264980865345   0.3967962383081409
 0.3349156603197952   0.8995215795727769   0.9555228893348781
 0.9435833711825742   0.8379476371198401   0.0952527691004210
 0.4303508965069501   0.3221197304182505   0.6096961025147171
 0.3172008058316205   0.4008052425665188   0.0565181546617635
 0.4139974045056178   0.0059172748929581   0.3289942667406284
 0.8233380935879826   0.1052350542324413   0.9494177913343822
 0.3209737940136357   0.5980687988851100   0.4860202801358214
 0.3961345119302701   0.5484202086754565   0.6802762719418852
 0.9107883215401823   0.0016495758933034   0.1595881793269044
```

```
    0.8570582283771861   0.6091994661960900   0.0764805129483877
    0.2488478429373732   0.3556698425617424   0.9156880984660405
    0.8495658755404157   0.4400749649862549   0.7943683123344120
    0.7904620071205611   0.3546192089621746   0.5846309551151613
    0.4298337680936467   0.3012919653948491   0.1423573702669394
    0.7311936566398778   0.3666563432530568   0.4282810885624578
    0.0811232805543955   0.2538263096981425   0.8476619462606224
    0.6566859194609165   0.9239738169572502   0.7441677385195775
    0.7473501391352146   0.8389576954353709   0.5665621334767502
    0.0790555510163480   0.7554435247458184   0.1455572007081506
    0.5865594926598839   0.2820361564582732   0.6533976575901917
    0.1614078072120822   0.9214656288040769   0.7558057939827425
    0.1018919866046737   0.0372942944468062   0.6673674402229887
    0.5892989977290726   0.4898990498235893   0.1835302290720150
    0.0979930748123251   0.5322461135305655   0.3228210767329657
    0.5950551029506570   0.0490859709837386   0.8175140765483030
    0.5041607164472051   0.6702016224874997   0.3979577703251680
    0.1836003263566921   0.0933956137991909   0.4402578217828509
    0.0962298374518708   0.4697162572162398   0.1740523780077969
    0.0182513941939167   0.8314249807235946   0.5884957773761323
    0.1815848957887760   0.8965749229082759   0.0202168714376593
    0.5096851121517316   0.8438125112115569   0.9021356151678646
    0.0052143929980895   0.3332435849757000   0.3985285011470027
    0.6752150414719313   0.8987161042438652   0.4414355636673344
HBNO
    1.00000000000000
    13.4012792382624024    -0.0132363621207287     0.0686013211702257
    -0.0124653488874593    13.3653223161803574     0.0189482087274228
    0.0678363583765280     0.0192632975516950     13.3600507005812208
   Nb    O    Ba    H
   16    96    24    64
Direct
   -0.0036501412679643   0.0022153407972919  -0.0223075740251871
    0.5158312751117224   0.4713597413724135   0.5040332375052238
    0.5127856041776772  -0.0172759533371365   0.4929840861034853
   -0.0007369325319450   0.4935623892657325  -0.0079636621794283
    0.5005678881758339   0.5015025265005730  -0.0233148685891227
    0.0142541651231065   0.0163994439407582   0.5057251756107254
    0.0147128912254284   0.5109238308122689   0.4877624721382842
    0.4920262047648307  -0.0026181987301007   0.0066244636441548
    0.2221799833375014   0.2527642004734952   0.2578662359235344
    0.7548208109446126   0.7813045850955048   0.7344972799931977
    0.2684271338907366   0.7508648445463086   0.7685212125819967
    0.7106117271833123   0.2542925217030524   0.2401709691783169
    0.7769309832319178   0.7570876388102039   0.2167658177732337
    0.2670971821989917   0.2656722773637346   0.7604158456007944
    0.7674816287629643   0.2784215811038875   0.7339741786343502
    0.2726906879349665   0.7399035426070052   0.2662503215783290
    0.1001218446096289   0.1996856146196725   0.2825499681224400
    0.6149208003274428   0.7096865398284917   0.7763375939996103
    0.2741483135482181   0.1185294985931064   0.2009381552765954
    0.7856837780955080   0.6302394853148668   0.6844802032900981
    0.1896672397343621   0.2914401930785798   0.1153891438544024
    0.6984140069396799   0.7909992568404077   0.5937299357450581
    0.5906316321063421   0.1905901506963364   0.2019146982106954
    0.1158656092064560   0.7019277413281598   0.7311473085319955
    0.7744746637622143   0.1096419487941634   0.2898298795751819
    0.2782447479114745   0.6080892240782222   0.8213833822147204
    0.6750636032917768   0.2713047504593062   0.3733000765425287
    0.1940594598225899   0.7784248403522666   0.9108233347998882
    0.4023985561017921   0.7055581258732233   0.2899026254644820
    0.8994084366041073   0.2203152994779570   0.7880625278185089
    0.2225810571333356   0.6172267346392785   0.2267973564420166
    0.7292404508319068   0.1278092593895099   0.6910741719736428
    0.3244386066877085   0.7914287945557816   0.1264463082044477
    0.8189735847299758   0.2890984495119752   0.6063961322887104
    0.9125934453836023   0.6965037135375876   0.1996260698534988
    0.4038546862112556   0.2112381209274471   0.7015421364029982
    0.7323606631915075   0.6150993795152646   0.2849846296342667
    0.2268483176211426   0.1142818519681457   0.7979187407983432
    0.8109911811199071   0.7870895919270590   0.3744052748545571
    0.3213315852946819   0.2731473340694285   0.8852251633362470
    0.4614670951612347   0.3464154923767359   0.5368222885462082
    0.9526746687036807   0.8633939590721574   0.0287348518699249
    0.3728907660177368   0.5241046710420018   0.4475499032721608
    0.8688296598251209   0.0411380660118087   0.9434560029104442
    0.5357203733999484   0.4336626300471099   0.3572980848363619
    0.0349466038346437   0.9465432894454123   0.8583837556730189
    0.4444352294278006   0.8462107710266611   0.4740472162288894
    0.9397998125587360   0.3653919688730816   0.9836665713027589
    0.3680175146188052   0.0248609955609575   0.5600182717516011
    0.8498337599199713   0.5334863195893126   0.0492283328485240
    0.5311034416931322   0.9302355577293698   0.6435792096112306
    0.0457402224720841   0.4378064554948285   0.1360851679296949
    0.9544331503676113   0.6336558762659737   0.4932029392355553
    0.4633892395488203   0.1385691249732004   0.0516989101398647
    0.8713375402565305   0.4519737664054130   0.4460452125132351
    0.3686408177104248   0.9870708787744545   0.9565456496195481
    0.0568820764991825   0.5317672531628795   0.3366400572334579
    0.5474686435016698   0.0378640327453101   0.8658311270475534
    0.9577784905986133   0.1523802467969643   0.4616317547497743
    0.4332444623934820   0.6390854889835530   0.9691811588145830
    0.8649847886115678   0.9782870444031029   0.5458845624017810
    0.3571149782744615   0.4666404004318077   0.0478875654744767
    0.0362662584682174   0.0665246854512487   0.6319940852837951
```

```
  0.5386657241925015   0.5348247118335231   0.1357391611464732
  0.8865709190560537   0.8143834169047606   0.7173746242105027
  0.3730554159665217   0.2961289771157344   0.2268134485841765
  0.7092241368151546   0.9055865258318235   0.7795721015705920
  0.2088757708515446   0.3917315337962516   0.3039222056203407
  0.7889293315866303   0.7277352780104144   0.8767313756881571
  0.2772612801969347   0.2087166139818099   0.3935598160039119
  0.3943208536479270   0.7907710820949931   0.8073996027130519
  0.8690081929998720   0.2921542645782800   0.2798664690781916
  0.2152401417811004   0.8914120294876179   0.7345128915053959
  0.6981694625032158   0.3849039556254317   0.1903960246059813
  0.2988919209989956   0.7277041910792479   0.6362559698802419
  0.7826187664938683   0.2125886879950905   0.1003041698583710
  0.6201345564394429   0.3104772458045751   0.7095050964846177
  0.1317751678488716   0.8009382113120181   0.2033252033179451
  0.7912673916619002   0.4030088322949218   0.7894899173977096
  0.2788506633494460   0.8884001587839797   0.3194701939513715
  0.7108338669807474   0.2220147763772649   0.8746345232692861
  0.2058439758176915   0.7203452756239148   0.4053386853495328
  0.1211633714428125   0.2971508928842081   0.8070301853153320
  0.6321722351560037   0.7920134121104759   0.2737995477401346
  0.2818577985012768   0.3902106973105186   0.7028667483569577
  0.8026473046356475   0.8902852069401586   0.1869834658045851
  0.2083950896554708   0.2155283972328527   0.6195511849130549
  0.7204095560465478   0.7259537295916885   0.0969951633991637
  0.5505896806631703   0.6271644102438995   0.4604394663790637
  0.0568780446849537   0.1418240585287136   0.9646334600207924
  0.6423029794107074   0.4701195010069152   0.5501819539780407
  0.1412345332006592   0.9675668253517882   0.0447583526676147
  0.4621632988137389   0.5299390706870389   0.6389220073494785
  0.9608632470755560   0.0408873440398972   0.1343839572248738
  0.5497695647364859   0.1240009341339210   0.5400153802577133
  0.0280789630603149   0.6266111758823710   0.0469435130020598
  0.6437102342797214   0.9529928606787901   0.4609697383667535
  0.1255895174470543   0.4704753925701654   0.9378196397557723
  0.4727902225523851   0.0270022958731638   0.3689532694747032
  0.9512838971860907   0.5409234713391833   0.8527382759962334
  0.0628031565168151   0.3600376382858672   0.4606214671250399
  0.5462501546133687   0.8584066999760322   0.9651161342746177
  0.1381319159421869   0.5334981464311476   0.5415071490490194
  0.6359101538315003   0.0332730140005967   0.0587060944814845
  0.9690409040095822   0.4624414995361552   0.6287698932726457
  0.4667273383738675   0.9351033048138367   0.1406661141586404
  0.0398623909109742   0.8664889778154535   0.5431507428255227
  0.5435201279173796   0.3604255407998611   0.0261460015234383
  0.1384354686022649   0.0263994988462832   0.4471943798079882
  0.6262624241868616   0.5483429949487862   0.9431601903234765
  0.9660179549523848   0.9620981361528977   0.3608388381315542
  0.4725233446564630   0.4465959282960605   0.8562301117859414
  0.1233571018266641  -0.0022216119500193   0.2479554013029205
  0.6350172870900638   0.5110522299893312   0.7475280060491194
  0.2503890253331235   0.1346366699077702   0.0002428288100920
  0.7550212203544863   0.6232457290863989   0.4864834821537723
 -0.0200643664905354   0.2420714275946907   0.1248053114389812
  0.4920935346651835   0.7300894195624956   0.6281096532847862
  0.6296831021950178  -0.0116914457986410   0.2561744504788395
  0.1256094394105212   0.5060506724796778   0.7442617853078154
  0.7511369590938874   0.1365379949513210   0.4929578129518091
  0.2285325895800005   0.6165632068590339   0.0222141091628889
  0.4847504506076924   0.2266157598771188   0.3756696716378323
 -0.0092090983215873   0.7408760811448636   0.8773593585511203
  0.3809889300045481   0.4954157100832278   0.2477601549773961
  0.8825448202228742   0.0132618516857545   0.7454759725118087
  0.5234156859826288   0.7297724385197705   0.1218223555093659
  0.0133093435534549   0.2609857742343843   0.6288537019438999
  0.8712013510574292   0.4871889951562960   0.2427909471685042
  0.3837785024960251  -0.0031721456751218   0.7573228471753670
  0.0069725607496326   0.7641270003696846   0.3677492072281773
  0.5137164975114733   0.2460372148223745   0.8695489167663297
  0.2584985563008518   0.3758104534917159   0.5087937568617565
  0.7495081407717882   0.8902777536772286  -0.0139816824830864
  0.2422089001855389   0.8778188539578614   0.5246036904556535
  0.7466609542347752   0.3832498463274070  -0.0085028537365666
  0.1920466734052906   0.1447779730336997   0.6125138909436921
  0.1013638717626520   0.7098043455614649   0.6603241094955600
  0.6750049427454777   0.0967111332567539   0.7287124071723289
  0.1977929346057797   0.6552428093990120   0.4386811664184618
  0.7477701537519852   0.1659979317765216   0.9031087270670343
  0.6033334852715698   0.7010644076249959   0.8479247936057283
  0.1681435490496370   0.0839140546924657   0.7678260296193867
  0.6650098901648660   0.5914578612275017   0.2718270273115098
  0.7964326691801995   0.1448835789112519   0.0767759941051527
  0.4241168664994096   0.2575762202833760   0.6466696246106943
  0.9535469576267525   0.6726291071952544   0.1378910420004087
  0.8552959936237298   0.6120916446212139   0.6940970782851530
  0.3453215963107378   0.1039606559560660   0.2036007554446246
  0.2403120455719834   0.1521758685393833   0.4236139919399901
  0.7514839522924779   0.6715877754582633   0.9058116954126507
  0.9192409037172024   0.2531987773887486   0.8503555193473766
  0.3390594247838619   0.5678462371793943   0.8219139900645754
  0.8425415326276470   0.1008724281753912   0.2658000630539577
  0.8955743207652829   0.9583178424802825   0.3439685868059011
  0.4371836721895951   0.8172534365937827   0.4071858462880156
  0.8407135917416646   0.3977511665541703   0.4854353859249063
  0.4039527871922398   0.8948692155518987   0.1455855140756036
```

```
   0.9347191362899294  0.8553160572200570  0.0993968322010467
   0.3279935320052891  0.4063252960977122  0.0189441569356996
   0.8381420079853055  0.9118386579581927  0.5338929283703497
   0.4297672630008206  0.6859882062232375  0.9122667374865600
   0.9759999672762790  0.1684029722410507  0.3922343564853741
   0.3485111101545484  0.5855464559147623  0.4779652650644136
   0.4093710305593088  0.4897800808576033  0.6710492450365845
   0.9033582169787038  0.0015579900945455  0.1588149450227559
   0.5108797724306708  0.1645914864661004  0.1042980021134023
   0.8234780330272551  0.6014568123843989  0.0509782329899149
   0.3517282903812937  0.0958658081380649  0.5615298967298845
   0.7706498715580816  0.8420002271344236  0.4007199258487986
   0.8710056122895453  0.3057932541059531  0.3514514753475368
   0.3403799944902784  0.9281907137550459  0.3191355982987512
   0.8942442815624084  0.5012759233274449  0.8282204148438130
   0.8983471232075420  0.4493206532969872  0.6420543182672565
   0.2794154102299385  0.8284424341458884  0.0829811449715619
   0.4036703768682990  0.2792736997702488  0.1625855394661938
   0.1887254593066557  0.8495769152281911  0.9270671839811608
   0.6169707453640210  0.7622845553453245  0.3388331953017318
   0.1172512925303257  0.3403185577078215  0.8671589186771897
   0.1460733539010048  0.4292042696918069  0.3191019635149321
   0.6800487064387015  0.8529839904220221  0.5575193626403394
   0.1538875158504895  0.3561799471189039  0.1072196931769611
   0.0954263860088923  0.7528189644856058  0.1621995040966645
   0.5920659121054390  0.3129955310815759  0.6421366289722792
   0.1528354958382810  0.9118691698902904  0.7684967243317981
   0.5981131308022748  0.4969131107896123  0.1559628687256119
   0.5178785314510703  0.1493042504440811  0.6009124541445841
   0.6516874181553979  0.1031561329814278  0.0746900151795913
   0.1135791175614606  0.5698444387307244  0.3072095998598207
   0.6040134806774777 -0.0014167824990389  0.8380039566867481
   0.0591413912992843  0.1811534371037566  0.9022505392863377
   0.5025611870316093  0.6490895815988209  0.4101567455248123
   0.5956503033147356  0.9268154507188530  0.6776916396237233
   0.0791030663514603  0.4869526302565042  0.1780700288325750
   0.5597010456377181  0.3511713915257761  0.0968164907469160
  -0.0032782005539046  0.8407749063803659  0.5974018191712168
   0.1365707031135741  0.8987819967113667  0.0699224615036517
   0.6020705438828321  0.4130669045585880  0.3333374340054085
   0.5155370083922605  0.8353641833740026  0.9025150272588361
   0.0401011139253842  0.3176300100014176  0.4063579926870197
HBWO
  1.00000000000000000
     13.3630470830855526   -0.0424672776921544    0.0342273228831465
     -0.0411540553012025   13.3489211661372558    0.0156574072698481
      0.0320690944632696    0.0171238625417454   13.3122849928035230
     W    O    Ba   H
    16   96   24   48
Direct
   0.7463717912620523  0.2776244663068551  0.7399350517590744
   0.2615311687587784  0.7439911474534567  0.2333899282305179
   0.7517285421917763  0.7534912573617599  0.2381717984048279
   0.2427559389300215  0.2429512495723445  0.7772584863274201
   0.2474384398218646  0.7285471385489912  0.7301231967453664
   0.7304212143488904  0.2533081143338464  0.2313018964515794
   0.2558883439312473  0.2306925971591852  0.2723077190576607
   0.7744978880497618  0.7695594909061801  0.7620872458301574
   0.4839918234348163  0.4889875241033040  0.0095625330968283
   0.0024642816131883 -0.0206715108059808  0.5260924304514014
   0.5101500228393483  0.0028260398961372  0.4847051633820027
   0.0025275149757425  0.5349891333039999  0.0151092592359417
  -0.0077490805890742  0.4853151599817769  0.4885323901509157
   0.5239634725298251 -0.0205830323363415 -0.0211753732999668
   0.0144658873319197  0.0309377124436539  0.0168891793709824
   0.4998796477990313  0.4968081911702750  0.5109524040480293
   0.5529703432190830  0.3535511501776666  0.0421969236921492
   0.0253667663896478  0.8474465590338877  0.5618325375574889
   0.9527360080405340  0.6283993361436376  0.5417575459186393
   0.4641240688723392  0.1263239320947129  0.0298834126252748
   0.4631507807967617  0.8763351645109648  0.4781648783271852
   0.9601404600521173  0.3826377688137525  0.9734525809979382
   0.0670446639698089  0.1499615537111440  0.9745682443217842
   0.5379478306159206  0.6504774779091044  0.4691100993737207
   0.3668724747083971  0.0424320247611666  0.5260948385219764
   0.8720814803085745  0.5296270990661189  0.0590012274235109
   0.6361108671818265  0.5330798216548400  0.9689891508445168
   0.1254879687594708 -0.0024266114394900  0.4598615506317565
   0.8603670606595059  0.4618989748140906  0.4619217638901221
   0.3699784300423635  0.9425872599974859  0.9380195410989156
   0.1309779659701824  0.9723132528008764  0.0607315429753483
   0.6223076852793975  0.4952130121581453  0.5684075642494753
   0.0225787888985141  0.5426487423391173  0.3689572158738740
   0.5148041991069713  0.0438440377956270  0.8543646256765645
   0.5434269121814101  0.9696691443561849  0.6276282334053321
   0.0436671501290544  0.4827932180483841  0.1517635347729362
   0.4765139671243303  0.4382079016438688  0.8840082241366490
   0.9599371667416716  0.9380086715561801  0.3741644263732836
   0.9698484982545525  0.0724224843023232  0.1388105445509158
   0.4542259164859836  0.5668334758135263  0.6458285789287028
   0.9660939592198992  0.1295381346131397  0.4631433976713578
   0.4514301993778766  0.6183349770781632  0.9807900068495068
   0.5679208620341805  0.8535577188874168  0.9437079586553875
   0.0516076999027289  0.3644855403992474  0.4721379235257143
   0.0484413773463970  0.6613584746793875  0.0411446817816444
```

```
 0.5557252896529563   0.1415515554343614   0.5317363576865225
 0.4549901421051726   0.3720445031190888   0.5476297531001170
 0.9562687872192366   0.8829201515267199   0.0271642588455492
 0.1194977582376004   0.4882495288855735   0.9554865779221711
 0.6310808500296765   0.9827867389495758   0.4276220354150838
 0.8628823397622092   0.9737994861763529   0.5545496700834218
 0.3635473619005684   0.4540500776051838   0.0652680815869859
 0.6447226480025999   0.0278678479430432   0.0258445180941167
 0.1278240256494382   0.5257013496387379   0.5572485175403715
 0.3575837524265307   0.5414887663809289   0.4701621410533312
 0.8728140206128162   0.0502838474726066   0.9594561091786219
 0.4766113938351123   0.9404601875944049   0.1073748221916117
 0.9592054629419975   0.4415761930386154   0.6417751175977993
 0.9579323746812769   0.5776312616727585   0.8757106409121329
 0.4691406915211220   0.0615872635951383   0.3535807153552272
 0.0336353535184215   0.0427498811451066   0.6431791123320867
 0.5337996245652080   0.5326092997092330   0.1510012953354748
 0.5250015717198399   0.4660166891659830   0.3802848891386172
 0.0372659038142383   0.9794591509804187   0.8670452314313942
 0.3016327472376503   0.7814403115931939   0.1051785611589173
 0.7790402152883656   0.2865025797726554   0.5880901419218283
 0.7080286738493098   0.2754305773862685   0.3640008195061517
 0.2208772084829378   0.7590147527699183   0.8632874388978371
 0.1998642578736719   0.1988454582410027   0.6250250839113140
 0.7010877332114155   0.7179664828830149   0.1164905340576517
 0.8281234585495264   0.7256328171314809   0.8785570877845122
 0.3128708434458769   0.2202405790918467   0.3961874164482755
 0.0985663148824890   0.2750138169305742   0.8123332397559644
 0.6075624145300699   0.7970977502752908   0.2817064243266579
 0.3801767451016689   0.7051782202952277   0.2852468341023657
 0.8675242452915198   0.2173303230714897   0.7515567636305879
 0.6057767263968797   0.2041192599392874   0.2044959325105501
 0.1209432545069543   0.6887031031166280   0.6996349358562977
 0.9070240996829725   0.8077483412849746   0.7123963229677833
 0.3897948880887156   0.3046906701344577   0.2323068408863667
 0.7857165238953725   0.1128712709616520   0.2717671622662337
 0.2877154060718161   0.5931702619438721   0.7821764947740217
 0.2412446885311607   0.3655147103237725   0.7115972356750524
 0.7795472171754447   0.8829945945107636   0.2048178482755015
 0.2135833539266227   0.6193875638625214   0.1971922527530040
 0.6863490863147644   0.1238300786100875   0.6906679131959508
 0.7396876453012637   0.8922953640032065   0.8038209291062873
 0.2168110818978241   0.3819601482027352   0.3110120956942446
 0.6974137353244868   0.2392520755463410   0.8612766190485224
 0.1797280075294241   0.7638212004937772   0.3465559197864603
 0.3172751426382412   0.7046090487819687   0.6084270958221058
 0.7792538290545679   0.2037393724770363   0.0874013330172680
 0.7830404955371070   0.7889967221763515   0.3910210362298284
 0.2999726681550861   0.2713358225124938   0.8956406043357118
 0.2254266213250024   0.2774615287469132   0.1247585010994500
 0.7280648981872574   0.7964509649739349   0.6158247562866730
 0.8742081155056671   0.6965933970857726   0.2369233434186145
 0.3812885437891195   0.2068454647052858   0.7204740635650833
 0.6023403858308907   0.3187904514059786   0.7048337553396475
 0.1027121197006402   0.8030267521839808   0.1608493432827840
 0.3849549045725201   0.7999654344459958   0.7659689173195861
 0.8841931230547289   0.2930877911398807   0.2564082950819346
 0.1234669760383829   0.2017618112512322   0.2913873585054033
 0.6325846774028303   0.7275446956692216   0.7939228282703561
 0.2123822895093556   0.8699181801665249   0.6811099097923515
 0.7208019388255613   0.3794727185808883   0.1806775410496379
 0.7060060230522717   0.6141560973401842   0.3036810653756945
 0.2154702073699079   0.1084881243839005   0.8080132405151877
 0.7792386623762476   0.4060099199224370   0.7701392111470491
 0.2877962780692274   0.8978368478007909   0.2683726311425860
 0.2909768962907640   0.1095352223456573   0.2223052065354027
 0.7773828494784519   0.6248164771891394   0.7057192835417436
 0.2654332506552518   0.8636023242371139   0.4807700673205720
 0.7621521710054530   0.3811083389193512  -0.0157876594061721
 0.7521208017123371   0.6249468300033867   0.5028776724870363
 0.2719728632334012   0.1065014186559144   0.0134615303158412
 0.8615260766853401   0.4973262077907902   0.2550439778966101
 0.3611552055289965   0.0025420499923712   0.7388647374453553
 0.6074751924118821   0.5217488008824754   0.7683920418073137
 0.1143779914727570   0.0020551597194328   0.2664190514645022
 0.2577070239553114   0.6189233068009093  -0.0049331041948903
 0.7682187819540935   0.1211967161307296   0.4767701344438833
 0.7632213171509675   0.8748112569217548   0.0051612592389811
 0.2511481156397382   0.3690352677531567   0.5170035323225916
 0.6166304150920933   0.0012341105030913   0.2262671557507304
 0.1046926491909363   0.4891929365229093   0.7634077651200076
 0.8713321128578299   0.0262767958753191   0.7565013615749366
 0.3634575369788385   0.5067476366979582   0.2650567778541294
 0.5028950447807967   0.2431344158778274   0.8766233208909560
-0.0030952383450068   0.7428263842554117   0.3820289848852333
 0.5225898801303337   0.7672815252635009   0.6301568563604808
 0.0277761477540417   0.2742718478844164   0.1247350561965380
 0.0065607921140314   0.2445965147921957   0.6210775356301773
 0.5013145240482518   0.7479206358692478   0.1163960627424588
 0.0311346883406214   0.7800459517796623   0.8717880541903230
 0.5132050306393334   0.2675181798066427   0.3805564328786785
 0.1853245974332359   0.1274384115213785   0.6175364103858950
 0.4941705170212888   0.1491688667796341   0.0925075955947609
 0.6518097868348052   0.1024318279522145   0.7524893594990231
 0.6014867221565749   0.7727684142434382   0.8462421680125097
```

```
  0.6347271566160605  0.5993876126653497  0.3108609572099979
  0.7274976613239491  0.1574900574800288  0.0618371334377464
  0.4029148800013803  0.2433022329367663  0.6604556474660681
  0.8247099053868795  0.5779394937514971  0.7354321097720062
  0.8939918143902441  0.6147782833571895  0.8808065262436213
  0.3408225459059367  0.5549418554856982  0.7486537184105847
  0.8419276564889726  0.0907451303353560  0.2295717728938166
  0.8890754247772256  0.9350588851815861  0.3574855912484589
  0.0077832022684565  0.3557781974718504  0.9244873286911087
  0.3398738543595524  0.8955766759935807 -0.0134014834190802
  0.8898689810647072  0.4375393548172531  0.6655606632188450
 -0.0212354724291319  0.8491839263346305  0.0882725979762565
  0.4004222536766137  0.3137286805525708  0.1601820300687754
  0.3990359506015598  0.0684692329129198  0.3323052550093736
  0.0177007060002821  0.1528129152044056  0.4160940693688008
  0.8394374110047731  0.1038225337003668 -0.0021552069030698
  0.3291477121476263  0.5998338609824549  0.5051116823513759
  0.3951803463825294  0.6103568372155688  0.6322036549791219
  0.9863345861182932  0.6555940433755538  0.6007785339157484
  0.3374139001296850  0.1055560797184662  0.5027276317891723
  0.7486137527352368  0.8486991841628285  0.4147459628612584
  0.9096604338691515  0.2595647911510027  0.3161330614678837
  0.3867318355892230  0.8192686490967455  0.8372343799653501
  0.3355998041891453  0.9283996260807947  0.2212092516367847
  0.7683721025925881  0.3473933874731657  0.5485425445809103
  0.9458083151472477  0.8189798382496851  0.6460538402300048
  0.5863608958453360  0.7837801951302054  0.3505231395185817
  0.0761137779462084  0.2249337915507322  0.8633442511050090
  0.1456774458016835  0.3951339808080617  0.3201144994970509
  0.7648161468207053  0.8538386113603302  0.5858314828623858
  0.2591447569886386  0.3367730740956007  0.0962652225048258
  0.0795415281261675  0.7430649885052907  0.1193121372769670
  0.1417998029191055  0.8799478452096463  0.6635990239741995
  0.5985170181883096  0.5064163058492639  0.1718898465004064
  0.5879097352280620  0.1498503239216187  0.5987525900183592
  0.1364092766135352  0.5965147280168883  0.5739220416697325
  0.4918021431399346  0.6732819568171404  0.4171139146040455
  0.6628849294351449  0.5896109613049949  0.0072631365327446
  0.5942386124349981  0.0139437005961635  0.6585096450456128
  0.1020906798503464  0.5184506851213990  0.1802304692320750
  0.5696275260595068  0.3321375495266120  0.1102079824799955
  0.1264382364401866  0.8549399456795138  0.1108722662759303
  0.5746051248485901  0.3268186037129890  0.6375316983271644
  0.0922253972950639  0.0174473655584392  0.8342692031091115
HCaWO
   1.00000000000000000
    12.5799812676902683  -0.0246204525919657   0.0068396143005184
    -0.0233143945046229  12.5953041775013848  -0.0781126880172639
     0.0063248700842626  -0.0780566700586245  12.5193287761468799
    W    O    Ca   H
    16   96   24   48
Direct
  0.7484415983783790  0.2459540825136294  0.7618516806632761
  0.2463295563551574  0.7505738208912374  0.2636061034986869
  0.7638417747506340  0.7304218550940283  0.2306116015567528
  0.2409801015436618  0.2639538467024621  0.7251306521866796
  0.2632738571587187  0.7601465580579290  0.7663610687509679
  0.7602907967835976  0.2412880697729387  0.2568490864747491
  0.2764589144113797  0.2371913804463310  0.2485294371899670
  0.7586198006907503  0.7481007111488110  0.7232058332971396
  0.4845843105795059  0.5156962399937960 -0.0124319649858870
 -0.0060886364066945 -0.0013646148572342  0.5361807965240720
  0.5270243111877214  0.9759552874046667  0.5005935155657402
  0.0067374652456178  0.4804542342881762  0.9865375866224850
  0.0263321014537712  0.4984322992627436  0.4858927548399340
  0.5093016579052745  0.0140426919727536  0.0126991406392160
  0.0221101367588453  0.9785104914423871  0.0200179811421655
  0.5080666191911660  0.4689447029948826  0.4909069556184010
  0.5337606578806058  0.3619749942480779  0.0381723830593506
  0.0440176758466225  0.8631007467942840  0.5453220903097410
  0.9493797920617801  0.6436883337498357  0.5268166281537291
  0.4482584673178805  0.1436025820294335  0.0236668093683736
  0.4471421042283432  0.8502164311915916  0.4719166671151931
  0.9526310225629251  0.3461775923577417  0.9841743317997685
  0.0624901706717972  0.1331224709413382  0.9693055752319379
  0.5541932785569509  0.6327843308043321  0.4674621324152049
  0.3689458477803191  0.0199587113295478  0.5595110670193242
  0.8535098704853312  0.5151702560680393  0.0607907084631731
  0.6315135218682218  0.5248863043844010  0.9599518675046953
  0.1435312737722133  0.0353044589755149  0.4680240735420602
  0.8653489138377729  0.4636614065701065  0.4436337045620166
  0.3632521055392561  0.9746795850619788  0.9355691910365157
  0.1620579996307601  0.9582361747144109  0.0431796599946793
  0.6531327362503069  0.4590409905546627  0.5146246497781670
  0.0282275104384874  0.5585066132393955  0.3555092887920122
  0.5344253230777136  0.0574595632388572  0.8512416292426611
  0.5411910872446755  0.9234306632760327  0.6503435387843646
  0.0361938579415685  0.4488272151053050  0.1519076255320693
  0.4588328230257580  0.4641273307191939  0.8521755265222306
  0.9669904726685178  0.9605755066600323  0.3728501055793055
  0.9897713833641644  0.0463036356433269  0.1461374484554330
  0.4862412766364185  0.5323442145537862  0.6402506898300170
  0.9457569078892056  0.1504043579788937  0.4843091064320688
  0.4389442937322718  0.6536914055733078  0.9774247455033535
  0.5501772118873750  0.8583792767937990  0.9605092181379878
```

```
0.0613912231457738  0.3585174008405267   0.4666142868333446
0.0466762816208164  0.6340017340886529   0.0381488988199835
0.5607502936544200  0.1175969652022886   0.5273979456626858
0.4614268871650791  0.3393383105392485   0.5335144826940015
0.9680166393416125  0.8465469602914962   0.0354122365708517
0.1454323105203214  0.4651318252647907   0.9513708758187206
0.6554759860409763  0.9429987050914472   0.4484041228668337
0.8553964506536237  0.9801442343166868   0.5714120263306158
0.3564203170096554  0.4760445084414188   0.0580622373585632
0.6497185240711703  0.0320795569996047   0.0450331714316143
0.1527457356896282  0.5347856725275981   0.5480884377906188
0.3786537552239745  0.5112769656399728   0.4329873171276379
0.8676226882117907  0.0143842380943475   0.9551624393491911
0.4728855249624523  0.9505041822380886   0.1376171379818476
0.9716308543118425  0.4585750512222187   0.6386003577163155
0.9676435374074271  0.5372884900179336   0.8595619733402098
0.4780061194150418  0.0308027810581634   0.3568304603695915
0.0298139947422250  0.0648581435520402   0.6613274438841672
0.5228908801167934  0.5719557959960713   0.1531888884855910
0.5353688358103564  0.4195353891411832   0.3395583898177225
0.0316140615707033  0.9356395680439605   0.8556319447596941
0.2817692075822483  0.7912358066673798   0.1113644848978450
0.8003748074839553  0.2804805738451884   0.6254446573469384
0.7013361899365180  0.2722215259352773   0.4068172704206367
0.2013241084520754  0.7701473165310564   0.8969558794530198
0.2052065870304008  0.2139961980679709   0.5932393218216875
0.7022675831203028  0.7220689103760479   0.0997266395486409
0.8158543435004718  0.7158931212019936   0.8843869022312979
0.3152898253758768  0.2182559279726220   0.3879746110228179
0.1091326582829874  0.3086317970027182   0.7668775949971546
0.6160568332003827  0.8042738345228341   0.2700177614518939
0.3964582517556649  0.7350318526882224   0.2987124918484390
0.8966672463825679  0.2124111012513622   0.8096439840859472
0.6036031641084714  0.1812169829407549   0.2261205172807955
0.1124089376604923  0.6988854014161848   0.7244911119379420
0.9086088573321673  0.7976489092407066   0.7081180103324378
0.4029480546119813  0.2905583274827222   0.2067443048240047
0.7703780550305760  0.1074233796981829   0.3194018583160659
0.2790932803222118  0.5978321978902095   0.8082575297781855
0.2879123144278393  0.3941388756483101   0.6838059250210058
0.7866666479016365  0.8942414722835449   0.1854222421983575
0.2373077481332387  0.6165604794486398   0.2136608851818607
0.7273742864893681  0.1073032799372860   0.7251755452978584
0.7313371870487956  0.8771531335049275   0.7949990895064336
0.2298687196150205  0.3895807025311651   0.2965833603501796
0.6770762926577257  0.2457893411295139   0.8916867179089883
0.1896076435703669  0.7514345537073729   0.3959724579288817
0.3013626143274933  0.7120674133881819   0.6095266505304859
0.7794718107794532  0.2021688636890251   0.1026456767811664
0.8026544162274202  0.7799541172931932   0.3854955944710599
0.2985963863503590  0.2786505542463749   0.8841083229365094
0.2040934672608397  0.2809083322720219   0.1032211696974072
0.7052368046372859  0.7710032060770121   0.5934674064214466
0.9032077252162752  0.7000347813575630   0.2116139433157378
0.3972341119044187  0.2008890177682812   0.7167353513626267
0.6087570526468102  0.3158149473118644   0.7180417666583093
0.1001474412199070  0.8128144215901016   0.2293307182136203
0.4030874079990551  0.7887456456663070   0.7848727386060336
0.8973144002346840  0.2855766066687826   0.2612970966954785
0.1123119017513631  0.2068966954341150   0.2878223731910687
0.6118639266903113  0.6924568668277798   0.7717876863497386
0.2283438235158622  0.8886179961876086   0.7092357753196242
0.7105781243097240  0.3777732021289039   0.2245993023821881
0.7277273551174944  0.6032099397125735   0.2890743145509900
0.2246614215392869  0.1096608969551497   0.7893257654349666
0.7852158036887597  0.4006067949940075   0.8235482116422757
0.2827157622803433  0.9137557705474082   0.2953708272973236
0.2671832645445562  0.1053400397773645   0.1899685347597722
0.7856685006578564  0.6038114460554175   0.6937647282831445
0.2509029482583035  0.8821196758861286   0.5181291125777409
0.7250203841973750  0.3779583487491447   0.0168814038145679
0.7474229741089394  0.6147980766740179   0.4842535573387295
0.2549492302465606  0.1251000854170022  -0.0088186473255787
0.8731325918530344  0.4828235941517509   0.2524718147027675
0.3859350665574044  0.0084691982871959   0.7480376597971435
0.6368582217599256  0.5038871691667131   0.7710030670727398
0.1229322483721575  0.0152907964505050   0.2774713625088009
0.2499639384002043  0.6222290488311262   0.0091991168765030
0.7508701873250621  0.1270912664996902   0.5143897381900661
0.7479852485900245  0.8658027393506712   0.9874207707789123
0.2593025332705617  0.3756167074686354   0.4905549620154834
0.6381945003661085 -0.0052996806621596   0.2443913382601635
0.1314193117589047  0.5003636469393205   0.7473614250717049
0.8746670933171213 -0.0011084216389266   0.7666856383321937
0.3850720329029934  0.4885544573268886   0.2432893605845112
0.4916062518735406  0.2619125289915563   0.8745777104017997
0.9928243839845788  0.7595651750856290   0.3811450734438044
0.4934384909264987  0.7301071840987038   0.6203583249914270
0.0177474247457791  0.2421560825687073   0.1241440988386340
0.9874837662315010  0.2642063621389135   0.6300551914490043
0.5021192164347046  0.7563463420621089   0.1215702605217346
0.0056476237890374  0.7416203813422159   0.8758105560137484
0.5032429471834500  0.2239950615606137   0.3817857112055762
0.1729122887217730  0.1023407662035600   0.4979747841369216
0.5616397575437830  0.2110939064023459   0.1682849631479767
```

```
    0.6070109397322332   0.0559893013871434   0.8228216080055527
    0.5837999650874744   0.7215004621202560   0.8386123405529785
    0.0876061787525021   0.2559551675798312   0.3442754456408290
    0.1666559258078454   0.0747247634779112   0.7494579428193013
    0.5944880463157373   0.5592968178978406   0.1801148726302672
    0.7272477863078868   0.1485104476339442   0.0756641655191383
    0.2651736589775761   0.6500756162959983   0.5774467662479779
    0.4377227592712002   0.2292000083955400   0.6563217894851328
    0.8116679674558646   0.6403148630679786   0.8997111696250252
    0.9209704581213913   0.2591999061342513   0.8695131197720418
    0.4185620285524496   0.7853845122606471   0.3591651469574602
    0.3468395268430224   0.5590291499208371   0.8077370673571287
    0.9038125196632023   0.9963873771575479   0.3450456404479623
    0.3308586418561801   0.9057171991935999   0.9489168792911130
    0.8310218342882990   0.4088320627579034   0.4863944336549550
    0.9054576611505080   0.4930533265154686   0.6623570048416266
    0.0698447201751412   0.8084301358276639   0.1571489966704528
    0.4128501950537788   0.9945449518525180   0.3328007233268258
    0.9633738557407556   0.1749508435546831   0.4129836035310900
    0.8396092419955072   0.0812382422754344   0.9831878407248651
    0.4297934576410815   0.4947588667064476   0.6792918995738816
    0.9318241757719506   0.6512969123188287   0.6024972623840846
    0.8192630499059462   0.5840001189960370   0.0521531516808334
    0.3483791805137026   0.0937760676021016   0.5493768701783477
    0.2730419428890167   0.3439572148827996   0.9186927336334271
    0.7648185654504135   0.8402069477668949   0.4187650705165641
    0.3214367152594143   0.9428579484331777   0.2340412506487209
    0.8509812498050737   0.4408328981489810   0.8122505901762163
    0.9497833075411406   0.8090398802411416   0.6416188677564837
    0.8573517644763352   0.9259091570289615   0.1905540444155626
    0.6920981269049933   0.3457247689587317   0.4353457023884184
    0.1618609716527036   0.4129759819732072   0.2681078753935181
    0.2249064925178702   0.3519086308664707   0.0793859819609827
    0.5964679082884494   0.3200998013927649   0.6409550957737353
    0.0022898571355992   0.6630753240256975   0.0964785560513522
    0.0919621891320991   0.7036221682491109   0.6508493657774024
    0.0468060376987273   0.1558539963775842   0.8969626087645600
    0.5100673008023007   0.6603414835153967   0.4100607742428333
    0.6080517679040531   0.9312138093810152   0.6908217573223959
    0.0790212080285599   0.5065911100142376   0.1825789135537862
    0.4836656395904140   0.3334525889090194   0.0912618678088489
    0.2453735036775929   0.8545596307835287   0.0818348291010686
    0.6059602754192596   0.4144225803487921   0.3015283743172307
    0.0997484114033924   0.9415919846421028   0.8184620247551662
    0.5037851267654511   0.8335670133516711   0.9020858449726360
    0.5984219231732159   0.8031151861693818   0.3458309033771143
HLNO
   1.00000000000000
     12.7832662852604653     0.0577370547017078    -0.0145053690334510
      0.0573791532852845    12.7124280685054689    -0.0065590348538548
     -0.0144199052034196    -0.0064678438277969    12.7497788538103514
   Nb   O   La   H
   16   96   24   40
Direct
   0.7415636613568372   0.2371865352536927   0.7534436218085589
   0.2471824912577187   0.7402607062788383   0.2614794411683249
   0.7395412724448819   0.7618898724091016   0.2590660741976267
   0.2481774699643311   0.2519280737391331   0.7671707660114481
   0.2491858503167680   0.7446955537400506   0.7424716483691332
   0.7395658740744719   0.2496845194512230   0.2423548311619939
   0.2362325122599274   0.2358120552579279   0.2518491646156615
   0.7571254194010769   0.7395883114949703   0.7360865213492552
   0.5006289640960085   0.5074532077956783  -0.0139585189143200
  -0.0035186128697197   0.0146473350330349   0.5043697690794572
   0.5185164929397655   0.0105834028816302   0.4838218033493238
  -0.0120462164042226   0.5213843009514946   0.0065449784501576
  -0.0118054129656100   0.5013728670704156   0.4870628193733302
   0.4991055646033199  -0.0048703012108428  -0.0037822790670081
   0.0083216102213918   0.0103368639840827  -0.0204966045995834
   0.5049761049342407   0.4841863009403580   0.5069643247152755
   0.5566133313021957   0.3546849188450998   0.0330978283517219
   0.0534782076822893   0.8508464968637913   0.5374316004634533
   0.9390582318180070   0.6409024823000827   0.5233621988447750
   0.4400439402315723   0.1430008359725986   0.0276634568520083
   0.4453214528044149   0.8558292503009077   0.4690209963550996
   0.9420969480849148   0.3635130338684876   0.9562396442096430
   0.0470422467199872   0.1513673851317775   0.9619127638646077
   0.5535467468671145   0.6438541130507873   0.4615808158652986
   0.3605373118387920   0.0330978293172557   0.5565721708101065
   0.8462789712562823   0.5424516303119162   0.0466780421191646
   0.6385412966612605   0.5304092337395488   0.9413159708815569
   0.1348087162994207   0.0118936441630654   0.4377957340108775
   0.8524265364535940   0.4614951705710522   0.4430270954412549
   0.3562929024782258   0.9678280756378628   0.9400231337275008
   0.1387940745978053   0.9771865803186347   0.0542832054675069
   0.6451903774336595   0.4668301554412589   0.5556810356614210
   0.0277604917213209   0.5495252640092058   0.3541283393101576
   0.5284102322688592   0.0556606483818054   0.8532199718052526
   0.5335966607182998   0.9410538202775821   0.6413436053771260
   0.0136283904504103   0.4448491788728398   0.1354790905122465
   0.4679778116854721   0.4434667235655383   0.8564352251475005
   0.9620879096557707   0.9423198478073623   0.3503531971596859
   0.9633365176743397   0.0607413394155528   0.1388777826876747
   0.4770100134274737   0.5600896626301906   0.6417651584036284
   0.9434798757379010   0.1495915491612912   0.4697800165306945
```

```
0.4339763442926050  0.6395473743355147  0.9838070874031017
0.5566739233654632  0.8527431345211595  0.9646691930775465
0.0617786097143926  0.3652627427269448  0.4981405886858867
0.0529301300963376  0.6518478683871163  0.0326598145322211
0.5612588687464231  0.1447279553590901  0.5265622410302964
0.4412686093630397  0.3560738148478534  0.5356290578800437
0.9544487103250094  0.8560548077023101  0.0334627723751618
0.1432444655204588  0.4691312256956932  0.9506228630338837
0.6500830689413782  0.9721489598847626  0.4353783267441840
0.8589477149295429  0.9688591277757282  0.5457487692378635
0.3521742532668088  0.4654442188884894  0.0570846218463754
0.6457426255322428  0.0335566105335480  0.0464545378921686
0.1419144086289203  0.5409147256921276  0.5598043403364865
0.3507378877673993  0.5356613717746967  0.4502137858175373
0.8608538425326132  0.0162047986509788  0.9348501667783907
0.4754590383396299  0.9422239943957768  0.1396809204336332
0.9544410416125088  0.4522736121297216  0.6519710640614225
0.9801106561439276  0.5644778001604205  0.8580547902160198
0.4713350843125120  0.0563818916862273  0.3535664298782554
0.0283372463517638  0.0596242450100106  0.6425361597194879
0.5364402655488034  0.5499188495166103  0.1458209132613577
0.5169226701476193  0.4413950232015165  0.3603534717777820
0.0393230181974630  0.9477606952572664  0.8459473451163778
0.2985743958432187  0.7881656296925122  0.1057753059200304
0.8005453233092694  0.2824763596591794  0.6079701026433830
0.6977815057794015  0.2897594669609037  0.3935143325336899
0.2030526666763471  0.7857048457051917  0.8974468709172391
0.1963146455195408  0.2145138254337145  0.6040645363490029
0.6971205435524245  0.7165268927014814  0.1075171327664362
0.8043084539951795  0.7202215715328156  0.8992790304622641
0.3065858235989828  0.2394376567727239  0.3902910628921851
0.1069735276782605  0.3123879663765992  0.7665980247131492
0.6031471922572592  0.8064997232759875  0.2783593724077770
0.3958636021761281  0.6965597615664799  0.2746829038519236
0.8945528831681870  0.1938002977181806  0.7883607323913374
0.6016156161015790  0.2083667159161025  0.2200672979425381
0.1145256440191093  0.6950761968168707  0.7301497107934358
0.8900008072701355  0.8040603531289084  0.7283405579069165
0.3939758743069275  0.3003345212326617  0.2105815197061102
0.7746912573074385  0.1128757072508064  0.2955679079156445
0.2834160060287408  0.6006935561560243  0.8070980995821928
0.2819606252268502  0.3961804256815226  0.6983489043483389
0.7753135487941066  0.8954637931168160  0.1957818759414112
0.2259900424726720  0.6082511359701817  0.1959089770922584
0.7343110716398707  0.1058428483592530  0.6842397667485558
0.7146199575418304  0.8941641155179298  0.8008364236754180
0.2163102743342817  0.4030654762723490  0.3023800127073735
0.6922807054019190  0.2038712792729263  0.8992910208675021
0.2021823053338413  0.7242286857580769  0.4028581214421861
0.3078126050708179  0.7341274914608411  0.6103253838409229
0.8037924101153148  0.2183225548398148  0.0940367761394702
0.8052089955377737  0.7903587877497992  0.4032595107327061
0.2944296384582173  0.2788897525280220  0.9016680602194824
0.1956722279495204  0.2785108175163673  0.1090426791181696
0.6917097387438428  0.7670013595458216  0.6070099746215842
0.8988614332227527  0.6988681822507206  0.2208560898201467
0.3832137562988875  0.1926953243961479  0.7333509651206949
0.5949915049206306  0.2846483009295683  0.7137926872075527
0.1106248853069434  0.8062940386615803  0.2314458763055233
0.3996472529949830  0.7996242548783243  0.7848055868687400
0.8972153960668889  0.2994472645609937  0.2791309009859246
0.0987018035938200  0.1956040482541293  0.2776223488729875
0.6048337396323183  0.6984560672720765  0.7878405590790357
0.2165623808988920  0.8985954746720725  0.7030997548309098
0.7362542208377507  0.3915298180626718  0.1823852894424962
0.7362470352137490  0.6178046986574457  0.3108667220931426
0.2102664539757831  0.1013778891043809  0.8075730268873249
0.7550271032767968  0.3785118176439696  0.8050990995368933
0.2877423913017532  0.8985498915089950  0.3051144955297262
0.2728230860377333  0.1011521671940306  0.2021428296068937
0.7766998914300682  0.5983030222875242  0.7003702638179355
0.2476220036271207  0.8842177138056793  0.4981635957589450
0.7511381163827798  0.3843778832900479 -0.0049192296531226
0.7511134196908196  0.6179917003467510  0.4974176526333516
0.2562682296120858  0.1181352048500392  0.0134975704823547
0.8722693373617515  0.5000549766747884  0.2490348275673126
0.3842095773615862  0.0055220807402827  0.7564213280685109
0.6202820365596826  0.4966458506667252  0.7470418947061163
0.1182022943443481 -0.0035895229550717  0.2456886939177995
0.2420983241124005  0.6250266956736502  0.0126332127737180
0.7564551342844201  0.1204262701663214  0.4894762855795209
0.7509788795892899  0.8845524295774551  0.0046880839944591
0.2450390744648473  0.3762371235612029  0.4970175140802685
0.6200189034499454  0.0091676701881115  0.2360123843864075
0.1132901565205834  0.4994998117653617  0.7515934452363175
0.8763725805324117 -0.0079399298827785  0.7420465397998968
0.3789811367531187  0.5039849261637775  0.2536110894231585
0.4882932065113473  0.2417541512689035  0.8844117549246026
0.0069637005922770  0.7445904732649182  0.3779665383389800
0.5024515287033126  0.7442013702984692  0.6207576152871287
0.0097520672374995  0.2583842374527734  0.1185776436749590
0.0005385395133764  0.2522591983226653  0.6241539205357933
0.4957825991578921  0.7585190411319972  0.1246485602733369
0.0009245677776730  0.7521690747329236  0.8741020850963400
0.4904386391111330  0.2523627824328590  0.3754273173740872
```

```
   0.7250064923880124   0.6496000037654527   0.0803957880353458
   0.1988638985836721   0.1471015163858009   0.5674702617827855
   0.6867242303513084   0.1352088281206840   0.9367828303958358
   0.5830282242949075   0.7307552758839760   0.8548338813577514
   0.1399975198547869   0.0717136163694414   0.7993234846055267
   0.7939026585028942   0.1488166914611952   0.0622293185791283
   0.9402462501094599   0.7106126884504326   0.1576070059965521
   0.7931609601847748   0.6520193881529748   0.9334172477564409
   0.9348984027033034   0.1934522865314773   0.8545887038475054
   0.3525926451102100   0.5669709746250717   0.8009881248497791
   0.8924149150844914   0.9478665823539250   0.3184960815975118
   0.9698596630476261   0.3419378192336552   0.8881393135976682
   0.4536108865915617   0.8219145624116706   0.4002199531525695
   0.8974245735035002   0.4903657519365164   0.6880125690851528
   0.9966723632621671   0.8322869554681473   0.0934318360443794
   0.3219455054376539   0.3947335152673109   0.0566712750628669
   0.3195002781317874   0.6046667810092633   0.4649723109646457
   0.8967954616609582   0.0453841690919147   0.1731732811868107
   0.3402598393145754   0.1067290287510494   0.5463340325289451
   0.7967888705022903   0.8542329214870650   0.4470334224286970
   0.9356654585747900   0.2631085771162124   0.3355272623229801
   0.4288960530305958   0.7995326889796213   0.8562242294634893
   0.3459238054706752   0.9403040823071308   0.2776530490647203
   0.7925963764722391   0.3553420036090124   0.5836616404314261
   0.2757189378656513   0.8572598383003148   0.0801363521299244
   0.4211975486781475   0.2579755417182965   0.1517823706926745
   0.3541958344932327   0.4212228324817730   0.7021420666936912
   0.2277790255007257   0.8528335600875756   0.9265619854486267
   0.7207465939595931   0.3567201692382652   0.4241868577298737
   0.1567443528232956   0.4462673805832982   0.2786298036448287
   0.6441979713388355   0.9224744557549487   0.8142937141905201
   0.5580146528167497   0.2510491099050649   0.6544150590373091
   0.1519716093519466   0.9281731839321682   0.7345516434047398
   0.5971858153818769   0.5208699291950316   0.1832803758037256
   0.1673748978419714   0.6116354369569065   0.5451783335309198
   0.5368693916013364   0.6700520054219576   0.3912021894712523
   0.6023294415306859   0.9573380128007563   0.6710840917912181
   0.5531586434742741   0.3228176671051037   0.1031367833192179
   0.0205958670182383   0.8099993544349643   0.5933407772616280
   0.1729930555073279   0.4019697642953689   0.9750361798055859
HLSnO
    1.00000000000000
     12.9298831733328878     0.0308475225789658     0.0277040711345469
      0.0307312490859007    12.8903410294043752     0.0299348946273789
      0.0280135543869334     0.0301547251240284    12.9500600183862442
    Sn    O    La    H
    16    96    24    56
Direct
  -0.0099837553680560   0.0123533929280827  -0.0123409308292687
   0.5022414156268848   0.4991062605667779   0.5024429455165699
   0.4918206063693163  -0.0028071033118300   0.5055530372567351
  -0.0132100621267428   0.5132223431127855   0.0041764021136497
   0.4900853215171945   0.5092833162187778   0.0019274101578965
  -0.0100964981791244  -0.0072983840745873   0.4869188531789665
   0.0036780075172983   0.4986677012140949   0.4975786241651510
   0.5060410321242790  -0.0057675448524016   0.0043842580437316
   0.2566207256267379   0.2455885759458467   0.2384947433687994
   0.7575890832565580   0.7381960238645482   0.7450624792979751
   0.2548994523718898   0.7569811837421224   0.7597423035214267
   0.7550988271088164   0.2564816908000068   0.2673571881168905
   0.7521292382260598   0.7544232714354812   0.2552282896956895
   0.2537181300872625   0.2440500815539146   0.7440029825002852
   0.7514923255335862   0.2477538455054356   0.7672859498663185
   0.2487334657281700   0.7443543268822406   0.2618203542866778
   0.1047142713803861   0.1899006562842574   0.2801763525578249
   0.6066324296348126   0.6884088695982501   0.7964366599653506
   0.2743534822578876   0.0978165003344723   0.1793100917705697
   0.7885072832885262   0.5874792450248417   0.7008674688769810
   0.1841490710276421   0.2874666091251663   0.1012444098171258
   0.6855059445909504   0.7575428938422740   0.6070465719291657
   0.6146504646588579   0.1837182345352017   0.2503972794947381
   0.1070015847935814   0.6951984806177849   0.7191563161894473
   0.7851342294730520   0.1003758109374274   0.3290668135323041
   0.2826981461566277   0.5975769330889317   0.8002413820790181
   0.6933719991560822   0.2844080635107341   0.4215023310552216
   0.1973003681496625   0.7765255954237392   0.9043083531659127
   0.4024722485768076   0.7031077391510084   0.3053318265510459
   0.9107705954534795   0.2010810862400413   0.7865633029445938
   0.2445616144679112   0.6073851240432614   0.1910136874991734
   0.7320629440926730   0.1080326684026607   0.7021330986596836
   0.3131031534223542   0.7912926171215620   0.1157292958153170
   0.8074734867979286   0.2861644126782086   0.6171814951492944
   0.8990715378393652   0.6968376779440324   0.2140853388635489
   0.4027020131598400   0.1826281555784458   0.7111730957449559
   0.7202408971645318   0.6009080335882006   0.3044111635298793
   0.2170769215465983   0.0947696234592137   0.8027796597120601
   0.8101731865112838   0.7644409860211210   0.3990840986035796
   0.3180805881287347   0.2627854008158398   0.8837946508706375
   0.4511456852592721   0.3458511454392576   0.5377917485155644
   0.9391234336670695   0.8635806111281840   0.0234350493529929
   0.3617099231845136   0.5118699338449783   0.4353697325797308
   0.8378434283232251   0.0349486717950245   0.9310982865467019
   0.5380503385740897   0.4375986736376110   0.3493061217921758
   0.0253244907699963   0.9563316726173073   0.8421653042532470
   0.4342453591017491   0.8493450863762381   0.4724405739991038
```

```
 0.9362628699903980   0.3595308110848255   0.9757225777053973
 0.3391696490511931   0.0351703257541709   0.5483626164557508
 0.8460182278588428   0.5223423506738227   0.0711047999270676
 0.5125569844699901   0.9366627969087168   0.6474749595981865
 0.0386066576265563   0.4412345120135980   0.1439660235848012
 0.9412600756206894   0.6421872620046619   0.5439683363384404
 0.4461237719301135   0.1445225334853268   0.0273976372492857
 0.8510019649967761   0.4587395079798146   0.4548354026229486
 0.3576116426188388   0.9684713826841334   0.9455747673088397
 0.0119085116097230   0.5668100877404471   0.3595512160515190
 0.5332646066771422   0.0556772868120580   0.8562681646107939
 0.9328835607189820   0.1483343834860184   0.4807328773154421
 0.4317686557269067   0.6573978650951643   0.9673510168457660
 0.8416359223889496   0.9660553010423276   0.5503865353834507
 0.3386423494038906   0.4694065159153775   0.0436564164348117
 0.0220047061450336   0.0522205866805734   0.6371175237433413
 0.5093100876147517   0.5683881002223763   0.1423863420149399
 0.9061293321993953   0.8049907091645211   0.7067655162724885
 0.4005596407679106   0.3101780184269129   0.2091363180291814
 0.7087730786812746   0.8930207798715146   0.7872916536986142
 0.2124568649709942   0.3987708969301780   0.3032763760805006
 0.8199121410648538   0.7414899657738845   0.8863565157178799
 0.3068787920509885   0.2281389972152053   0.3893720855986997
 0.4035928539691321   0.7974490355419243   0.7868792393384680
 0.9026826692993526   0.3054888385827251   0.3075262857501367
 0.2315282819190158   0.8954954578215361   0.6881382377861376
 0.7189389544986526   0.4041082550473359   0.2177662939250770
 0.3040832616572347   0.7140619105730505   0.6049946676540392
 0.8072193854075111   0.2320160062810514   0.1198967775234781
 0.6084106266878788   0.3115518994804999   0.7482778490091831
 0.0992474425970650   0.7908671352842823   0.2144441552167584
 0.7814779239290072   0.3954814075957722   0.8250446563015932
 0.2517504635873543   0.8863985413571628   0.3226833773097106
 0.7100569750370481   0.2075098726094912   0.9176236943992552
 0.1889058125151868   0.7039741273590496   0.4088717522261302
 0.1067122969221212   0.3047972300907749   0.7903517466134051
 0.6088775398406322   0.8153652240516875   0.2648383460918140
 0.2871871022447179   0.3997641700064852   0.6970656167039267
 0.7891044845221258   0.9031424808247276   0.1953133353305196
 0.1848353135524644   0.2347948988625898   0.6054935914903778
 0.7099265955914128   0.7101022015774292   0.1027263657382986
 0.5536759113993650   0.6519867084331781   0.4637229751784231
 0.0480642575810027   0.1629333054846395   0.9565983469124059
 0.6487308800169449   0.4700986835072271   0.5526641626467936
 0.1497046932385672   0.9662007881036003   0.0349421429866215
 0.4750065432994787   0.5583650356329986   0.6481073444167272
 0.9878992634173266   0.0602753970405223   0.1377162525519970
 0.5463408959437852   0.1422961160024875   0.5199518193148962
 0.0514222456499699   0.6529434290046296   0.0262292922586018
 0.6417694139628758   0.9608521707590133   0.4494575580182160
 0.1361157724087859   0.4743364286065319   0.9402526724056559
 0.4620695366353221   0.0426701861931592   0.3515958469732173
 0.9654486762883380   0.5538810375223973   0.8461327759131967
 0.0418268534822065   0.3426325892655522   0.4605915666040071
 0.5679852783328651   0.8526951687387642   0.9968863375584994
 0.1444488862271621   0.5064719344413151   0.5625917731466828
 0.6501556728867663   0.0360313158615109   0.0649427981679010
 0.9636839492268967   0.4439990467895208   0.6507369940855521
 0.4609914731712713   0.9608315556787404   0.1549950173632860
 0.0479312035492557   0.8505734625691181   0.5277968394258301
 0.5408516484538793   0.3557185632272352   0.0367703784114820
 0.1405980467184676   0.0381773445922022   0.4444727626092971
 0.6392133426904166   0.5237622152770792   0.9440738967282449
 0.9726354500527323   0.9459294147592685   0.3413802517443590
 0.4629986004589904   0.4472203307554017   0.8557470805151658
 0.1242755103322651  -0.0058336195322500   0.2389165207751491
 0.6269429833152749   0.4956465723251579   0.7383806394183182
 0.2347199091982883   0.1392934638333963   0.0005335501566658
 0.7465056433021654   0.6214309294532093   0.5037631112747788
-0.0106167993500202   0.2375871925555718   0.1281857664864622
 0.5015685833557341   0.7477844166377907   0.6350271115313688
 0.6307555542028158   0.0031529505020044   0.2557496369981649
 0.1259694366757882   0.5028685890923644   0.7438374829294052
 0.7262852829927322   0.1266147764384308   0.5234482655710614
 0.2338643331566346   0.6339240529848442   0.0131102031168927
 0.4919180145288738   0.2447444296286670   0.3662743203581917
 0.0056230511671736   0.7708257901390307   0.8847504602617925
 0.3770422110988468   0.4927672229304519   0.2487844031020224
 0.8704634834477106  -0.0036180056963364   0.7550134229238296
 0.5056532616456291   0.7506290541610998   0.1342748258381581
 0.0054382194778978   0.2470561553266529   0.6272554402549861
 0.8828962135057912   0.4949646069720890   0.2519950552279026
 0.3737861192956090  -0.0146838648697424   0.7499310206985528
-0.0073866212582247   0.7530088859845830   0.3783884536532295
 0.4980308633103867   0.2495966737086888   0.8736502469395442
 0.2518086176469403   0.3699884368453527   0.4916458535990158
 0.7596917715531685   0.8758249657861471  -0.0041161264460438
 0.2331658182463048   0.8714371242561453   0.5070942093060398
 0.7412872691304161   0.3737453569517698   0.0324118019514974
 0.7529586045692690   0.6487165029946519   0.0851156471599701
 0.0827111531317472   0.7291548830119423   0.6556170492393183
 0.1961982426941784   0.6316597908975381   0.4313450358226626
 0.7562367468910722   0.1451837995018469   0.9331392988459496
 0.5818169117985268   0.7319425895077442   0.8533665779379509
 0.0862989725734839   0.2100080658478765   0.3502990790627597
```

```
   0.1495433496496978   0.0587876744498971   0.8014232514291584
   0.6719542802390676   0.5611692271770007   0.2618221543968367
   0.2928491121900140   0.6431163514643522   0.5813741464396827
   0.4342815400574122   0.1829017316498587   0.6417506642743849
   0.9218445468164889   0.7184088161746752   0.1447936239903996
   0.9449089260390043   0.1865042431980523   0.8546454933344255
   0.3975110890490415   0.6713144117772456   0.3743361084581587
   0.3472708570661140   0.5799570190008538   0.7631992727013530
   0.8510566414030046   0.0640562720939542   0.3152986200123387
   0.3881176314152807   0.4484614248615581   0.8449484652788064
   0.9558885712581389   0.3276922191070119   0.9104213797973535
   0.4427695689442668   0.8255876609294291   0.4009088941906440
   0.3380905272041978   0.8948246471599831   0.9441978393003225
   0.8502996551335099   0.3866039213435166   0.4290718252908813
   0.8988059090203810   0.4799942266012693   0.6714802630005990
   0.3970226046291254   0.0010116586045804   0.1711545534233223
   0.4982937510970293   0.3130723858680565   0.5873597471624917
   0.3171800170213572   0.3965328462408044   0.0392795241728772
   0.8147024025885018   0.8994324503553647   0.5258924086562047
   0.3944817087336646   0.0197433919287359   0.3266749119733804
   0.8962565343392064   0.5722848868514750   0.8187348839968996
   0.4333770220007457   0.6932951184524371   0.8992742954066552
   0.9245030251784044   0.1835148488857228   0.4130685060807183
   0.4501657440403208   0.1688424903056006   0.0990509285933472
   0.9310814779483928   0.6495312493964257   0.6186547349883351
   0.3182925329079642   0.0999315744125818   0.5134755823886159
   0.8511910154732424   0.4225581879828764   0.8155616534453193
   0.7837808167384737   0.3529018300898559   0.5888092081123859
   0.2949628816738538   0.8630131265307610   0.0987842712251305
   0.9389910451763127   0.8114471306568661   0.6379474619014884
   0.8603749689985557   0.9264088039023400   0.2078099282745193
   0.3583206063916179   0.4254702616683353   0.6896569551235834
   0.6891237422225688   0.3543676159045824   0.4526200327091181
   0.0848468985278171   0.2628885996998478   0.8520197504497762
   0.1517981491723219   0.4370716111990123   0.2782734939081030
   0.6524826941429448   0.9153350743650085   0.7411205491709334
   0.0727291047332856   0.7558201930903591   0.1525143658200734
   0.0765867110459979   0.0101732700430190   0.6699232914824382
   0.6783813066700127   0.1057639264542521   0.0540945552473530
   0.6009393120823395   0.0426038590042677   0.8240512798805708
   0.5484728701367142   0.6795858078954508   0.3933990809577203
   0.6696517661249584   0.5926279997261655   0.9511682230518386
   0.1662286558812084   0.1032518678806192   0.4731927712483255
   0.1065496990644234   0.4056771218856755   0.1318092340324231
   0.5038037387050009   0.3317423186584292   0.1004603925734544
   0.1703633527021028   0.9082499134789950  -0.0103406709247817
   0.6061517621473375   0.4350159952475626   0.3130355901698964
  -0.0045737363522395   0.3248313027975676   0.4000504981323207
   0.1626529652077466   0.4077776705808619   0.9664009147266391
   0.6649539685671486   0.8902884944942651   0.4665309051336630
HLTO
   1.00000000000000
    12.7787266900668754    0.0157902006903083    0.0255477920065521
     0.0155843800550068   12.8372844925478073    0.0375284737219901
     0.0256301019400174    0.0383206042909972   12.7741463715231660
    Ta   O   La   H
    16   96   24   40
Direct
   0.9940321074867238   0.9899647023502192   0.9974651803903036
   0.4940321074867238   0.4899647023502195   0.4974651803903039
   0.4984446563341511   0.5042717592540858   0.9998876886296068
   0.9984446563341515   0.0042717592540861   0.4998876886296071
   0.0028658406616687   0.4944281310112181   0.5123174787651759
   0.5028658406616683   0.9944281310112182   0.0123174787651760
   0.5132997806216836   0.0150779047425579   0.4890914792714415
   0.0132997806216831   0.5150779047425579   0.9890914792714416
   0.7666080754224157   0.7633225957232230   0.7631104540406696
   0.2666080754224154   0.2633225957232233   0.2631104540406695
   0.2673902713882947   0.7375387321511290   0.2349134959539261
   0.7673902713882950   0.2375387321511288   0.7349134959539264
   0.2396119929098342   0.2695038253664557   0.7384831779581960
   0.7396119929098344   0.7695038253664559   0.2384831779581958
   0.7398503187768979   0.2397193025153483   0.2616861817099982
   0.2398503187768979   0.7397193025153413   0.7616861817099981
   0.2835045463849596   0.1100132196264827   0.1990924875462715
   0.7835045463849595   0.6100132196264826   0.6990924875462715
   0.1961505612160084   0.2798994049393634   0.1060289130957080
   0.6961505612160082   0.7798994049393635   0.6060289130957082
   0.1084430328975002   0.1970701869810978   0.2791269660152831
   0.6084430328975000   0.6970701869810977   0.7791269660152829
   0.7832996654551310   0.3929683155918322   0.8002008953595980
   0.2832996654551340   0.8929683155918323   0.3002008953595981
   0.6989220555621556   0.2234794121783919   0.8924546777997603
   0.1989220555621631   0.7234794121783920   0.3924546777997903
   0.6118741638292857   0.3019770280505217   0.7177283917830202
   0.1118741638292855   0.8019770280505218   0.2177283917830201
   0.7220679261697769   0.6098477016347651   0.3037460594057278
   0.2220679261697766   0.1098477016347650   0.8037460594057275
   0.8017528434799162   0.7832418904553110   0.3936640189263529
   0.3017528434799165   0.2832418904553184   0.8936640189263528
   0.8953784787428977   0.6977079647317688   0.2220508585400344
   0.3953784787429050   0.1977079647317687   0.7220508585400270
   0.2249223960567688   0.8977533155647077   0.6963141424325974
   0.7249223960567686   0.3977533155647081   0.1963141424325975
   0.3066010780531011   0.7244814151481387   0.6047110180845516
```

```
0.8066010780531013  0.2244814151481389  0.1047110180845516
0.3974394108149048  0.8065497852009933  0.7761533671942795
0.8974394108149046  0.3065497852009935  0.2761533671942794
0.7205810667338588  0.8986357883212307  0.8005194248329510
0.2205810667338590  0.3986357883212304  0.3005194248329508
0.8100836810190367  0.7294416686481512  0.8987842786504545
0.3100836810190364  0.2294416686481512  0.3987842786504545
0.9008522264367288  0.8085100259711973  0.7239435148135653
0.4008522264367291  0.3085100259711974  0.2239435148135655
0.2287274547885313  0.6024944789736479  0.1941664703527131
0.7287274547885315  0.1024944789736483  0.6941664703527134
0.3140865146801853  0.7731043246400267  0.1000959887033112
0.8140865146801857  0.2731043246400271  0.6000959887033110
0.4024437275829939  0.6961757353068332  0.2799728740869851
0.9024437275829940  0.1961757353068332  0.7799728740869852
0.2759042392641988  0.4029960287490894  0.6916411638928569
0.7759042392641989  0.9029960287490895  0.1916411638928574
0.2010853121289699  0.2254306662381179  0.6037831630964624
0.7010853121289701  0.7254306662381177  0.1037831630964621
0.1028019697287210  0.3097520148049335  0.7746244075948090
0.6028019697287209  0.8097520148049336  0.2746244075948091
0.7787440673312078  0.1042592719455119  0.3006113716986426
0.2787440673312082  0.6042592719455119  0.8006113716986498
0.7016309821078823  0.2778141687139837  0.3977662083391583
0.2016309821078826  0.7778141687139842  0.8977662083391583
0.6042059708398624  0.2004655579575363  0.2242459536031164
0.1042059708398627  0.7004655579575362  0.7242459536031161
0.8581729483810180  0.5336684071306583  0.0590771343035643
0.3581729483810181  0.0336684071306583  0.5590771343035642
0.0568327749305737  0.8464855555684563  0.5420009121027807
0.5568327749305739  0.3464855555684562  0.0420009121027809
0.5378696431510552  0.0456513513887633  0.8490314721462207
0.0378696431510552  0.5456513513887636  0.3490314721462135
0.3481171553504001  0.9578511702709821  0.9421447905430803
0.8481171553504000  0.4578511702709822  0.4421447905430800
0.5439396649576463  0.6517632462090514  0.4619867163559597
0.0439396649576464  0.1517632462090513  0.9619867163559598
0.0295389847530451  0.4426805313732626  0.1431725286127065
0.5295389847530450  0.9426805313732632  0.6431725286127066
0.1466511235670452  0.0164612441518181  0.4446723939751624
0.6466511235670448  0.5164612441518185  0.9446723939751624
0.9447766958705009  0.3602962514403335  0.9710926492915730
0.4447766958705010  0.8602962514403335  0.4710926492915730
0.4573423953437101  0.5599861916695362  0.6525824268341529
0.9573423953437100  0.0599861916695364  0.1525824268341529
0.6315254171552086  0.5008731890291536  0.5691598973026305
0.1315254171552088  0.0008731890291534  0.0691598973026304
0.4310983348855291  0.1316282218898630  0.0001835591393883
0.9310983348855296  0.6316282218898628  0.5001835591393881
0.9808235247606844  0.9310418652952686  0.3625424648686552
0.4808235247606847  0.4310418652952686  0.8625424648686553
0.1460732746132985  0.4712581294601100  0.9432703988856186
0.6460732746132986  0.9712581294601095  0.4432703988856186
0.9509736500512170  0.1410884777098476  0.4632815801525487
0.4509736500512245  0.6410884777098478  0.9632815801525483
0.4646516650640654  0.9502242753851637  0.1483158123046414
0.9646516650640654  0.4502242753851642  0.6483158123046412
0.6389512762688397  0.0345932496482079  0.0555414021181982
0.1389512762688394  0.5345932496482075  0.5555414021181985
0.4489458500187514  0.3548475202794645  0.5362743908222619
0.9489458500187511  0.8548475202794644  0.0362743908222617
0.9707020192866299  0.5606887017799039  0.8561717482423373
0.4707020192866297  0.0606887017799040  0.3561717482423373
0.8587051615973732  0.9690557433118940  0.5536240256480192
0.3587051615973732  0.4690557433118942  0.0536240256480192
0.0595263980557047  0.6466223386476389  0.0319939562774055
0.5595263980557045  0.1466223386476386  0.5319939562774052
0.5343932035969451  0.4453943999176819  0.3612453102688601
0.0343932035969450  0.9453943999176825  0.8612453102688599
0.3557554728250362  0.5300863449238682  0.4485541011989944
0.8557554728250365  0.0300863449238680  0.9485541011989944
0.5534108609018853  0.8575291800213473  0.9715946332759500
0.0534108609018853  0.3575291800213475  0.4715946332759495
0.0339070397739847  0.0530330735795169  0.6425585362716894
0.5339070397739847  0.5530330735795170  0.1425585362716895
0.1189243512511703  0.9980547570065138  0.2560578571837243
0.6189243512511700  0.4980547570065142  0.7560578571837314
0.2486789950992483  0.1209113245265925  0.0011761197669777
0.7486789950992481  0.6209113245265926  0.5011761197669778
0.0024183001470769  0.2466307971597707  0.1227673289095936
0.5024183001470771  0.7466307971597710  0.6227673289095933
0.7473393107121731  0.3825161913388725  0.9975861125608357
0.2473393107121734  0.8825161913388724  0.4975861125608361
0.5007626927816281  0.2512893259711900  0.8800387641793588
0.0007626927816283  0.7512893259711905  0.3800387641793588
0.8776275910612282  0.5040118486117782  0.2536735485348087
0.3776275910612282  0.0040118486117778  0.7536735485348087
0.6210753078665604  0.0044341501312114  0.2465261479135894
0.1210753078665607  0.5044341501312117  0.7465261479135893
0.2594453147350282  0.6184077101479043  0.9978308437630355
0.7594453147350283  0.1184077101479041  0.4978308437630352
0.0072574044242073  0.2439775501350644  0.6168731053519555
0.5072574044242073  0.7439077501350640  0.1168731053519553
0.7478497097432214  0.8832353010857209  0.9979336238592507
0.2478497097432215  0.3832353010857137  0.4979336238592508
```

```
  0.5041476616667414   0.2559237650809009   0.3810494810310645
  0.0041476616667414   0.7559237650809076   0.8810494810310644
  0.3853751709333872   0.5003727460734977   0.2473645723474643
  0.8853751709333874   0.0003727460734974   0.7473645723474638
  0.2176102744344474   0.3440099086546998   0.0695773128950315
  0.7176102744344471   0.8440099086547000   0.5695773128950318
  0.0760522811872673   0.2064376961319507   0.3484168542204161
  0.5760522811872671   0.7064376961319508   0.8484168542204165
  0.3526211293649780   0.0786775344344584   0.2108252164439041
  0.8526211293649778   0.5786775344344586   0.7108252164439040
  0.7141183546639277   0.1566184284641659   0.9269082062582685
  0.2141183546639278   0.6566184284641660   0.4269082062582682
  0.5801077788938148   0.2910009577328731   0.6488747273954433
  0.0801077788938148   0.7910009577328728   0.1488747273954431
  0.8473688081708716   0.4275141085652604   0.7742952227063561
  0.3473688081708713   0.9275141085652532   0.2742952227063559
  0.7974772693588094   0.8536971976581674   0.4229587315236485
  0.2974772693588019   0.3536971976581673   0.9229587315236484
  0.9296195086235902   0.7161176428659398   0.1555301917198275
  0.4296195086235898   0.2161176428659395   0.6555301917198275
  0.6548600053340301   0.5771173240544842   0.2868693684687901
  0.1548600053340301   0.0771173240544844   0.7868693684687905
  0.2969375373791753   0.6565981282367936   0.5709403836445440
  0.7969375373791752   0.1565981282367935   0.0709403836445442
  0.4319972411358914   0.7951594470714295   0.8439254573576014
  0.9319972411358914   0.2951594470714293   0.3439254573576014
  0.1561347258609805   0.9315696763846547   0.7040942773443389
  0.6561347258609805   0.4315696763846546   0.2040942773443317
  0.5328121543385419   0.3096054754612468   0.1053221474053720
  0.0328121543385419   0.8096054754612466   0.6053221474053718
  0.1037173896209966   0.5170069408081309   0.3212860020819428
  0.6037173896209966   0.0170069408081308   0.8212860020819428
  0.3150647491137872   0.0940487362356231   0.5379620881482970
  0.8150647491137873   0.5940487362356229   0.0379620881482971
  0.0125219788093954   0.1784922037821796   0.8968034881115773
  0.5125219788093952   0.6784922037821797   0.3968034881115770
  0.5967877449744229   0.9549022957520090   0.6780192494381727
  0.0967877449744225   0.4549022957520092   0.1780192494381726
  0.8108915898634460   0.3927600857851456   0.4590090010991201
  0.3108915898634457   0.8927600857851453   0.9590090010991201
  0.4594383999378095   0.8274098790168537   0.4034226453958641
  0.9594383999378098   0.3274098790168534   0.9034226453958645
  0.8932380173553426   0.0418336560652530   0.1916273294299004
  0.3932380173553426   0.5418336560652528   0.6916273294299000
HLZO
   1.00000000000000
    12.9991981517329478    -0.0199985789479347     0.0099517523984225
    -0.0202145972371557    12.9920579900787310     0.0087244721712911
     0.0101625467470594     0.0085160787738996    12.9167637581932162
   Zr   O   La   H
    16   96   24   56
Direct
  0.0023447494604816   0.0096092388959256  -0.0058081851600839
  0.4972001907220583   0.5072055646608740   0.5033987882762375
  0.5071283458179928  -0.0092176719718434   0.4966344753474945
  0.0041983612821218   0.5040781591847208   0.0081459330661746
  0.5021168909719910   0.4986795531568715   0.0106951724657548
 -0.0103160558521258   0.0071650765150078   0.4898220657618212
  0.0033724470062073   0.5149368585518167   0.5009195576993271
  0.4860022533536022   0.0096310124559120   0.0035663895785818
  0.2478579619394158   0.2486159732394706   0.2643164164915265
  0.7518910148411193   0.7497240600450321   0.7636619978034420
  0.2577708678678433   0.7631847029466970   0.7501089018382191
  0.7362053354483764   0.2560379034944372   0.2480259177709681
  0.7473098770138630   0.7732232273694030   0.2502039308848376
  0.2504501899115654   0.2543412906330998   0.7345861871593489
  0.7409731637503661   0.2416025446852786   0.7402736010458327
  0.2474832721287975   0.7576056355892998   0.2517458159956156
  0.0950720907079791   0.1863816479567589   0.2837264031668524
  0.6043812032594343   0.6902438673851000   0.7918891128892505
  0.2756997421467703   0.1037797043748254   0.1906145067088373
  0.7808527817301171   0.5982064635767551   0.6962328759952295
  0.1927870286568458   0.2821731012198409   0.1084891769036071
  0.6943332437039672   0.7782274626593423   0.6099652158020469
  0.5969230608170801   0.1918962038801133   0.2279294468095213
  0.1042467748716994   0.6940282400026389   0.7228189287762360
  0.7822754989197938   0.1063686232482513   0.3009688869717537
  0.2846315511300808   0.6049733262628303   0.8092962232681188
  0.6867675279659834   0.2821966666339727   0.3985156100450170
  0.1941655178407959   0.7810762274114521   0.9030877165586618
  0.4011893153205964   0.6934320825899223   0.2929419968937220
  0.8982059660286740   0.1866350615366666   0.7795797348246041
  0.2138612924616421   0.5960069224231039   0.2068109115696618
  0.7147507787217281   0.0941038375012755   0.6948760128871293
  0.3195408258423694   0.7499059969036193   0.1086912081744467
  0.8111602272448428   0.2693236350168751   0.6040859824633498
  0.8964513354572686   0.6981170863657455   0.2172685120407599
  0.3899530140148954   0.1839126428894924   0.7384233533714694
  0.7252828679624344   0.6122650728235068   0.3118882718119883
  0.2044129143200140   0.1049319144048573   0.8054698386170752
  0.8075518691388761   0.7880657526756213   0.4037940092881854
  0.2971040422777631   0.2900299711909381   0.8980987533863513
  0.4526007022185716   0.3542626634583215   0.5455670242593865
  0.9408241147227926   0.8550244710104365   0.0253391622131416
  0.3610262406916325   0.5108800471096110   0.4342417028243460
```

```
0.8503098183228389   0.0390680197881805   0.9422855008244467
0.5405165375981327   0.4409959945107881   0.3528822424458655
0.0254331736781429   0.9418832412340254   0.8446393696285353
0.4376381500888676   0.8518600012641722   0.4948926946643508
0.9457732408023224   0.3550465552522030   0.9603794036126432
0.3540957211504107   0.0375157846015678   0.5536823020087511
0.8543299628676803   0.5344793412294796   0.0595227618814544
0.5427566964436490   0.9501338777790385   0.6534610802386541
0.0195540725204435   0.4293048064187760   0.1497520016590096
0.9352485393549277   0.6527902663206123   0.5302065553285952
0.4329368066077537   0.1550807813911917   0.0222702049524692
0.8507935446310061   0.4689661578769029   0.4466012737716440
0.3465319437046384   0.9846072773996551   0.9307655519358606
0.0203166069926945   0.5640542850820650   0.3485255994179139
0.5353401090014899   0.0584280856425001   0.8515923446169984
0.9477461771889074   0.1551360332342346   0.4594346624467464
0.4517134684607807   0.6480720730371345   0.9603414819551502
0.8495122519319840   0.9780225024462759   0.5537996531832047
0.3550826015993460   0.4769815621136671   0.0672374914270986
0.0321602816402054   0.0574634882443043   0.6464317475846020
0.5300757700556377   0.5634160171903368   0.1501934018080765
0.9017927536630745   0.7953683017913288   0.7134159153963068
0.3961054084188987   0.3046436332264958   0.2183769917139206
0.7434775427449425   0.8909072305980418   0.8184941084500111
0.2084265659963693   0.3999653635386536   0.3027351835720274
0.8101771930603812   0.7092835183904888   0.9086275446909269
0.3153246410897289   0.2441876741085545   0.4006278394867032
0.4025520653517249   0.8090022872520091   0.7766675924573306
0.8957977531225138   0.3032189514327490   0.2938956318359233
0.2234760655308532   0.8996050559725824   0.6875578875549676
0.7107958163276684   0.4136081746774195   0.1974313220140692
0.3070411377608490   0.7114641135847047   0.6028462612860080
0.8131769828050985   0.2511295479881613   0.1060457719771405
0.6000343148450812   0.3096333279470245   0.7296220290941149
0.1047492202115903   0.8060827041092679   0.2080947215588423
0.7870167980489465   0.3942157997775875   0.8030755683164525
0.2810649584170131   0.9050164310053074   0.3012540882419524
0.6911366132774494   0.2205799185153776   0.9044427592567579
0.1890241345688917   0.7315133954234834   0.4006553213132316
0.0957751862980801   0.2989604416601906   0.7898343498349961
0.6009601428087713   0.8148174530694200   0.2667534101272354
0.2678825481539080   0.4042233094378510   0.6925241003354864
0.7809225638223684   0.9088055303495149   0.1857065526436583
0.1860555238582878   0.2288436284792413   0.5956347125933432
0.6972613603017067   0.7214345404392983   0.0995496011565681
0.5636578533177573   0.6567260232581268   0.4648596622534568
0.0595675889587578   0.1492381730361327   0.9806434386963764
0.6541953008041390   0.4674561281176359   0.5502561721171580
0.1483515412493597   0.9660751275564888   0.0538686952485030
0.4911338150356358   0.5700581800414695   0.6457438219497855
0.9578392267415710   0.0435938643773690   0.1481230880792920
0.5523358688193813   0.1487324504769856   0.5364060445574547
0.0601888074350312   0.6584323122409852   0.0456255910281293
0.6450787551985928   0.9902319384485372   0.4304501951217978
0.1496297817855607   0.4753492229693803   0.9480299517624402
0.4571434713764326   0.0541443789163535   0.3505312314734090
0.9798603063663228   0.5699458387946379   0.8617471041252265
0.0589333400956196   0.3601464006053371   0.4606055081507413
0.5449597253242172   0.8520508383522977   0.9666221890442430
0.1589143427486310   0.5577592668381479   0.5331372570541780
0.6352389745147611   0.0182415340657478   0.0587199758543587
0.0016990062403630   0.4406678769879850   0.6425930611026779
0.4502827859961684   0.9471792301957844   0.1532741959665731
0.0602161341646677   0.8581668026243386   0.5380982053507308
0.5591369018549954   0.3462742026481517   0.0397486855980849
0.1527421583540996   0.0377780901038537   0.4444336199395430
0.6397304701957864   0.5105050611489571   0.9307987024345639
0.9884764568238428   0.9332655858983230   0.3527559592365808
0.4590301852776334   0.4404854615413492   0.8523114596653103
0.1244863885784962  -0.0124427391177894   0.2519364680936404
0.6204486888386278   0.4897308199644558   0.7512933626351751
0.2535900055739877   0.1304169265374603  -0.0059009151980461
0.7586027633982974   0.6260577225050564   0.5020306710066005
-0.0099827029382029  0.2528532452746540   0.1269271373753732
0.4985225617463893   0.7508999582054237   0.6321436653298799
0.6227675845782252   0.0054156485498923   0.2441512190736135
0.1235916565122027   0.4959659690286710   0.7651571805890208
0.7507051722778738   0.1314161768732820   0.4973291776550156
0.2552082600672371   0.6120699085719542   0.0084389229829905
0.5027366308946092   0.2474574092328171   0.3777673303671784
0.0016969241743543   0.7420406699208523   0.8790608244118379
0.3791900914067578   0.5037295495457392   0.2495155856334425
0.8636633123072461  -0.0013597430167373   0.7334528589524698
0.4977540844952860   0.7479415100888583   0.1294975886017845
-0.0031203210644343  0.2570802490980331   0.6147580628377273
0.8839828700063385   0.5013569538727477   0.2474577802929363
0.3716085745218598   0.0014417707404729   0.7507712651921563
0.0095204625451664   0.7486723933172067   0.3734310182842945
0.4946097221694570   0.2501622036246399   0.8727318075827353
0.2505559192460715   0.3835740755130950   0.5074887503912793
0.7384789019100145   0.8872233314207894   0.0031522009041411
0.2541041286572827   0.8758461037127528   0.4938568644164972
0.7550559022883855   0.3885544639736853  -0.0009789566886124
0.7219560157242434   0.6547491597073718   0.0743362888846535
0.0740328972957152   0.6919157749322660   0.6534290763959685
```

```
  0.1502659050562122   0.5736360223535210   0.2417558015838450
  0.6877562263811700   0.1650448403412717   0.9556883931046856
  0.5823815786819434   0.6801244372617029   0.8639522206249282
  0.0532499719625490   0.1790874233613754   0.3479895203083244
  0.1298337704193554   0.0975009425978500   0.8080709689373795
  0.6593116134204892   0.5828713055320284   0.2903872678261477
  0.2813946028019139   0.6459449415291419   0.5750755651816014
  0.9153294178392558   0.6860758366610289   0.1449474316460727
  0.8498987401753763   0.5702004272776598   0.7071433031473489
  0.3458857577584368   0.0759854885559886   0.1910583069770579
  0.8134060332551549   0.6413083849457595   0.9433948472265452
  0.9328704447249740   0.1873185608409325   0.8465344997308107
  0.4217965195919009   0.7139415631240207   0.3626118254970743
  0.3392731423607250   0.5694346274879808   0.7715103978413506
  0.8509079148188963   0.0779138738346331   0.2916885716744242
  0.3920938864731753   0.4367667673515426   0.8176092515437553
  0.9561721805144864   0.3357198314231664   0.8881838582780400
  0.8193460886182514   0.4010672167756395   0.4571754898152623
  0.3887274739234659   0.9314816451885263   0.1959832315722055
  0.9534666445196813   0.8329380125760023   0.0968101110524784
  0.4827087409136461   0.3284430083214773   0.6103417348096724
  0.3896902118673816   0.0235253742481397   0.3340086002766123
  0.4354694499717727   0.6792323994062914   0.8932695571052275
  0.8211498756625000   0.1046557413542290   0.9650463363944151
  0.8976924005831444   0.0007436589903041   0.1700442968690830
  0.3288648582562581   0.1074839387508205   0.5419507544338674
  0.2721457685128472   0.3534248999261184   0.9309370172047604
  0.8087186478518792   0.8563576245540362   0.4368093000072749
  0.9137212829324529   0.2665588430816678   0.3585444159753445
  0.8511028170176799   0.4263399755145181   0.7802519905435420
  0.4144585378510223   0.2876874690901951   0.1467848651119551
  0.9248364807322077   0.7652515295420077   0.6470719188117848
  0.2305622485253921   0.8394669966132927   0.9351826446534923
  0.0866353953678296   0.2705188013657523   0.8595693876196141
  0.1479385444740006   0.4299223476886058   0.2686216792142817
  0.7099917912899967   0.8444229481702196   0.5767862472602673
  0.1984878722882093   0.3517174878718760   0.0793094126611741
  0.6471159142348926   0.4542891014038921   0.1952402954609649
  0.1047534997891461   0.0449583709352306   0.6620325660799358
  0.0713700650990878   0.7038035922388655   0.1069354908219924
  0.5302209184142097   0.1764520817302667   0.6032300036037880
  0.1779442890034666   0.6118543586160069   0.4809780008457278
  0.6034068884206566   0.0498267713705525   0.8202017760616185
  0.5644025663614330   0.6989441549852707   0.4018060121742614
  0.1805200563768819   0.1019391454234267   0.4734526784749866
  0.6031421137494377   0.9896104419432228   0.6784988353106528
  0.5549767249509026   0.2948582276361641   0.0955055880618173
  0.0752465064155954   0.8553829857991047   0.6121658007206959
  0.1661609864306810   0.8937161159650406   0.0656921384933719
  0.6804553742425359   0.3998664522461053   0.5286193302436257
  0.6136561625190623   0.4230818947058045   0.3467209191026667
  0.0889424557014268   0.9436023952380057   0.8033245774360153
  0.5108281324082541   0.8285281102506638   0.9030392527514132
  0.0519720156315015   0.3388088295852699   0.3886625251574858
HNdSbO
 1.00000000000000000
    12.6456293219429359   -0.0441687624575245    0.0744127766035236
    -0.0441541617270129   12.6133845491187575   -0.0242266344020685
     0.0741599885171499   -0.0244578361239537   12.5613893960706307
   Sb    O    Nd    H
   16    96    24    40
Direct
  0.7543904855656961   0.2624122005405218   0.7742668195607365
  0.2514589684665782   0.7607024269966820   0.2559186390079465
  0.7616096521403632   0.7539940264253957   0.2477000096463120
  0.2633908216722305   0.2462132276037587   0.7436789940105282
  0.2448059211468304   0.7442148277928572   0.7631841193233962
  0.7451434413372896   0.2459271334151192   0.2799209761736360
  0.2544892487773425   0.2592125617450720   0.2440921285739000
  0.7552779378023985   0.7551745340258920   0.7541486697325817
  0.4914953581420836   0.5279860502281768  -0.0096028690153867
 -0.0127486483580925   0.0208078892895586   0.5140728069846127
  0.4860225777966952  -0.0072880158112754   0.4940034662310165
 -0.0081782585034471   0.4849335682269856   0.0113232390161475
  0.0183271973757405   0.5194765244182046   0.4971839813464570
  0.5172356284872103   0.0273765113483291  -0.0129471663934954
  0.0173993859046006  -0.0125428762275948  -0.0003559250381085
  0.5106397349971542   0.4943060820851813   0.4911551700335661
  0.5424957190883936   0.3777894430256973   0.0403193791499951
  0.0311843506210167   0.8659421380404880   0.5472963249784434
  0.9692207758015721   0.6612453929358577   0.5312041737049280
  0.4570480946474886   0.1656021231640288   0.0205191861148660
  0.4308311373439686   0.8390559449630793   0.4638660735299652
  0.9392344766174479   0.3338810016892922   0.9750045193295911
  0.0746706142507516   0.1316288391819688   0.9911445925477069
  0.5738746818580140   0.6365779219110054   0.4882483587285903
  0.3448710194128078   0.0230032476596115   0.5558480234453337
  0.8487101196277946   0.5155985385799170   0.0639774943308550
  0.6503257858104796   0.5462948256840390   0.9517749356026723
  0.1310777335430916   0.0260847932801612   0.4545777381600657
  0.8768010300417109   0.4882130581475964   0.4435437576467250
  0.3745437246604273   0.9844543310215552   0.9376844028215103
  0.1652051254738940   0.9608374714019553   0.0598387969957512
  0.6615050666241871   0.4519328987075827   0.5357680110879859
  0.0460667415808468   0.5673910149020803   0.3419528788388430
```

```
0.5474908821757604   0.0710364192825677   0.8382622260994937
0.5137991679328983   0.9267394343660835   0.6348567833299714
0.0198421964460385   0.4225552458356299   0.1517760744266064
0.4855106904005326   0.4584817136953748   0.8538257081241315
0.9532589690607035   0.9620953944623337   0.3638440600269513
0.9775714224201444   0.0333543268680829   0.1555502422684726
0.4770387665988139   0.5393763967971209   0.6467538353831517
0.9352311583907021   0.1639604641157242   0.4790229551058485
0.4282761889075529   0.6676540502452977   0.9655693723002492
0.5644922852241425   0.8657866282677581   0.9715271079092156
0.0643029239638923   0.3626033568961180   0.4733866157736184
0.0315510312243499   0.6385334054748070   0.0464892604500015
0.5363439362885847   0.1426582387639300   0.5426762042495483
0.4533280323862861   0.3535019582857880   0.5151033784458809
0.9623934721474335   0.8452143244858923   0.0281225607904888
0.1354850848035671   0.4771774648088600   0.9548578273296202
0.6348377727460625   0.9745283611819604   0.4489748507147058
0.8456900152970200   0.9865227452176991   0.5711403545718856
0.3429586371411309   0.4757564288759384   0.0362744457194032
0.6614299389400863   0.0417670740778110   0.0448448296088696
0.1614958402590590   0.5350710019559879   0.5501867150558830
0.3703844886838065   0.5302299727941517   0.4349466041348089
0.8766247389926922   0.0237247236203988   0.9485365029258224
0.4780813210817055   0.9644224960643417   0.1396671778974376
0.9778350576638424   0.4595109970725516   0.6469367919457418
0.9584091299912407   0.5339296006133140   0.8597871192050158
0.4681518458975275   0.0531465464608928   0.3519686900798155
0.0275997566903760   0.0731076261964358   0.6573924150975677
0.5091609446116361   0.5786825384649074   0.1368790424550195
0.5445770525511568   0.4515605839482008   0.3359321189321170
0.0454076713322927   0.9390897170381776   0.8462202539956905
0.2927884385074961   0.7914220754547415   0.1068866698543891
0.8046782585218429   0.2880036710356405   0.6185190515148207
0.6847425422301785   0.2798759764770533   0.4217407379321046
0.1718579624732994   0.7662129373196473   0.9001359167552885
0.1976220699917242   0.2187226430854402   0.5968909320030613
0.7047875896081796   0.7188864550132543   0.1012111146267497
0.7966545063324281   0.7141824258037925   0.9015846184783972
0.2984008777179935   0.2378253332880969   0.3907447019166626
0.1084969386600391   0.2927832654201327   0.7836398118047253
0.6052700611300531   0.7984038661361252   0.2730993316795355
0.3975393089452111   0.7184421159421928   0.2842897594413477
0.8983252230291103   0.2125692109496354   0.7891791221077932
0.6084285477596095   0.1931102678398995   0.2321942939769584
0.0954555679490175   0.6812931788871827   0.7184701986834083
0.9025376443987786   0.8109369620805664   0.7263875944818859
0.3938283230460551   0.3275906868493585   0.2189767296970709
0.7779853402068535   0.1003390516279972   0.3280218742239281
0.2705462024628993   0.6006856286677701   0.8127307838031234
0.2730584858991796   0.3905299396090108   0.6880641735848697
0.7792453411868379   0.8984051244922012   0.1952738583697101
0.2158551673694937   0.6146621763980973   0.2103902533426946
0.7161690936977250   0.1123992536149129   0.7242645213954478
0.7285686560921519   0.8984449909585158   0.8121887559899434
0.2086756104921063   0.4104272934575371   0.2897435306033809
0.6942901441126941   0.2249651115195293   0.9232932417739473
0.1851464520206522   0.7270021110244660   0.4076873120935600
0.3001513318095010   0.7125082264721149   0.6106333312062098
0.8030435393556121   0.2139531732781826   0.1233751999053796
0.8033809168275695   0.7706281020374716   0.3960142443420412
0.3042734702438029   0.2759752827629214   0.8904631415205870
0.1947378987908886   0.2889245576144404   0.0929305413625729
0.7052306754495079   0.7895278642808197   0.6006782020390308
0.9012071975626547   0.6883172716526137   0.2230520777167194
0.4026762038396876   0.1883062860562051   0.7168513942065516
0.6165653295419000   0.3141430810234220   0.7353749704624385
0.0959182801814016   0.8179503446225177   0.2281280785175301
0.3982227959850108   0.7849570400313850   0.7915737796018305
0.9004349820272416   0.2855093080319024   0.3034787510052677
0.1081329843387576   0.2070580397824096   0.2727388824763636
0.6093487810457152   0.6965346145026127   0.7688093915701820
0.2273684147941544   0.8875963211239060   0.7113618271556442
0.7310030547922820   0.3906500488437270   0.2214755123856707
0.7159890246607126   0.6031755120473979   0.2937864030005131
0.2152798247265527   0.0981582967977480   0.7875851410941610
0.7804880933022181   0.4052394068777027   0.8362402850216969
0.2631083999090787   0.8990858844705281   0.3228749677799311
0.2809186240025925   0.1158939356679018   0.1859297945369049
0.7862470508637794   0.6085676879982709   0.6887611114432919
0.2348544968411968   0.8867831963521233   0.5172852787371921
0.7335624391112189   0.3793375180283143   0.0304160882451639
0.7621822886402743   0.6257483217974731   0.5057584789279337
0.2674021107812342   0.1239366524050919   0.0013373051678425
0.8813284425749270   0.4972482520337176   0.2562294279971299
0.3896150390113169  -0.0017615909464125   0.7494435555762095
0.6408918357418090   0.5036719457189586   0.7530968641636295
0.1398881444773692   0.0192061256032273   0.2686664599642908
0.2348559620931203   0.6358164089739197   0.0099176162464023
0.7308737985812229   0.1186944164485425   0.5171677174345519
0.7629661878271168   0.8844855559989022  -0.0007193779215662
0.2665944497535530   0.3830210984339785   0.4997827632034048
0.6312817615725964   0.0067782685133972   0.2592486563429070
0.1384306059586886   0.4915000184382938   0.7630820598166277
0.8772808486435560   0.0050868115960673   0.7604217367986367
0.3757782304212298   0.5149726886133044   0.2511630297274189
```

```
   0.4934332875125900   0.2648081312398100   0.8676004402172453
  -0.0134144704764045   0.7665412337087099   0.3793010796123896
   0.5055788393305629   0.7454840297285947   0.6254030114444904
  -0.0011821185711394   0.2337467756204005   0.1300626527343277
  -0.0052497396280031   0.2539037665330270   0.6272656271384431
   0.4798406908899804   0.7667065562395029   0.1181906829298116
  -0.0154226193224344   0.7387325274296671   0.8741363880268480
   0.4865974008244495   0.2396497924559700   0.3670851888927159
   0.7448982092225571   0.6585501955909822   0.0717503801664574
   0.2203305590067503   0.1533880977492611   0.5593827715847309
   0.0593714484881263   0.6885270881567052   0.6485061786154717
   0.6495508035627297   0.0853242249554736   0.7664232843521813
   0.1980205347583788   0.6562029644344447   0.4377946008715187
   0.6954084415832961   0.1512227308610940   0.9531656954607383
   0.1506830891739964   0.0777553529496772   0.7470476725980494
   0.6477263060674346   0.5814928614943484   0.2639433419005661
   0.7658791392761719   0.1511418189866214   0.0936059342580297
   0.2714811895207372   0.6448613435347668   0.5826031745806176
   0.8865440731391044   0.9972061988336556   0.3415119666261137
   0.9320359484698660   0.3016485313221058   0.9023698605337591
   0.4233883849973048   0.8044524604254593   0.3919727076716414
   0.9147644999043459   0.4990636062848827   0.6723145681786664
   0.4189495173461470   0.0024504129320155   0.1747272743224592
   0.3219237285415520   0.4144699282674365  -0.0081180089530865
   0.8885679653549258   0.5003908365805335   0.8409360765871685
   0.4204218168641845   0.5004801317439985   0.6850561996218453
   0.9144106287074394  -0.0071838429214392   0.1796541556564262
   0.9221293804369481   0.2388624185650439   0.3651197372483586
   0.4212921877164490   0.7402244038787327   0.8542674857125705
   0.7696269033580450   0.3465530761251464   0.5822624165895894
   0.5754160435699940   0.7671360456700045   0.3384250670473234
   0.0839654539899202   0.2540249245091872   0.8472682715656866
   0.1441955367143287   0.4293983055776583   0.2503556417274184
   0.7308134971063923   0.8567395471644785   0.5693267398613687
   0.1943804085830397   0.3601848374362174   0.0616803736995924
   0.0598192324949625   0.8080728261126143   0.1592349954606092
  -0.0162419125448868   0.6671674212132405   0.1041577607616721
   0.4939481548307524   0.1752085326209174   0.6015997807204896
   0.1147723045602345   0.5753087865451560   0.3008298047276401
   0.6900146301474047   0.6151703911543597   0.9435128514537910
   0.5004864772904065   0.3455137795519657   0.0998896573650129
  -0.0141194043260043   0.8364650935026682   0.6067803440857301
   0.2039479890700229   0.8920656163237080   0.0539267843573708
   0.6748200058043043   0.3843578518803328   0.4893060086301013
   0.6166722257467838   0.4365703380257454   0.3071034771815662
   0.1168701457271857   0.9318592195149717   0.8135250982325445
   0.5644219719131192   0.8389848443357084   0.8979669457484505
   0.0513895932576161   0.3271350353046341   0.4047204043801971
HNdSbTeO
   1.00000000000000000
    12.5726499419551097   -0.0786505658927354   -0.0373283672317491
    -0.0784952758644129   12.5984067622433713   -0.0172590547233196
    -0.0368732457187332   -0.0173756472894937   12.5546091095303467
    Sb   Te    O    Nd    H
     8    8   96   24   32
Direct
   0.7627137939610402   0.2528506017406501   0.7673325448681779
   0.7632632077823822   0.7454589024957120   0.2453669638338907
   0.2367647773814420   0.7577018552726180   0.7583623281350965
   0.2584324586342511   0.2519439077841805   0.2345037715815753
  -0.0000725988122661   0.0198899575643053   0.4896036293163668
  -0.0071839637018913   0.4910400759394056  -0.0014694974943165
   0.5051580518741877   0.0092523969857515   0.0021554732291026
   0.4995904784424456   0.4867294068611434   0.5113749329316162
   0.2613904140927288   0.7411135135195407   0.2627078193259122
   0.2571676994276447   0.2591720362763397   0.7450054678674551
   0.7502994157498714   0.2366262338063987   0.2698590374086666
   0.7364602631895278   0.7578715286213550   0.7516535301909291
   0.4944687038072225   0.5032085917368166   0.0094172723344045
   0.5047851725586233  -0.0125733354960669   0.4948345208091158
   0.0070954025841718   0.5032973016624110   0.5071957775371001
   0.0074405741659520  -0.0059106904434029  -0.0162456214278174
   0.5356702023094659   0.3551867822664647   0.0545118072013008
   0.0586366994704648   0.8750696824958840   0.5010287462702224
   0.9488142673710597   0.6509322660828567   0.5390800141479009
   0.4441878521942872   0.1500155369375763   0.0164152161047786
   0.4502572442793638   0.8459578206283765   0.4809117832318878
   0.9356422064571145   0.3399069745225830   0.9696234131190163
   0.0606595539493033   0.1360393819010669   0.9639736636225604
   0.5422946832103577   0.6402692960207667   0.4670225904647253
   0.3663198064167392   0.0193650003667074   0.5531860623926427
   0.8562908756408597   0.5111790851982195   0.0681266675178885
   0.6389158606498130   0.5229447425717708   0.9643737743748848
   0.1541755617224139   0.0428459896827151   0.4310870721263584
   0.8538561969230726   0.4699640591157326   0.4570599800175209
   0.3690891041108442   0.9705946270941085   0.9294795165605888
   0.1500651920526187   0.9601647417714885   0.0313847022851395
   0.6475929746840124   0.4795933836509594   0.5635760927898439
   0.0177262595274587   0.5612390856661257   0.3689551310674810
   0.5414819613317747   0.0617994732997537   0.8452547025012999
   0.5357432013063078   0.9383964715448748   0.6513364808631077
   0.0301606335511647   0.4268114470040689   0.1502178488614091
   0.4740712805276061   0.4465293351384607   0.8682608622126040
   0.9710907328052374   0.9721244021699991   0.3336403251097049
   0.9717379148174659   0.0347069373392624   0.1280045668188383
```

```
 0.4647266787996258   0.5432592408175718   0.6522299859023835
 0.9443723797556766   0.1648577496389950   0.4638955498336778
 0.4531341559436470   0.6529425268197793   0.9676612267322496
 0.5606899529257829   0.8554900265571205   0.9671005702981530
 0.0529256288850743   0.3654842870325022   0.4749937964520277
 0.0380134707310287   0.6438998199940575   0.0413469361389070
 0.5550395811635570   0.1308445379968638   0.5281458691545385
 0.4504265159047096   0.3321031199534925   0.5354789391151348
 0.9493196788450776   0.8544054371769865   0.0078054754530601
 0.1391665244225787   0.4798775871772841   0.9549794506076581
 0.6464648201286000   0.9630080939695762   0.4469393221337454
 0.8561530981282633   0.9718391818474118   0.5346494458520145
 0.3511024251028993   0.4834033411317489   0.0528213060130918
 0.6541800718156746   0.0302573721404937   0.0536789340106600
 0.1557105054330052   0.5448088396147519   0.5442231467999616
 0.3622970080495242   0.5007236146076366   0.4422858670095795
 0.8712968027005462   0.0244245418189940   0.9195650045596614
 0.4758450977455956   0.9492044409622069   0.1431421824402587
 0.9821722599469527   0.4637331847635695   0.6483522967179322
 0.9579466849739054   0.5442502005942130   0.8550893958498882
 0.4677664872906798   0.0344673624307240   0.3526788146436613
 0.0242524908131189   0.0684252139556483   0.6375053090963422
 0.5218967110309267   0.5628653150853427   0.1468331009438417
 0.5382764654406126   0.4294833665326511   0.3617125793825108
 0.0409328907986535   0.9384618307569059   0.8325714949344175
 0.3125629834538804   0.7740925686305919   0.1214105500707366
 0.7992353137550323   0.2966878649249072   0.6124945242139220
 0.7118668622439812   0.2772209915420585   0.4094907596309576
 0.1803139508521880   0.7856325188695931   0.9131150969594420
 0.2024044047331369   0.2342655598300320   0.6031973966463375
 0.7064080913272952   0.7204228199240309   0.0939981010959996
 0.7891950252130446   0.7277277550759234   0.8912158812624723
 0.3009125958602983   0.2161766720734830   0.3822991751180279
 0.1211488583259819   0.3177742509059162   0.7736739429033448
 0.6236668790534831   0.8057472122483198   0.2647593435126528
 0.4041996416460995   0.6942675228557813   0.2938211108760981
 0.9130442416385081   0.2396529266850347   0.7938009135086463
 0.6092406274022252   0.1900850439025313   0.2363916153909169
 0.0933609588468482   0.6962865731214195   0.7272161923621210
 0.8802466648366905   0.7993377100239525   0.7162753353423953
 0.4052126470921966   0.3066998969246083   0.2140200689684959
 0.7827089447356256   0.0939082373327762   0.3106906618846259
 0.2692600006123875   0.6200319918789711   0.8240907447664698
 0.2880052494673421   0.4119038578988660   0.6852283542707612
 0.7956805294308985   0.8961926754291660   0.1860806884005889
 0.2227367937588297   0.6039728625162081   0.2089699734192173
 0.7448716104088451   0.1116378539424441   0.7041917508047320
 0.7064125428522107   0.9012764957224111   0.7897826246320311
 0.2149230657954431   0.3923684819649465   0.2843091184654673
 0.6899401444039968   0.2411408480226813   0.9035301102487774
 0.2057153040387142   0.7212097981043805   0.4050435427075592
 0.2996508985176063   0.7318317578373935   0.6182332558461162
 0.7938617674232321   0.2032371267573171   0.1143273986758706
 0.8172216829358799   0.7889367445670982   0.3982624352448901
 0.3075075297663725   0.2848935662750661   0.8871858540288727
 0.2047618169753996   0.2790728792292826   0.0807629622574664
 0.6821738136661896   0.7635525898853496   0.6090323289070481
 0.9072878948960998   0.6948004798573354   0.2200601906713591
 0.4023611024989142   0.2189227034738532   0.7097477213868402
 0.6110393502889505   0.3047029746530426   0.7208588781092751
 0.1243064878614711   0.8045605967749269   0.2317670390287115
 0.3908406281250446   0.8122870349455615   0.7988936776966294
 0.9050593001575536   0.2751302565043661   0.2900126494438295
 0.1190331509109332   0.1795377323187180   0.2395211826231132
 0.5969557523498034   0.6951674798139381   0.7850508994593025
 0.2156388070008114   0.9059826296290504   0.7127352247477174
 0.7384350361851850   0.3777714092500877   0.2134913649512755
 0.7362300411987475   0.6071987319016546   0.3139203870339266
 0.2304221171373321   0.1156839411629699   0.7901258751523470
 0.7756570739652560   0.4124408364314660   0.8154042784841283
 0.2919355398109557   0.8968245114367228   0.3128077305849708
 0.2939352692811247   0.0972839780584745   0.1775468701896679
 0.7795948606762420   0.6052542177018910   0.7036816932917470
 0.4961856984975734   0.2573100602565104   0.8788336169793074
 0.0129549811249394   0.7534421509964508   0.3705594522381155
 0.4823064442431925   0.7382349089881367   0.6314578242682723
-0.0026825161932761   0.2266999367527675   0.1148052959363160
 0.2450221011699811   0.8774929357524972   0.5136561144415422
 0.7367108451561039   0.3753854462123479   0.0221683010108175
 0.7414932668310904   0.6306525809688044   0.4991602490612866
 0.2537638444661779   0.1188666042148460  -0.0221502852124909
 0.8789279446734368   0.5008346264901650   0.2536875910866828
 0.3747486564905448   0.0156946082438594   0.7377591211600794
 0.6165332590260543   0.5075224632696972   0.7615635743109489
 0.1363811368020869  -0.0079153722392236   0.2280566292927774
 0.0090133638792306   0.2530915802855773   0.6315017051769716
 0.5000370158460072   0.7561840222156549   0.1352199923729381
-0.0177118619887332   0.7360555010609726   0.8683073668256607
 0.4893043868712421   0.2287257171997274   0.3749045521000773
 0.2410715349638584   0.6308903307260939   0.0126018193505087
 0.7472289681141177   0.1179739042784733   0.5099808689982102
 0.7576543698546183   0.8898710256107257  -0.0111159369803255
 0.2424017455818870   0.3652929038962131   0.4784018491872668
 0.6271186480112653  -0.0066036375015019   0.2520668303838236
 0.1334293156974211   0.5063248743625961   0.7656385143532392
```

```
   0.8836631612720620  -0.0117694314391856   0.7284135909143538
   0.3733233152821493   0.4986860807444655   0.2534776785254325
   0.7250875236708684   0.6560942946634178   0.0535240008972184
   0.7493601454541711   0.1433900001891553   0.0834690480880297
   0.8455211165649863   0.5804936408060554   0.7383324059291534
   0.3664515102144050   0.0681516821770227   0.1823688990598716
   0.9088112192561388   0.0142687815498847   0.3096031948003178
   0.9385036188796502   0.2998807738983810   0.8941719061657195
   0.8256205120218481   0.3970594029501725   0.4677226404736817
   0.7515606011668798   0.3545196121277599   0.5901196173069214
   0.4389276539923693   0.2924510127729838   0.6054723029675499
   0.4816922304780913   0.6753068549686196   0.8978834330913503
   0.9764575764801013   0.6814993157182960   0.6074467141799527
   0.8045955935994936   0.8579240408809329   0.4354461257822741
   0.9318028821964361   0.2323612897454269   0.3569498008190730
   0.3837101440902463   0.8570667892680970   0.8662133430451769
   0.3645309496822953   0.9294298312048712   0.3096392291289568
   0.8480591712687576   0.4452703278682018   0.8125303118171679
   0.8690513832055593   0.9259513027187868   0.1834503663603231
   0.3555818474770587   0.4533259488379059   0.6969042123175858
   0.1839294323623457   0.8549902581132977   0.9538089612973568
   0.5469323523540198   0.8278006479892632   0.8950816658338554
   0.2327269423537896   0.3413013784280262   0.0415264689159707
   0.5858541897413569   0.2700968942795746   0.6550294198219329
  -0.0013975561735204   0.6770194793314055   0.1014438169594155
   0.1872046120318577   0.6065851259232503   0.5046670806365058
   0.6136839281745947   0.0509546620329646   0.8179345148992014
   0.5001788282095202   0.6719811254250330   0.4072181292049974
   0.1942754397355878   0.1108621204579379   0.4372015074757805
   0.6078520765058611   0.9417460605198841   0.6846490751728336
   0.1012920701799176   0.4339902461195065   0.1862289098664636
   0.4879990199812735   0.3248886379493582   0.1146613283109207
   0.6108891782970596   0.4418281780153492   0.3356411686210539
   0.1109132548359922   0.9461325829413121   0.7923216304885824
HNdTeO
   1.00000000000000
    12.5233472462372415    0.0623132200975952    0.1333490389599520
     0.0625237330969940   12.5348514841708756    0.0113238298698791
     0.1346361909082642    0.0124762435220896   12.5002694919393420
     Te    O    Nd    H
     16    96    24    24
Direct
   0.7407587250893353   0.2581244478022119   0.7474493498267937
   0.2470913887084781   0.7635103521133260   0.2381951966391228
   0.7551412811228959   0.7336155880125637   0.2559624042704901
   0.2520751656837790   0.2425879217465554   0.7568142994155517
   0.2460912918709622   0.7376673140053707   0.7356936949263690
   0.7520751200952562   0.2399588442651770   0.2366504131712107
   0.2471587348765406   0.2609649095734956   0.2486255407708493
   0.7567782704329331   0.7624269160086905   0.7590859065462596
   0.4868637098329915   0.4913664962405945  -0.0070327954759875
  -0.0089466163333616  -0.0087824661872684   0.5003438049825093
   0.4893763235100009   0.0056836665576726   0.5018961636787391
  -0.0260487142872057   0.5047280096926318  -0.0106276317633417
   0.0132510107074429   0.4936677592304796   0.4902658702786921
   0.4991999258414062  -0.0120695469355469   0.0065930414188861
   0.0016250583143722   0.0025062349837355   0.0062841751547968
   0.5186458962743091   0.5108350325808465   0.5035474836248917
   0.5273215638386207   0.3488388378595563   0.0313461033035272
   0.0429625270576684   0.8429236717220148   0.5381824522827325
   0.9639739978864937   0.6374907059415431   0.5079994514188004
   0.4488530948593962   0.1325575545987135   0.0182834004285104
   0.4444597357557513   0.8637899238688271   0.4651084964742818
   0.9320541874137900   0.3617041037870949   0.9637518225686086
   0.0510409634561866   0.1548785544464048   0.9641923716931392
   0.5701854294698520   0.6526423199412752   0.4704233539237065
   0.3456512949823276   0.0344577145372660   0.5554012227571634
   0.8332270998557337   0.5338204695073550   0.0495679576896223
   0.6292862465955386   0.5245740805950565   0.9474874423282456
   0.1354692900165802   0.0133309512384465   0.4529524381717053
   0.8729833692739593   0.4652655774520211   0.4397899790414740
   0.3585511612787239   0.9591902026705610   0.9517314618534165
   0.1442328107871558   0.9825035201580645   0.0509692713579564
   0.6581067187991837   0.4751212441802704   0.5521254753340341
   0.0477879018649672   0.5341342760813428   0.3378991034908381
   0.5358778114178829   0.0347311676634784   0.8522005849812455
   0.5185892535968705   0.9498066802987898   0.6429344528164320
   0.0142711060892918   0.4574721302208197   0.1311710499448064
   0.4616496701732932   0.4430369028558453   0.8504867928381391
  -0.0335385824583504   0.9397862049354728   0.3602522724510840
   0.9755167375332385   0.0677057529081478   0.1444858978481874
   0.4769645303368170   0.5598396160384528   0.6423167028829659
   0.9442214393985050   0.1347985709918112   0.4840529660104148
   0.4237607033944231   0.6323937338947927   0.9673231328518547
   0.5554093494708660   0.8470630613304683   0.9715127968821615
   0.0661546318490137   0.3402975232206308   0.4599006051388511
   0.0259677964203835   0.6554326278473245   0.0312181210288604
   0.5410478251226657   0.1566680340463341   0.5366377512111006
   0.4596315716817017   0.3716394946961434   0.5066838831351284
   0.9439133160336580   0.8647477752402103   0.0312799671911020
   0.1214193811149310   0.4878174264002657   0.9383838237072111
   0.6351978962527582   0.9827406059310986   0.4509227359895742
   0.8529006266385963   0.9650249281127123   0.5638075636364859
   0.3324924758943176   0.4626346026102779   0.0531657934310534
   0.6390665949827961   0.0180748331487975   0.0570469460399004
```

```
0.1536741135825339   0.5159376478811621   0.5392089911964202
0.3661475975740114   0.5488806416419588   0.4521661796311052
0.8497363403096181   0.0465083259668476   0.9653529113205288
0.4655301998851848   0.9332024146865522   0.1505238811342966
0.9790795334280815   0.4375600253042246   0.6297641901433272
0.9567225052970059   0.5666760290362456   0.8477036602503542
0.4687782929769866   0.0653712578794103   0.3620274713668390
0.0310559315309063   0.0412121221612432   0.6521106162926089
0.5074495859878521   0.5457293144543627   0.1357336574875509
0.5406569777158040   0.4731799793563415   0.3473053509758934
0.0152498951086031   0.9605272792148247   0.8595852929956764
0.2998815652721764   0.7891616689627623   0.0848052932072498
0.7985067801560606   0.2903339143584488   0.5939540866628283
0.7007996759621870   0.2591549688593906   0.3834721361221233
0.1984187369060968   0.7604481302934187   0.8809597250122948
0.2071040026624993   0.2050452926811601   0.6007241904866579
0.7043094966295979   0.6988940058727672   0.1057831707974566
0.8042251626413405   0.7302736751869722   0.8981748475365051
0.3123981079394875   0.2356890501234451   0.3839309675940327
0.1044652877570843   0.2878761927550828   0.7738494430773245
0.6001042152825258   0.7740689164985992   0.2835251029220515
0.3885650496995182   0.7111611977178397   0.2607394890979055
0.8856130948485880   0.2030412045419424   0.7648151780503111
0.6124238043372233   0.1804854346143072   0.2168697921652552
0.1078112103304508   0.6853454406012818   0.7106714446992695
0.8979559592687454   0.8133945968662414   0.7227510611639092
0.4037461582993407   0.3175312461049337   0.2120233258268399
0.7894225308322036   0.0883332723935892   0.2882925646415143
0.2828179563118186   0.5876504385293747   0.7853610743724322
0.2728602061289289   0.3816702534185277   0.6940913123946678
0.7587243821830549   0.8747414577282651   0.2000356937567704
0.2170319066082737   0.6233174254808300   0.1867962729544519
0.7170730450330688   0.1181812946618366   0.6872792668194483
0.7180510266703615   0.9039523507549225   0.8073128509371722
0.2262240746669995   0.4068249494252647   0.3019559881077564
0.6904915258695562   0.2216652470070047   0.8883831757269518
0.1991003230489358   0.7395939581913779   0.3822985841771938
0.3157284394158529   0.7173024004376677   0.5983643816029328
0.8064089698600119   0.2128223955949862   0.0930052416762208
0.8054247719441252   0.7679549549817578   0.3939901363022018
0.3004122468572543   0.2761395904565370   0.8963915129595904
0.2054759866513740   0.2917986673609623   0.1041602108998407
0.7205436304409660   0.7891126380806314   0.6024743321331246
0.8973565707315948   0.6843808574552871   0.2178422877072177
0.3942994754865253   0.1862232150822265   0.7230352064432879
0.6002493330297531   0.3089526618451311   0.7141941126314569
0.1101288531916237   0.8236529870194305   0.2135789254273938
0.3965696939065936   0.7833779340701409   0.7803697300507886
0.8925743642640198   0.2922601648524888   0.2678988230779318
0.1035129800619522   0.2193933022275713   0.2852772646738458
0.6178097169114137   0.7030066501403653   0.7603002826017553
0.2264179738486342   0.8832284226996249   0.6922476807040775
0.7156601088846936   0.3821389944966312   0.1918563750065252
0.7219019283479567   0.5928019539888397   0.3078791596510118
0.2188866741032143   0.1003994936300882   0.8057081054776858
0.7716206263062192   0.3988031718438614   0.7966360385453880
0.2891346109003702   0.9116336361589698   0.2814436509428297
0.2736081946207062   0.1176176145056730   0.1990692313807970
0.7977507224804468   0.6121372499872098   0.7066238132220052
0.4985379794643224   0.2247065134764608Ø   0.8690664179370240
0.0032113011797815   0.7510072105956050   0.3587463709958325
0.5102458982954639   0.7599722032412700   0.6145696792168226
0.0053506349175348   0.2668313612371701   0.1169115658114336
0.2569663728361370   0.8749006580668892   0.4983524060894453
0.7382902645717645   0.3699973737564344  -0.0019525807960512
0.7719212015901022   0.6177075719082404   0.5042615362815094
0.2595032888828712   0.1255420366696779   0.0050581681754389
0.8665741100998551   0.4829443596272005   0.2477539993587698
0.3673178183831426  -0.0063078638684785   0.7565491249280124
0.6204745210172169   0.5042604970057158   0.7563410528252467
0.1196986773843381   0.0180965098833675   0.2550006715363147
0.0005564571868539   0.2481304630089720   0.6240956484912803
0.5021068834191327   0.7353649605095824   0.1086233626718199
0.0031778638928891   0.7626103597475280   0.8684414063116185
0.5075418928210005   0.2506811628739356   0.3613542313749620
0.2332053825474794   0.6189600994187981  -0.0007846552911500
0.7484774272556460   0.1208909607961988   0.4998759970553925
0.7519206975726086   0.8750141060574297   0.0093389465966627
0.2694336444699079   0.3719521185513231   0.5063090299288208
0.6129608622252958  -0.0062668794407939   0.2531550285785099
0.1140093396868698   0.4889820664908743   0.7479836079593003
0.8615208928950820   0.0136043938733549   0.7573829627924514
0.3655864512622348   0.5182667436949107   0.2532282930325615
0.8203210773408376   0.1093066793015280   0.0100188237990312
0.3325584059115668   0.6082718120442142   0.4946645869964946
0.7627709090728393   0.8501234734267528   0.5715833844361101
0.2943509435752339   0.3931737529646722   0.0443030138214442
0.0945387220058773   0.0112563290496419   0.6875298496917615
0.6039909460758669   0.5107248777186602   0.3146172728859990
0.8558414975482571   0.5814991466643853   0.7535444271852627
0.3559234944940035   0.9359942938766571   0.2418421651954222
-0.0129192422656938  0.6820236439412087   0.0966466072518899
0.5147421020438303   0.1889108558559890   0.6049015321347839
0.4213597155425206   0.7505018339258777   0.8478328439840043
0.0612108628413829   0.3005193433100018   0.3891929790888290
```

```
  0.7675393338423110   0.3547489785244209   0.5601915018439535
  0.2910096073189383   0.8584870194194575   0.0459234238360508
  0.7461802895024032   0.6365277204928473   0.0719065416244144
  0.2486016478164188   0.1425859397870260   0.5704959442860090
  0.3497362111041658   0.5535265554371243   0.7574736728776059
  0.8549149976395881   0.0590260044974974   0.2529389714784033
  0.5983308146027492  -0.0042193728313564   0.8199509899491410
  0.1123342663324201   0.4954867023158795   0.3076993102763649
  0.5708655552079701   0.7358093481460403   0.3480520979094990
  0.0400948374200391   0.1844139444534495   0.8915111184641591
  0.0049855505417733   0.8093486809823500   0.6001693648692520
  0.4430772997634319   0.3108684682288073   0.1406199214462997
HNdWO
   1.00000000000000
    12.4206609367129754     0.0341140947389500     0.0311340717854473
     0.0345253248360360    12.4125810670796941    -0.0194555837461067
     0.0310102571681808    -0.0197775029561429    12.4343366179337611
    W    O   Nd    H
    16   96   24   24
Direct
   0.7542925958331324   0.2557699197590177   0.7553701684428645
   0.2601485207435656   0.7521190980412752   0.2575717836380368
   0.7533359396463191   0.7504496737758890   0.2479870167058999
   0.2403676252585124   0.2460369939853895   0.7536335895156363
   0.2562789550264656   0.7466215146154934   0.7386688136352009
   0.7477384006863678   0.2630980492006734   0.2504155284684037
   0.2496436269650668   0.2535134902271883   0.2434383165315828
   0.7476016868475238   0.7663527555184729   0.7535415839987636
   0.4873010830717097   0.5109143629862372  -0.0109459117016152
   0.0042613346217100  -0.0076910071323130   0.4899038619968021
   0.5188967031179146  -0.0132869069214811   0.4945405666188907
  -0.0213254642132429   0.4967385035698464   0.0016693116682578
   0.0003549220374286   0.4925155892763177   0.4917365198569387
   0.4894692644415259   0.0029677858391766   0.0123227005704960
   0.0062254083004579  -0.0091909273001758   0.0084900823327616
   0.4901665128733639   0.4887971148378835   0.5036304405794767
   0.5422665540079740   0.3493966321614628   0.0331117192507277
   0.0507664522442627   0.8565657236387112   0.5270864005456081
   0.9495016000013803   0.6553530203880210   0.5286795246286451
   0.4480190236880344   0.1436203365751619   0.0267667597238642
   0.4436090481998589   0.8537135023135215   0.4764057207955956
   0.9397610348727363   0.3585673021879197   0.9733217528752216
   0.0495488550384729   0.1520954569427031   0.9656435227265636
   0.5527156282558249   0.6437530078799905   0.4645141625315501
   0.3531522102022132   0.0330288349146266   0.5495092788884526
   0.8453657422918155   0.5372776148836780   0.0543514003056293
   0.6369342508598802   0.5190527936221013   0.9495769351791620
   0.1460493082360923   0.0275650443747132   0.4450353193482648
   0.8604784264438285   0.4767039500363948   0.4479991242125920
   0.3584839372500390   0.9726962266822095   0.9409398435841225
   0.1479650216979555   0.9741751171596559   0.0461667758803502
   0.6518839255616899   0.4664982959055279   0.5402061172495233
   0.0328105975332941   0.5614536726278456   0.3583516357143212
   0.5342396384525410   0.0487533929622021   0.8522633830389056
   0.5255291505209632   0.9443620723240003   0.6417478365123284
   0.0254905305773336   0.4587454161342039   0.1412901345571285
   0.4682045116339543   0.4447451566226835   0.8523500461600523
   0.9673869461392575   0.9533391278974568   0.3526166462501008
   0.9793954144365294   0.0610964761093790   0.1424857477005201
   0.4842439977221356   0.5484249150981100   0.6373944787047761
   0.9616686947917842   0.1564045764156862   0.4671769490997167
   0.4422455644479926   0.6489230356604719   0.9620161745414874
   0.5498973524755081   0.8459786100571854   0.9663361197898638
   0.0573157448336546   0.3591064530584236   0.4678694005372097
   0.0485684324656339   0.6554580153455247   0.0326723136327313
   0.5512168254882818   0.1343729836140779   0.5234302573186146
   0.4512456038359352   0.3488186002342944   0.5233110553245097
   0.9460874997705471   0.8583788442679515   0.0321050172631691
   0.1424293610054662   0.4679476913361179   0.9540196285582289
   0.6562331121321335   0.9586331322908433   0.4460900581375659
   0.8668245379745964   0.9976935812995887   0.5556960146032680
   0.3497452515914906   0.4756234436755787   0.0490006761161888
   0.6374814513679835   0.0160128180492978   0.0498457447156317
   0.1549249022339729   0.5363589535117399   0.5464969883793198
   0.3619113826199964   0.5239413888501574   0.4421569495699458
   0.8491946122942527   0.0381366173010802   0.9572968394712387
   0.4649696771017382   0.9409515582598155   0.1432611185375585
   0.9829427363252005   0.4602272064597683   0.6396973027454010
   0.9847599107319968   0.5613423899376677   0.8641013874616477
   0.4747308249976583   0.0342209965994184   0.3521409716204605
   0.0390632029977166   0.0580868738503007   0.6473361897189253
   0.5238906264628677   0.5546671426055779   0.1358936128378387
   0.5353920141131671   0.4441662908138311   0.3481944331859823
   0.0169990759225140   0.9555898946660594   0.8602561212094110
   0.3071700478355911   0.7663207323878160   0.1130091090955326
   0.8107932910635685   0.2857605829828320   0.6109500815136658
   0.6912743248880242   0.2798216790385040   0.3936338038182027
   0.1944977850918832   0.7918475724726507   0.8972211153215626
   0.2000627350932428   0.2310197696441137   0.6079315405146206
   0.7036451695758099   0.7069788164646732   0.0992056496558539
   0.8064295142276907   0.7255846061976539   0.8890495749453605
   0.3067095611838875   0.2162246417756400   0.3865530388672574
   0.1102125792708705   0.3105330733394799   0.7835326501145622
   0.6105997691695175   0.7982456107053280   0.2707845106574559
   0.3962120082757372   0.6930726173450649   0.2865864089768994
```

```
   0.8993295521352570   0.1974553403924041   0.7835739965008996
   0.6145634758818559   0.1873256814105088   0.2226536548398703
   0.1094253299535289   0.6991487877056382   0.7213115615615040
   0.8894788545846951   0.8088626986953481   0.7127662292375951
   0.3956323402561003   0.3033774099835165   0.2128506328059718
   0.7866098069165767   0.1021476478842047   0.3115986157781102
   0.2747292377664552   0.6124366672577369   0.8123733988351379
   0.2914378080065085   0.3998427833315457   0.6983555466076102
   0.7804900681823909   0.8881222657236436   0.1899546589466467
   0.2121713809651517   0.6032680336103222   0.2047096076931987
   0.7201648808150196   0.1139177873254341   0.6999222222265395
   0.7151946403917379   0.9022956960081296   0.8056571367266551
   0.2230832091632804   0.3944801892252374   0.3025784350081372
   0.6961836550407866   0.2208603599521177   0.8956571986638135
   0.1999375101355173   0.7222438425895469   0.3952457794124146
   0.3071375826613905   0.7104306747454238   0.6046165029159890
   0.8044824415951024   0.2215127825459688   0.1125586290569925
   0.8026288355979689   0.7952222011167526   0.3971772962082258
   0.3129078397284831   0.2755060866869354   0.8836189732422564
   0.1918080063431634   0.2860967240753267   0.1031643783791034
   0.6959604042639123   0.7741110323119978   0.6070131222278481
   0.8967939799881052   0.6984719565560318   0.2237977665731254
   0.4003864608828155   0.1921403850664941   0.7110525221629843
   0.6130279245862188   0.3112899922231594   0.7208848748652326
   0.1224860506311516   0.8172688880344449   0.2272482314535610
   0.3942814606019641   0.8017616852017159   0.7794384061436235
   0.8902896526896771   0.3092717459924446   0.2841283436168613
   0.1079953511483448   0.1997157432042510   0.2817979081608791
   0.6144098976661098   0.6940261632323457   0.7787236588529331
   0.2226394360328644   0.8910023553441457   0.6920633095728781
   0.7133687352925085   0.3995343479092315   0.2014051223756647
   0.7272385419687841   0.6126914272267141   0.3109023488549520
   0.2187615486262743   0.1065523974316978   0.7977755670939387
   0.7846132633771087   0.3963257359463347   0.8130129886320313
   0.2982113678956906   0.9017415014491897   0.3063575524032112
   0.2790807945667740   0.1104770284308681   0.1860524946806943
   0.7896830672585597   0.6017702280515694   0.6934410182848419
   0.5071749249559486   0.2505351596851890   0.8662685352152156
   0.0092054047794154   0.7544060698001503   0.3662152871100959
   0.5042944168297721   0.7507699138196839   0.6297036035662800
  -0.0005478912123122   0.2545872376807595   0.1342746575571544
   0.2519997796975109   0.8750208457475389   0.5045856113394641
   0.7411071202909274   0.3685510536041502   0.0029493867861481
   0.7518796721987996   0.6315861692886188   0.5017138375334587
   0.2523402178972476   0.1342164534149485  -0.0008606282367031
   0.8760308476598105   0.5021230871501503   0.2531618366177765
   0.3750514784267083  -0.0063328049206882   0.7483716038040962
   0.6346199505184126   0.4998192621639924   0.7603836468118061
   0.1343323253786932   0.0108233682970382   0.2494525220182698
  -0.0032131572015712   0.2584333243573870   0.6325696894105077
   0.4989600086709212   0.7406184355756269   0.1265220644065303
  -0.0025057375715191   0.7582209623174294   0.8635486591438916
   0.4954087807760215   0.2300953984750879   0.3635276575281571
   0.2449041587809795   0.6299749444144849  -0.0006616191086615
   0.7456694350723134   0.1396541749220940   0.5096385179751366
   0.7511994630833503   0.8741927688892790  -0.0017590917029443
   0.2521701927595836   0.3654753696811378   0.4886076661269508
   0.6310935971635507  -0.0004351778771702   0.2429809589275594
   0.1296636682321516   0.5075320505393591   0.7519804735334741
   0.8722638001740863   0.0095205954002202   0.7478493371332110
   0.3773467648243259   0.4909772363222640   0.2387767723555923
   0.8128211945508368   0.1013122340008213  -0.0103193141894522
   0.1926117646644693   0.5984932909030731   0.5151251371596218
   0.7656677746732946   0.8589601844666042   0.4301980385922037
   0.1799696801754649   0.4094533557713799  -0.0070386946942631
   0.1047372182657894   0.0369035945177201   0.6857662939037282
   0.6035654224245479   0.4614593688795237   0.3115533653089956
   0.8539987427519194   0.5619182170378657   0.7164467446312945
   0.3688840900461954   0.9402698207720178   0.3021323904816743
   0.0250830193071670   0.6942461446465990   0.0976246335290244
   0.4379915138520767   0.2049359927146998   0.6425139557372317
   0.5187656260174884   0.8079376114607687   0.9041986738357304
  -0.0008505956377901   0.1943081390615191   0.4081797728701181
   0.6896796749396600   0.4029620526531206   0.5096636597325056
   0.2199645302455736   0.8579387215955596  -0.0676357153898269
   0.7357409238820357   0.6423046321359017   0.0635319032168800
   0.3174124396060389   0.0957350768698540   0.5169377938003451
   0.3534822322980666   0.4313614539049307   0.7365685094250283
   0.8542091489889869   0.0655089817452671   0.2919245274510668
   0.6022317498084292   0.0256604119113387   0.8180295689782937
   0.1523398775992367   0.5644081710828233   0.2401334992893844
   0.5248293536401492   0.6795751276986505   0.3995854541446040
   0.0141397095248908   0.1825106530346149   0.9013918570700489
  -0.0301903949305215   0.6953591261106717   0.5939979105044408
   0.5056375927392336   0.3134751792418362   0.0938019425813309
HYZO
   1.00000000000000
    12.4821883024794928     0.0038565964062407     0.0117368416639903
     0.0038303626474657    12.5108515236093680    -0.0386386521490523
     0.0114916710195917    -0.0384402543350188    12.5640479834789573
   Zr    O    Y    H
   16    96    24    56
Direct
   0.0114340105165149  -0.0167285787970404  -0.0171076366679254
   0.5191290575250032   0.5052597616978801   0.4768044330609000
```

```
0.5173563063094517  -0.0097575666952824   0.5222860839292448
0.0136202093345668   0.4742249908645093   0.0135848256732973
0.5120336219568798   0.5198657645464819  -0.0254999474776690
0.0151149182758844   0.0189520040152851   0.4924070875277017
0.0186536568675494   0.5273423087627021   0.5197964397491682
0.5049378732091355   0.0191832938144787   0.0229457654016824
0.2229924486230375   0.2636586983249385   0.2506126971649903
0.7548244147623953   0.7364079231470552   0.7470190951567847
0.2439134048534332   0.7581291243015660   0.7559387451871477
0.7422704394537590   0.2738712248670347   0.2535031907419130
0.7810445723952801   0.7669751178327257   0.2262950160676709
0.2400414131493319   0.2474854447603382   0.7671912431196518
0.7333542536999506   0.2404024729283267   0.7522576349846397
0.2709569550272871   0.7694127034960846   0.2590470475713588
0.0744642234755666   0.1820609014067217   0.2871579464629979
0.5920822674984110   0.6704152783921045   0.7655368051666428
0.2777654398127835   0.1079559396530490   0.1935298904646439
0.7652936232042519   0.5914601475206471   0.6629452184277409
0.1582334353454853   0.2598235408609856   0.1057231926268887
0.6843227345653735   0.7826978108618892   0.5894249392561588
0.5887221054315718   0.1768579168338676   0.2310688163356678
0.1028384178213166   0.6761526619255759   0.7469260494583355
0.7834110235619783   0.1116051609459027   0.3138879686240355
0.2845440353469296   0.5986067107449583   0.8255747699938628
0.6729979442595144   0.2731863339342224   0.4081540532332225
0.1890443519975198   0.7859742873051662   0.9192073438943412
0.4190126873738555   0.6863856063789400   0.2841086487647768
0.9036957553710434   0.2229360416626088   0.7922858621295696
0.2243584945893951   0.5999230510892651   0.1917434868309184
0.7483100967236846   0.0955318006102677   0.6806372440063059
0.3343087673973750   0.7663344175180165   0.1040908809753911
0.8057273285191895   0.2881943072330297   0.5951044994041697
0.9274032739796954   0.6945855212535639   0.2256704193553452
0.3997827853899970   0.1887507355922192   0.7342215483714737
0.7419716869706168   0.6078006366813956   0.3021689754843673
0.2302427199951702   0.0975793532862474   0.8259830752264409
0.8237460945782027   0.7907206661602111   0.3974054542610345
0.2989737550589623   0.2941113213794924   0.9122398310448364
0.4371341446849943   0.3500895948674045   0.5179022702532503
0.9330347155659248   0.8454846070729536   0.0064054975220210
0.3521324803408394   0.5351199915623499   0.4323015855250504
0.8588491406451311   0.0255018344603009   0.9097126579403425
0.5154948900679599   0.4236363018135099   0.3300079469582818
0.0358813926112188   0.9210029974483228   0.8258221326834702
0.4354021451528469   0.8533663892473773   0.4984662969871635
0.9312248671576463   0.3387510058891544   0.9893520872477193
0.3603531986344787   0.0319211108739038   0.5857999260534787
0.8504084561452443   0.5208709343325936   0.0719078501564355
0.5410888334967654   0.9297201430150240   0.6762817097768136
0.0490910685656837   0.4621029208233855   0.1650881777099432
0.9446587792822473   0.6748711146739418   0.5327290152447098
0.4383112125409603   0.1672142256789152   0.0345226280091534
0.8673443369694098   0.4905010183203722   0.4379286886993741
0.3525633742373872   0.9702672200782821   0.9614143034138852
0.0492152581166443   0.5711564326364862   0.3605515158728529
0.5216362913528787   0.0638250939975091   0.8551562934893804
0.9498802441492314   0.1676761329796613   0.4753507252965865
0.4503282253718285   0.6708089283742389   0.9499641203790611
0.8648547805800468   0.9709025175565523   0.5446454206976599
0.3574401637965768   0.4830949957660999   0.0500001753402169
0.0233272941296356   0.0767964802707871   0.6634535789751947
0.5305027103195153   0.5658910715620465   0.1417694034510381
0.9039522433118614   0.7887797256413411   0.7082221697305451
0.3859137310338407   0.3038754315018296   0.2072208818125094
0.7191639456846954   0.8906658227635648   0.7975098889011126
0.2342764093734915   0.4062444162156989   0.3217738856188800
0.8034954622715017   0.7057667147108472   0.8963602806890933
0.2965069443502474   0.2123169400700913   0.4012787920068865
0.4024423096488135   0.8008493857572841   0.7876312053365912
0.8899925462710123   0.3304133411698142   0.2537549820625003
0.2234003896257798   0.9040943938312341   0.6977826354904526
0.6691349785583723   0.4081712598837425   0.2051775641516815
0.3026519198519503   0.7096461760283587   0.6034032416872904
0.7866223305408251   0.2173873731256676   0.1003046306246493
0.5889221540568665   0.2996845896412429   0.7013591382819161
0.1130174190071115   0.8149358713623465   0.2091812007013548
0.7672314349372874   0.3978754295295250   0.8069799986412133
0.2845570226355471   0.9206449958487242   0.3103595267732117
0.6762619920481441   0.2217258964898957   0.9032485766705700
0.1988674071158894   0.7392290634455390   0.4007553496984371
0.0989051115254805   0.3257870395100902   0.7625607333015592
0.6178421859618668   0.8044387932569240   0.2882535547693147
0.3015390826778064   0.4026818573872533   0.7019923941288793
0.7804846065405088   0.9228701753245662   0.1815599019180199
0.1919459302232053   0.2254202425090201   0.5974849508375369
0.6965497540900014   0.7374110200130737   0.0929471418460088
0.5659680224501452   0.6561870717570866   0.4631456160132485
0.0477055670781619   0.1494949410782247   0.9605480369395050
0.6625060883847209   0.4565189999709633   0.5291139664110971
0.1530330968522002   0.9838184604637685   0.0579989914405062
0.4715082044804278   0.5484960015116026   0.6345206679946724
0.9577277549484075   0.0456263778702018   0.1392143575388707
0.5480412259824073   0.1593086367917717   0.5380730959789657
0.0583420547912851   0.6458828476944376   0.0272007468958621
0.6596621086940471   0.9827318994321101   0.4486116355710723
```

```
  0.1656220261985249   0.4655682076182467   0.9435927145561438
  0.4641620407524142   0.0466200528879520   0.3656394115171506
  0.9706820503953902   0.5415222179875319   0.8586809474196669
  0.0592842161765112   0.3633238128821037   0.4791226847856001
  0.5591168025899719   0.8603528731238554   0.9684659915971122
  0.1660503610945022   0.5177091422064788   0.5785846861318423
  0.6702407388321413   0.0523499107603018   0.0247051427835727
  0.9570014398573203   0.4582870892227292   0.6568781416418751
  0.5105369683425136   0.9417570922224577   0.1606975602806397
  0.0711451258031706   0.8581032664148655   0.5374486466986206
  0.5664365561887801   0.3667759222626303   0.0267143779450103
  0.1761324566809213   0.0496208623953396   0.4654263330056256
  0.6707641149111134   0.5423627701969438   0.9348622853711068
  0.0050124279690904   0.9436073281219254   0.3484707106237677
  0.4822757914493045   0.4404693895539313   0.8398880130141431
  0.1307582916790327  -0.0014305067536368   0.2387198357336789
  0.6350701269275969   0.4826279496656072   0.7250573167509553
  0.2467188403359004   0.1356435132961676   0.0035290089243853
  0.7508338734681408   0.6380891509483304   0.4915826218057588
 -0.0201050202412073   0.2312402441838871   0.1239794259351315
  0.4904198232276441   0.7431113137601540   0.6243243769553883
  0.6405353445677086  -0.0084480559347391   0.2724822057713210
  0.1348947435948394   0.5013071812471664   0.7596967695534220
  0.7703694529981290   0.1200175393192104   0.5046472111772552
  0.2518712045280262   0.6405278965565137   0.0123358848892200
  0.4825784960244306   0.2319021799812564   0.3777876240320669
 -0.0153711145175999   0.7331272579948713   0.8703935618746248
  0.3706659190580350   0.4952620004068745   0.2405829970774560
  0.8705222339574646  -0.0291669134085495   0.7234348983770443
  0.5183813269664314   0.7607797365532624   0.1284870289720233
 -0.0081307401266053   0.2629560342649969   0.6331473931640120
  0.9044183521219715   0.5136002019074463   0.2535581531018831
  0.3673111420089941  -0.0092556207518186   0.7685126257427731
  0.0177856774046157   0.7667117728864226   0.3725714221682819
  0.4937948458534543   0.2584109619029082   0.8835324347421589
  0.2518866659900040   0.3779031940419968   0.5024869104129369
  0.7521313538803137   0.8745251633176268  -0.0183356148864689
  0.2541801810362125   0.8884751113978338   0.5034894711526164
  0.7546662793309175   0.3745461331418649  -0.0204711354406133
  0.1858909946947036   0.1585234768559202   0.5527808102216455
  0.5481411050039466   0.1845405321719452   0.1644862543590746
  0.1584578908703088   0.5540350124002973   0.1900297233620973
  0.5522033279864029   0.6848551796521940   0.8310804042726404
  0.0299977921028070   0.1822272795466476   0.3536913744562874
  0.6815986104261978   0.5659132477520205   0.2760605020474380
  0.7492888953805267   0.1501349073730808   0.0783785962652914
  0.2615279322635841   0.6473787308919747   0.5777580172464127
  0.4518380802298173   0.2008603309482909   0.6776327956561292
  0.3376353232280381   0.0654435545047299   0.2197598155615780
  0.2587813686132494   0.1432070997224377   0.4243485664289384
  0.9223377218239517   0.2810563915666652   0.8422918717558067
  0.4576498932685028   0.6906823076243119   0.3525510585651385
  0.3537666344680927   0.5624782368456985   0.8202294899616823
  0.8489502691720479   0.0746239745755786   0.2922249819667246
  0.3314212964036038   0.9047172185952194  -0.0008061919106791
  0.8199744619109535   0.4324552669783459   0.4593633965534149
  0.8850949530060056   0.4884543210409436   0.6738882693187616
  0.4671187757903773   0.3299554087510349   0.5874503603529380
  0.3291471720322585   0.4167606983571194   0.0172900512844614
  0.4021804183101960   0.0020084793255053   0.3412437443698552
  0.9067336019927918   0.5076733670359326   0.8280766135966132
  0.4276998977761632   0.7240601814894855   0.8923883248950407
  0.8375183707534630   0.1005241527858979   0.9096418251906393
  0.3109610100844506   0.6018557554869066   0.4367003982267245
  0.4030349003850391   0.5165800588115717   0.6560864726813959
  0.8924009802444883   0.0025031111549066   0.1599979859183281
  0.9357616641762323   0.7191949551112857   0.6012800235979224
  0.8180454574269522   0.5900429479992663   0.0540287427486000
  0.3285833001766516   0.1023287500281085   0.5779339925227912
  0.8213767881885425   0.8578465277356199   0.4398204358086908
  0.4085341411206794   0.2641400194562303   0.1417549078724773
  0.3668215455581396   0.4105538914704203   0.7463939145425511
  0.1932039045619792   0.8519312727256601   0.9620127176879804
  0.6695710166197589   0.3390808051663163   0.4541235175646671
  0.5981940616221219   0.7706373268819551   0.3560388821931084
  0.6993838273398970   0.8523258821021050   0.5570251339565964
  0.0673885737976937   0.7990419885113885   0.1481809870976117
  0.0872227416962144   0.0634522150073128   0.7065718745989251
  0.5947808431609553   0.5241021725382240   0.1634272642129805
  0.0200980229415193   0.6700849505959170   0.0922290535382554
  0.5696792867463247   0.2109923497810729   0.5960164870100184
  0.6757873017155424   0.1116769028790201  -0.0308255636966077
  0.1168666234940128   0.5424672548964531   0.3324518966434356
  0.5884218505474993   0.0421773322080585   0.8205000178151204
  0.0165343657206565   0.1819165664943921   0.8967378525153985
  0.7193475623440472   0.6072964203114833   0.9278496354189221
  0.6089381110309315   0.9223337538662695   0.7197950718190170
  0.5851805976385878   0.3664224505019077   0.1042064813281144
  0.0482095703680256   0.8267635572149159   0.6048115562242382
  0.7662056103550664   0.3531525733222131   0.5746795912177189
  0.5833694031456919   0.4166995071167241   0.2764175788487955
  0.1030302241499792   0.9261326464388359   0.7834141012112708
  0.5294122155673135   0.8424964041079923   0.8984554149789380
  0.0184158390783664   0.3253863041581445   0.4242962704194064
  0.2095738955826064   0.3977170016412988   0.9399942236318903
```

```
LBLNO
   1.00000000000000
     13.0450538889671357    -0.0105780960734758     0.0034091054600339
     -0.0101602712093738    12.9243876253405539     0.0010130472951017
      0.0045653991352013    -0.0002478201443051    12.6615736040127924
   Nb   O   La   Ba   Li
   16   96   16    8   48
Direct
  0.7537883310896982  0.2557169461937867  0.7467428311863722
  0.2513978751740050  0.7540939883103694  0.2492010388530261
  0.7561192590713566  0.7495610314685224  0.2484875551471830
  0.2557222258081205  0.2511470862625562  0.7494743448193970
  0.2509500583542057  0.7482862271165298  0.7466398303337982
  0.7526428218234138  0.2534844144374658  0.2432746733012023
  0.2583956126859646  0.2415528490883664  0.2470080800993264
  0.7553042008493092  0.7458662053669922  0.7454380300419613
  0.5004151079699674  0.5040653188104105 -0.0007040462031278
 -0.0032716575269571 -0.0008372389710203  0.4979326954823257
  0.4960843397361228 -0.0000759493472041  0.4965338376791991
  0.0001235575981137  0.4977447853550285 -0.0042307099617213
 -0.0037673565366977  0.4973471944603295  0.4999335807583364
  0.5011323086206519 -0.0029568099486354  0.0020459178650299
  0.0022948377176360 -0.0024572636976356 -0.0006768159933692
  0.5002979947411741  0.4967238078037745  0.4983714807668489
  0.5416466957604434  0.3631969590675035  0.0390764449161452
  0.0331399650768562  0.8597213656160870  0.5432341139517887
  0.9429225578826971  0.6339510826984240  0.5296454718837121
  0.4433577777430532  0.1343364539842479  0.0280865139274247
  0.4365740642755987  0.8633762821430090  0.4684709847013228
  0.9454106637179083  0.3597333061163549  0.9653027308150235
  0.0467876498586796  0.1386833760784540  0.9539137364563054
  0.5381172877187779  0.6373563283216205  0.4537206586728451
  0.3497857916029229  0.0340516789324129  0.5551057510041437
  0.8589375057575172  0.5274493908012291  0.0526665730902611
  0.6447865344231183  0.5368495473646779  0.9579003562815647
  0.1425576277735881  0.0269119173911228  0.4583743244800356
  0.8540199195957084  0.4619854502716087  0.4442264167193302
  0.3588318550424386  0.9579102110204208  0.9461238335631630
  0.1469205088743944  0.9752830417199926  0.0460655460973460
  0.6436745928098744  0.4750846142078175  0.5447193055738250
  0.0319168708600958  0.5446706097746733  0.3469589447483919
  0.5330785879335502  0.0436356297624439  0.8477692550222851
  0.5309362950972036  0.9509034119368600  0.6491954632103135
  0.0363945782842282  0.4503671556769913  0.1474096055545124
  0.4696397322303054  0.4590312237612633  0.8506417811320407
  0.9763287631377627  0.9556178912754290  0.3501465797844820
  0.9679049417133262  0.0454436294110510  0.1476309297726666
  0.4686407201426900  0.5433300237231282  0.6466130467835597
  0.9402644391406751  0.1420927587709983  0.4537822260627141
  0.4498965703800837  0.6471940314837684  0.9581849004022592
  0.5548946778817638  0.8539729318023221  0.9566660363115090
  0.0466683061560786  0.3533775745870655  0.4520702915652588
  0.0590429080828932  0.6352336966943063  0.0425713708863890
  0.5391633771891510  0.1410372791591306  0.5460787202173625
  0.4418350480995522  0.3574011249330843  0.5400372946470332
  0.9461552289267471  0.8620248490588528  0.0410903471337266
  0.1482266509778294  0.4696923141819934  0.9467086930876335
  0.6431881064430087  0.9760668030414332  0.4596506850855566
  0.8520910491441744  0.9821193863441198  0.5580655721968372
  0.3508172778806545  0.4794067677798768  0.0453526800442817
  0.6484068770828483  0.0284760704887927  0.0414412740607989
  0.1450110369754894  0.5241528570998543  0.5406077336377801
  0.3507334515640360  0.5200657364303574  0.4465052124058967
  0.8532333908464536  0.0248387460423731  0.9510039715541432
  0.4813916366315852  0.9535902066628857  0.1457843244072611
  0.9769993380843089  0.4509414651185914  0.6435021731828133
  0.9760419736145928  0.5463718271209805  0.8532980786918205
  0.4781528289813955  0.0452957605026826  0.3532377722740559
  0.0324569319996870  0.0479227691956840  0.6498054199980485
  0.5270196570274127  0.5464388115957637  0.1514030715549440
  0.5321822322696151  0.4502310778859170  0.3494643933217475
  0.0318181896016915  0.9509122451373471  0.8496053401268463
  0.3024544436399172  0.7727093877766587  0.1060600507727353
  0.8034935001794863  0.2735808305374617  0.6025960253654371
  0.7117972821351710  0.2879573225242468  0.3910767000177601
  0.2071831637425879  0.7847092799899530  0.8922560177267009
  0.2134775315666768  0.2127765719598204  0.6018198838664386
  0.7006467432780698  0.7111586370969155  0.1058169893172011
  0.7958744302410981  0.7252978680397316  0.8917490592989383
  0.3053851026134080  0.2236942631888292  0.3926367632939272
  0.1165998556575348  0.2972141034842781  0.7931223608442289
  0.6177833244385907  0.8067133641683115  0.2992133537918751
  0.3877246757630067  0.6930222292297067  0.2951437152883761
  0.8886419890415919  0.1937008401168631  0.7918817299414821
  0.6141852499594980  0.2079319763861138  0.1982254089789481
  0.1132394439093508  0.6958961428501034  0.7028832230154684
  0.8908361087162382  0.8064001770231519  0.7068383933233519
  0.3915749905850269  0.3077174322477877  0.2039953507208868
  0.7751328703065750  0.1086775690609025  0.2913618540560333
  0.2799017116084273  0.6032978798209384  0.7973197519471927
  0.2761506258611994  0.3981114272368308  0.6980697562306982
  0.7750257921076495  0.8933340609561842  0.1894580341555683
  0.2215970691323527  0.6018182511489650  0.2045591982202774
  0.7183849291933314  0.1066520413126118  0.7012378424806740
  0.7209191801482253  0.8966571905506160  0.7915701123295315
```

```
  0.2195478123157639   0.3915352313570397   0.2934564419388546
  0.7056840511172625   0.2188806035749672   0.8989155527512763
  0.2035307865399460   0.7176983940947687   0.4021084660942043
  0.2994005468400659   0.7127716032748819   0.5987316502270053
  0.8013509763345800   0.2187682164524668   0.0947616887783844
  0.8026365131266462   0.7731718104569724   0.3972392283545085
  0.3026338133888939   0.2846812494867776   0.8973190004832682
  0.2090968699554261   0.2786249123049084   0.0955927808601700
  0.7076440224573688   0.7833850424351249   0.5942689158101644
  0.8916544621465110   0.7000395550615814   0.2123395620429797
  0.3961600400741507   0.2056222450261803   0.7143344109685638
  0.6173553626681636   0.3128597583484626   0.7146625219647433
  0.1143005634598000   0.8031034147887894   0.2122240174180631
  0.3881128733433241   0.8060737263749130   0.7750905802018279
  0.8894944209023651   0.3123508512581387   0.2725422618981058
  0.1209440452819185   0.1853168622640124   0.2798549896054449
  0.6160682397116706   0.6894714954766389   0.7777189266982127
  0.2164551097116240   0.8966337167730090   0.6960601386293788
  0.7185390188528114   0.4029405533461453   0.1925679823661754
  0.7198278034674532   0.5993550423366985   0.2981957348926368
  0.2263584095889483   0.1008081928647908   0.7988092144491460
  0.7810376504975219   0.3974011599077162   0.8028536179583899
  0.2783376785493750   0.8954789365014908   0.3035652208688397
  0.2860246478666074   0.1019055237653468   0.1927356433242019
  0.7825122227847686   0.6045925736546766   0.6912298917451096
  0.2557073098820674   0.8781431306784383   0.4963236860884492
  0.7592562178629880   0.3814920187525528  -0.0066707487325023
  0.7577234358851791   0.6205968960174423   0.4991479669418574
  0.2598755599254972   0.1177853940301801   0.0007874074963297
  0.8773531343893115   0.5026057492788520   0.2434022110941604
  0.3778559179824416  -0.0051076723192584   0.7499162732806238
  0.6357644236931984   0.5017264899433945   0.7467880387612526
  0.1346710824549877  -0.0024217173085048   0.2465686496800141
  0.2439098431782883   0.6288108491357540   0.0032558158005428
  0.7515250862439887   0.1254584402349755   0.4984780934911118
  0.7515190778904117   0.8699926314803128  -0.0012666123297747
  0.2483495685472199   0.3719966458712122   0.4941725854041685
  0.6283083364152673  -0.0049232271724358   0.2527167648097930
  0.1203120382119449   0.4993102854253392   0.7491093999983793
  0.8706807531863341   0.0020501413462227   0.7503953573335391
  0.3701023394734776   0.5002172227952106   0.2499317647164109
  0.5059918590681078   0.2514393499886166   0.8739714346582405
  0.0022228108897307   0.7512409399824358   0.3736505141058465
  0.5047198330338661   0.7478922286854976   0.6220099686010986
  0.0059014786995904   0.2469885637907092   0.1242723057121392
  0.0019416551142204   0.2498282132856309   0.6287611887195574
  0.5001559206205493   0.7530895862483400   0.1350720532846068
 -0.0018186875747812   0.7461838486661737   0.8767747623582728
  0.5048254136110479   0.2472042309342632   0.3703242817961305
  0.6662418465575562   0.1725963811380037   0.0526688594611803
  0.1661341077054400   0.6669520347919671   0.5531057706294954
  0.8456263475547052   0.8364704154024312   0.5529422842146879
  0.1802231631309433   0.4241070192213686   0.1300460670130008
  0.0790124584551628   0.0935018664738742   0.8109142155688530
  0.5717374110407977   0.3989202731894702   0.1835542440676721
  0.8784242341840544   0.5003063699272816   0.7474729684519799
  0.3806412081980818  -0.0025419617471552   0.2489807103047092
  0.0037762742595655   0.7505129375295977   0.1197511214843205
  0.5265183763492284   0.2661311807888420   0.6173786547888179
  0.4155107853330086   0.8019400109150265   0.9185253637170228
  0.0560390871173646   0.2141855500443314   0.4015333307645022
  0.7455387839459772   0.3748907937247364   0.5174221009518953
  0.2493759555048641   0.8729063126105971   0.0168021476018539
  0.7676468937787785   0.6269040349415819  -0.0051989562363768
  0.2475002896016486   0.1209071712331454   0.4749141573592195
  0.4213631892171945   0.4073048673933551   0.6918241398207888
  0.9177731550728054  -0.0871181646720574   0.1981087195264550
  0.6775892124528917   0.0704257586215060   0.8730741275468596
  0.0815557666316453   0.5860928262855892   0.1909238879880146
  0.4946552830049643   0.7651282507551274   0.3741764069896569
 -0.0045497517245480   0.2601223674605060   0.8711762792255688
 -0.0089335828669681   0.7322962629965761   0.6281613446563454
  0.4917429913106495   0.2342240640350327   0.1243884065926618
  0.1761380109079709   0.9288264468997713   0.8493869643606038
  0.6793206252529485   0.4360373906640542   0.3467226450516626
  0.6725738158568310   0.5643720307669016   0.1458780365551123
  0.1794593999570701   0.0641536178192545   0.6407127714808328
  0.0755678879768209   0.4047285965138053   0.3058835528190473
  0.5787671339797038   0.9009385556539861   0.8095314645687071
  0.4272843040893117   0.6019713649498065   0.8112970704726937
  0.9160968205543755   0.0846151527031824   0.3079094712625143
  0.5771625979045363   0.0866209267820072   0.6921922015935132
  0.8424334363928198   0.1695549796802706   0.9395455917752067
  0.3419151689193466   0.6655758887467147   0.4421870192253348
  0.3436432815195523   0.3340346988137167   0.0545892674508433
  0.6806293869640334   0.8379462890629479   0.4416040288776449
  0.1665882402361743   0.3259750304395260   0.9424889377967737
  0.3221887276597679   0.5650190959774362   0.6341127867779560
  0.8213743867341231   0.0698388161631489   0.1149514578214626
  0.6774890339192270   0.9291418378846543   0.6284586988146825
  0.1763437653063751   0.5710236361256177   0.3717853193999525
  0.3201661688344942   0.4374143458604920   0.8761960729367722
  0.5735063081388487   0.5937426999316883   0.3123492763241590
  0.0694339352978485   0.9033083949630446   0.6857647953419727
  0.5887875621653387   0.6983274452149721   0.9204464895915402
```

```
   0.3813199759263566   0.1796048972167001   0.5714569096671904
   0.8954988378928698   0.3141456469574471   0.4183514549658827
LBNO
   1.00000000000000
    13.1958000783397580   -0.0086815051162873    0.0009777573823656
    -0.0089143313550527   13.2133211070443011    0.0017134904598489
     0.0015992614769326    0.0019350276296462   13.2003751921211929
   Nb   O   Ba   Li
   16   96   24   64
Direct
   0.7442742152640826   0.2522978366914606   0.7509108124782450
   0.2450572453547111   0.7455949784180956   0.2499586597091251
   0.7493293305302796   0.7501977699828457   0.2576884731912303
   0.2527607998958923   0.2437131278787539   0.7575544077932327
   0.2529225859731445   0.7504319055972127   0.7436191199585428
   0.7529729681156370   0.2538243023420491   0.2477607521228320
   0.2424547896395682   0.2440003569203773   0.2513129216900533
   0.7478968943797601   0.7431002261681998   0.7475547197507441
   0.4951183713790381   0.5032548800753884   0.0014402665759083
  -0.0027638813155175   0.0041426658012555   0.4993682463009621
   0.5022014223825638   0.0004411026564506   0.5007964622714798
  -0.0003752045161238   0.4996968375029160  -0.0028487447797658
   0.0044407362731851   0.4982307166081285   0.5048748262152581
   0.5053352775302246  -0.0014731022234257   0.0047929103227025
  -0.0020493791905501  -0.0072070028958464  -0.0054872208308359
   0.4947851313268608   0.4905396091329038   0.4958735614676957
   0.5445058471124836   0.3676003287538016   0.0427177560895874
   0.0379110727303092   0.8615809706609481   0.5419657345473295
   0.9592577212129902   0.6402513538117892   0.5370743488439107
   0.4599105113257998   0.1381487932104058   0.0404428823410044
   0.4651636004305985   0.8535941802842077   0.4589626212080455
   0.9545828027706852   0.3591342702067706   0.9666288889818763
   0.0451340022207539   0.1311170731641912   0.9613733838891632
   0.5441116479192514   0.6273893846302491   0.4549921913175368
   0.3599440492173046   0.0417765185472479   0.5371798992344751
   0.8539090549774824   0.5422933388492252   0.0347684623999246
   0.6320801107627455   0.5441058176022807   0.9515953464896115
   0.1375561262308455   0.0400733786289177   0.4564237272524720
   0.8601143087186595   0.4542860275563452   0.4667650573196175
   0.3648041072829368   0.9523466941603964   0.9643435981960848
   0.1390910276336235   0.9552760065471753   0.0379491110332800
   0.6343091513501868   0.4514688789215750   0.5402687488594742
   0.0428648584436737   0.5406853620816852   0.3651676326210035
   0.5412643512538229   0.0450145504506720   0.8669154787379664
   0.5346132005082491   0.9554888269434124   0.6411745703047357
   0.0416545139880146   0.4644019802165167   0.1409185850581351
   0.4569078348802786   0.4572537081943222   0.8628905011049604
   0.9629159076919216   0.9596414525942508   0.3577821276161529
   0.9542205856739927   0.0337930917681646   0.1353437478760165
   0.4556267270340223   0.5370993825739164   0.6338519599907811
   0.9545575654746967   0.1451974997157724   0.4693027753852504
   0.4502240961484416   0.6473272396838329   0.9654550558935958
   0.5494711407177550   0.8547489992908364   0.9656485287939816
   0.0444785554966518   0.3544969199536252   0.4715393945471368
   0.0419237733429908   0.6435277647387998   0.0347188554504883
   0.5455756870884982   0.1409603743780787   0.5327425403848443
   0.4508938390759014   0.3476191124356426   0.5324585840832593
   0.9534389364219432   0.8507771207698839   0.0354826242533886
   0.1400004407520333   0.4649928001411991   0.9556423312209279
   0.6441559219951895   0.9629135200488360   0.4583248612307991
   0.8536792994751123   0.9608395166016711   0.5384721252092449
   0.3529144779846055   0.4647603725784973   0.0446431451870209
   0.6474080124852549   0.0328588876659469   0.0487146170735106
   0.1457582706558420   0.5300303419971003   0.5488338346407923
   0.3517403322828431   0.5288275769188814   0.4511933984444683
   0.8546480702892642   0.0315253841129129   0.9504348791003335
   0.4653603981885117   0.9533447533441740   0.1469230135616847
   0.9639790292288523   0.4594589772907087   0.6486176457517376
   0.9629056472070640   0.5417960707546726   0.8557919351389872
   0.4626289296528488   0.0388069344846716   0.3566431027270064
   0.0306077394578591   0.0446499852805017   0.6440722482956603
   0.5336440357182942   0.5485025371270587   0.1459929360778525
   0.5304301769937841   0.4460235875826414   0.3510547500129905
   0.0297886096929335   0.9507210842051812   0.8519546480330035
   0.2822243922893476   0.7858621493233886   0.1043220349931498
   0.7925514316103591   0.2850141800823908   0.6106554675627920
   0.7024101192039047   0.2885673767708645   0.3856052265527161
   0.2037127035062561   0.7909419583551380   0.8802554399410499
   0.2035769498667083   0.2042983449311030   0.6205088725850861
   0.7112061023272734   0.7087518812443051   0.1142759560189394
   0.7889955845101454   0.7122413529061669   0.8896967442461809
   0.2870946872945647   0.2089094331594672   0.3925305366615404
   0.1137359544391321   0.2885802420326148   0.7994567998171394
   0.6075347913682837   0.7919880747760739   0.2886767780833260
   0.3867015287781278   0.7079127007391589   0.2913023421365468
   0.8883856798591514   0.2137963462138267   0.7973118533643189
   0.6136559131520333   0.2094262748685180   0.2057768878842104
   0.1154843620880112   0.7004591870018422   0.7062973401486032
   0.8926708909688270   0.7813672726602291   0.7080277724993350
   0.3850424327731159   0.2824024553221076   0.2040343045904597
   0.7934574385288042   0.1177623277707464   0.2968154255961777
   0.2923201115289795   0.6116039119033828   0.7888382765007512
   0.2917206602606627   0.3821854076187851   0.7126632859655810
   0.7881429112222212   0.8947291485518127   0.2205479432006717
   0.2129925263281242   0.6047514269013003   0.2072623277870544
```

```
 0.7058570886277091   0.1124462552677113   0.7089552050833304
 0.7051840892048170   0.8872297922256867   0.7833040785123081
 0.2061255489457811   0.3807185121536060   0.2975992187250772
 0.7029842860042481   0.2183056714549180   0.8950261705126171
 0.2022015351846113   0.7147666924815003   0.3918214739979140
 0.2967433910093337   0.7150180585043998   0.5994252184002209
 0.7965772000287636   0.2160156787190398   0.1051436306752127
 0.7939107826186084   0.7841827638267090   0.3980695069626040
 0.2974749215559191   0.2790452229881168   0.9011885405574616
 0.1991671693962202   0.2803026093571112   0.1077982355488197
 0.7066765437316662   0.7815328021433127   0.6040838190263363
 0.8924490796887822   0.7103612044773931   0.2151073158038851
 0.3955718344792766   0.1998163446643831   0.7188474545771328
 0.6037283841228143   0.2954639264279108   0.7163953396181759
 0.1036742253849557   0.7876729915621030   0.2122851235900175
 0.3956872939433727   0.7938189026377892   0.7819276754932547
 0.8967634418656496   0.2999938809042801   0.2859225182764782
 0.1008224265854650   0.1980721694153058   0.2840040869426129
 0.6058284247587503   0.7017055236403590   0.7790880218798192
 0.2156395042563455   0.8942022915661449   0.6984367945703313
 0.7181831260917033   0.3981424737294344   0.2058182600694653
 0.7172410934265142   0.6076659183829851   0.2980441519474852
 0.2165203555885641   0.0987491362978349   0.8013169717632426
 0.7837468064271561   0.3954372547095941   0.7925633613321779
 0.2859573986299696   0.8896286494274531   0.2884931774530835
 0.2848667669748706   0.1004761390233384   0.2112800584656588
 0.7848785714248774   0.6019951879910450   0.7066156026708080
 0.4999975159243133   0.2507531803934430   0.8824252170853273
 0.0010806671513798   0.7482042773681025   0.3770086506082156
 0.4974666955943011   0.7414399566609976   0.6173916212780444
-0.0023514005037874   0.2453533417450459   0.1208715355018350
 0.2480680167894688   0.8790046594505455   0.4986302748211336
 0.7445886298414076   0.3813062632146280   0.0022198160464370
 0.7463577910342202   0.6174067310981063   0.4991995567089575
 0.2494877392441308   0.1154345251413725   0.0018818572319327
 0.8805734986796411   0.4998727923663405   0.2544716045347866
 0.3796862341308516  -0.0028555556740358   0.7540914907324248
 0.6187465124882275   0.4976690364367343   0.7481563969606102
 0.1218692421631280  -0.0037250795478168   0.2476566117591606
 0.0020029232095853   0.2476156252545745   0.6352973686889407
 0.5008115302948103   0.7473958727587823   0.1286059094252210
 0.0023842047687318   0.7484753898358372   0.8708012438634358
 0.4991303522218816   0.2469177632709423   0.3672615698899344
 0.2500616083740437   0.6273286029811975  -0.0071245707776350
 0.7495124868045721   0.1266326806184583   0.4996677390360638
 0.7480705358689653   0.8701323303362878  -0.0015013905836058
 0.2483107402243039   0.3653091185527633   0.5049362256213904
 0.6273915144791183   0.0053163394459817   0.2505531983624585
 0.1292864051036271   0.4963050929070013   0.7522933678340464
 0.8671330454032283   0.0007248672641187   0.7489029300507543
 0.3683724258076939   0.4926657410412776   0.2479824742343588
 0.6801181074178868   0.0792130442270898  -0.0895692412086014
 0.3190167822825360   0.5731863752545290   0.5906454457658191
 0.7470367550102478   0.8720151661796413   0.5013214301312761
 0.3358168322559098   0.3229613298235265   0.0674084552871548
 0.1721449385604085   0.0649387392060021   0.6645650366994921
 0.5767899849161786   0.5797672616635113   0.3157517201229117
 0.8737192437875452   0.4979811187659931   0.7521274209284207
 0.3757166777042846  -0.0032876985830461   0.2501677365692079
-0.0024075146445218   0.7471601454092995   0.1242886223461094
 0.4281710336003980   0.3389461574647921   0.6746133096253201
 0.5692372318610411   0.8404426232358096   0.8227260637115922
 0.0639189188566576   0.3361143555549926   0.3303777621936404
 0.7419199353248853   0.3669683094552661   0.4948655260334842
 0.2500828964159622   0.8733869043640925  -0.0189354646420622
 0.7299670483298691   0.6282928907951580  -0.0003521625628354
 0.2422478446795294   0.1259759770667825   0.5109410248363763
 0.3741659066592309   0.4974750216733690   0.7487827215676427
 0.8756760696689132   0.0166447497941049   0.2509352306209452
 0.6199449661202675   0.0065671563111254   0.7576369023357914
 0.1273793854296991   0.4900214082310660   0.2595189266106728
 0.5006942994829490   0.7278123634681753   0.3715056650938034
 0.0069119273206595   0.2416200697851146   0.8837740756274096
 0.0159534160705487   0.7481268811280902   0.6230561674336544
 0.5061813590627247   0.2581230293702585   0.1208977687240604
 0.1697138877313450   0.9297788031240884   0.8336842617038766
 0.6700661851252895   0.4234106573781676   0.3437621103750638
 0.3285236021751678   0.0759751656556234   0.3836807393337098
 0.8306488767986088   0.5778825954707152   0.8867609493552174
 0.8329892679666562   0.4195861094900745   0.6156835249177723
 0.3324394777691453   0.9160872184758710   0.1156142488747610
 0.6753446274986171   0.5703275527272127   0.1615054344475445
 0.9187779760933320   0.8830225754803112   0.1694165989288896
 0.4261434281452564   0.6551276500833337   0.8250050461226078
 0.9328473740560657   0.1648906628256371   0.3304887276469647
 0.5661273808532167   0.1578966629595396   0.6745267303117112
 0.0801299525090340   0.6117273697753832   0.1650400337567977
 0.8398878569456969   0.1742117410701697   0.9303472805846178
 0.3387007362654277   0.6702461129420403   0.4302565930935764
 0.8851009970188629   0.8271008933143459   0.5739089076869505
 0.6085774087823547   0.8347208844415700   0.4213704918100278
 0.1612952274585096   0.3246159834348896   0.9351942755483404
 0.1613222011929633   0.6681298493657598   0.5698544887960794
 0.6593605742872630   0.1732524364052676   0.0714883311097538
 0.8178479621705540   0.0749571263286039   0.0867944556421490
```

```
     0.1770603527622178   0.4227426959032353   0.0940454691911044
     0.6675373793246031   0.9119020051268089   0.6369945050609276
     0.1832241499342251   0.5724230486771664   0.4102750974413897
     0.8296037325989860   0.9190046517344352   0.3719589122342191
     0.3234290214561554   0.4185958349122972   0.9054385460105450
     0.0764283369570144   0.0900453941726870   0.8226637713648027
     0.4137160135290409   0.1369214064360057   0.1720219100698518
     0.9190861892396625   0.6292168526699915   0.6708150930191852
     0.0741260176641066   0.9090348679716420   0.6773886700122592
     0.5776678846049752   0.4122269476815552   0.1820130096278237
     0.9137096778637804   0.3607911234011297   0.8319877477143628
     0.4195606918130974   0.8603058469954117   0.3257600405187944
     0.5883310660375867   0.3158601818348569   0.5757245628931311
     0.1351392316855022   0.8218274155067051   0.0802347919532573
     0.0945303493781996   0.1767283366603843   0.4242411073256738
     0.5842570232382099   0.6802591925209706   0.9171281710755655
     0.8596441361902800   0.6742346949661971   0.0831801101734032
     0.4074456659411478   0.1798869303259129   0.5762584068202743
     0.4140814487784599   0.8148762123107594   0.9235120840704549
     0.9068281936091417   0.3203717167480692   0.4288969698319414
  LBWO
     1.00000000000000000
         13.3464304657994450      0.0100919657819750     -0.0157110278611169
          0.0104792945823801     13.2346451902574422      0.0155782916826926
         -0.0174169405609743      0.0173799053264452     13.0463226584953045
        W    O   Ba   Li
       16   96   24   48
  Direct
     0.7513057463341323   0.2494859069816251   0.7566567720202626
     0.2501789647919862   0.7509971833813698   0.2536941955341548
     0.7456751448817941   0.7462548857708567   0.2530838539134717
     0.2417934986945536   0.2486211052707240   0.7578597217099424
     0.2427627945019468   0.7475458147388857   0.7461640036169728
     0.7426296867004685   0.2481376448045133   0.2468512222995287
     0.2549088271851990   0.2475713334211905   0.2411734243657189
     0.7542837697882608   0.7472054751647633   0.7440777230494025
     0.5004960380793625   0.5044718969224002   0.9955847501026089
     0.0020407987671307   0.0062659120498532   0.4977462356168441
     0.4956775241991735   0.9996107062006316   0.5021125045767562
     0.9966842134754563   0.5025622237427606   0.0044099288546923
     0.9962807525454810   0.4931388954590532   0.4948539185637402
     0.4963417697839334   0.9901974443172460   0.9961196411952128
     0.9989471836002795   0.9892404191235168   0.0016399165711830
     0.4990245303612620   0.4963601313904886   0.5013529169973461
     0.5497386030264099   0.3714857001281481   0.0206283926201261
     0.0514905797336763   0.8747340571825046   0.5223541526173379
     0.9455675950584460   0.6249688168649025   0.5193160823359996
     0.4458498503030465   0.1225796095624891   0.0206465423161560
     0.4545938450176811   0.8665001832392499   0.4717286118867063
     0.9461933378138353   0.3698616793520468   0.9817192042847350
     0.0491982739430023   0.1215181902338048   0.9769195904633552
     0.5412120238675513   0.6294401559770533   0.4712112198982898
     0.3615080549396696   0.0450365234119285   0.5489234688273371
     0.8661105765925199   0.5519856729968233   0.0481462066378912
     0.6311228849056515   0.5523936744656825   0.9511178122640375
     0.1403222212128072   0.0540952324276110   0.4571155415163106
     0.8657627064988389   0.4431420552767401   0.4511563870041673
     0.3656332009217214   0.9416460744577834   0.9511616484118305
     0.1304686543023130   0.9417308718459184   0.0466439774930375
     0.6301979043830250   0.4492051434386476   0.5474352123677461
     0.0293480840002048   0.5401618066540168   0.3526002743943231
     0.5299730004764763   0.0379448144883560   0.8524039325964782
     0.5280799838903406   0.9537902687468879   0.6462397533950419
     0.0307879029584190   0.4557949559395314   0.1475209944120785
     0.4665081652665770   0.4572323082862871   0.8522121139725877
     0.9683980584943165   0.9646706918301958   0.3577977564956207
     0.9642856786760620   0.0327731160096009   0.1435041378764575
     0.4667913645305555   0.5425533732520471   0.6464047823312923
     0.9546257323786654   0.1459689607231269   0.4569753905554467
     0.4499068743992806   0.6429381796876579   0.9536324448851813
     0.5471620398406687   0.8523209360045949   0.9536282581556569
     0.0472117476344598   0.3550232089643321   0.4520386938420283
     0.0468691338497667   0.6406317259365100   0.0448466283242835
     0.5438602568219624   0.1391139024212137   0.5478618456771653
     0.4498099474584241   0.3564620229804310   0.5479995202528904
     0.9498463157249563   0.8495702967215137   0.0412151727937254
     0.1338426605024771   0.4653616341221682   0.9691944626099852
     0.6328209686552537   0.9682572976129356   0.4666636220577023
     0.8697323441035141   0.9721940979673881   0.5408102582750189
     0.3629910237422552   0.4660917075266111   0.0298189920828832
     0.6327892312430099   0.0287749632495232   0.0316912587615376
     0.1330677013948491   0.5279267728964683   0.5324846195832277
     0.3601325236261715   0.5284834231461096   0.4682124369567288
     0.8634162784053894   0.0266351961831688   0.9623884513246894
     0.4702636551197633   0.9509658026541145   0.1332209094696033
     0.9693679892996018   0.4549469007486860   0.6335471814800615
     0.9701923598379441   0.5402450826435793   0.8662017510208220
     0.4629512148297683   0.0473849445747617   0.3683086568931025
     0.0365706743223481   0.0492249287108171   0.6365177785013696
     0.5260118096106077   0.5426206474136449   0.1330804139064413
     0.5248103226629727   0.4506247558623465   0.3666121094843416
     0.0258377513070166   0.9471803344693869   0.8645417970766258
     0.2934984500169504   0.7755300554372055   0.1175840945866470
     0.7901336198343127   0.2758639210602863   0.6188687175602203
     0.7033732680074574   0.2743456879963281   0.3841712060487628
```

```
0.2008055461716350   0.7736850963950794   0.8819570472738980
0.1916980726902046   0.2286924615429862   0.6220263883100341
0.7021927454908171   0.7210489328121680   0.1171148682900369
0.7950965540487878   0.7213278158262929   0.8806869289839042
0.3019527448038254   0.2263930823186170   0.3780501612805798
0.1052432518171041   0.2960164330781284   0.8041360800613636
0.6150688864931432   0.7958071982144128   0.3002041723484673
0.3815406546610192   0.7018377107100134   0.3005109832335021
0.8887285378812592   0.1999668440393884   0.7997940646276919
0.6065453982000388   0.1984674307891367   0.2027549887393275
0.1065176150721697   0.6987701738619914   0.7002555750340356
0.8907082329835531   0.7960708790785521   0.6993417149263350
0.3913017517023794   0.2953899917101256   0.1967383961423115
0.7792866398906964   0.1080140935010671   0.2872616187276505
0.2795015302908017   0.6093773681846765   0.7823072427774777
0.2810684515094095   0.3848876299834949   0.7207593771355568
0.7816213107706085   0.8888476581146401   0.2120773112243522
0.2136740520232257   0.6103362235768984   0.2171134001576410
0.7146385544136068   0.1110773110258585   0.7205636678787231
0.7167677869714242   0.8862280471473526   0.7786116476838479
0.2168550785978239   0.3851960582180989   0.2777313635846566
0.7040518213457339   0.2153172387065844   0.8993365933241498
0.2005128560672764   0.7133339959833108   0.3978005429142581
0.2909369345822184   0.7131378820782268   0.6025452501815953
0.7907799557737345   0.2104502748553445   0.1051805308678647
0.7968431802349746   0.7874414819326555   0.3957476563665039
0.2902633342482082   0.2795311066074121   0.8989847149631630
0.2054869606747874   0.2794133563238061   0.1000754759794711
0.7056239595440233   0.7819082280733278   0.5998711187845848
0.8810172283519352   0.7050101515378799   0.2247384413377386
0.3712243972659574   0.1972788054220565   0.7269864554411642
0.6199615091489423   0.2991168584473715   0.7298029546856047
0.1155267378509927   0.7945596188853097   0.2251194494322181
0.3746583254155280   0.7947925891947124   0.7734378963693820
0.8759932063860839   0.2934204013024637   0.2724101227315259
0.1239281740669897   0.1989269274627357   0.2732758689745847
0.6224088609905470   0.7002495226917003   0.7723359916048127
0.1976507445905067   0.8807085911902057   0.6995191929624063
0.6967292129104639   0.3798087831711048   0.2031693780125345
0.7113360169130850   0.6139842897749799   0.2968893138878373
0.1958521296478979   0.1119293380141886   0.7944692710556394
0.7994847209543083   0.3811171888326329   0.8008434011928355
0.2843332117568062   0.8841513368021429   0.2983128890649914
0.2992637882218594   0.1124939654433599   0.2021929582275441
0.7993899701500036   0.6151677295699558   0.6988486328577739
0.2521369163473016   0.8796253953815050   0.4974287861032673
0.7476125978563880   0.3896953246664441   0.0018537779558438
0.7452044418183258   0.6090573530505067   0.4959125750951163
0.2479552018574769   0.1025647413877269   0.9997303643129516164
0.8607786990796272   0.4970817436295795   0.2489315342825178
0.3598018033050480   0.9952603548270806   0.7515962213419117
0.6386718336478638   0.4993204692919926   0.7498136320714749
0.1334855203838477   0.9973715932826921   0.2466971198934704
0.2482065571454513   0.6214143194677696   0.0014702596640861
0.7468911859796014   0.1222257982671930   0.5024160997919581
0.7475051457529063   0.8736540161101723  -0.0007215221262047
0.2485958743161903   0.3712845515637606   0.4994745266848884
0.6201317095891968   0.9955523974046314   0.2527047076980967
0.1229296233611808   0.4972834921208663   0.7497715385728209
0.8749189123344280   0.9992328092295633   0.7496462129516164
0.3731386067206274   0.5007747882513932   0.2504800155898512
0.4971692737846139   0.2476444280929386   0.8750972319318128
0.9982802458910576   0.7494851464914702   0.3760582092462106
0.4980192527031252   0.7468052851673914   0.6269292870532153
0.9978666866731798   0.2445417985162552   0.1246816767609103
0.9910358099292480   0.2556917020448852   0.6285058075337755
0.4982917125740634   0.7468797706522791   0.1363528160081380
0.9971742527157474   0.7446752619811610   0.8716441541791672
0.5011589925889564   0.2506318210768096   0.3752665099899446
0.6596299466050530   0.1667679798159311   0.0577338004731838
0.1591713725797513   0.6661803205403239   0.5549369977527658
0.8372258149748731   0.8304309141675759   0.5522233526949815
0.3284488777681872   0.4207054230692663   0.8742532618688696
0.1199506433595692   0.9891469772182270   0.7622711111017414
0.6178235352338532   0.5061457505515294   0.2551758213483311
0.8765253131279556   0.4977416035598268   0.7498683526855734
0.3774622716691352   0.9860822853696695   0.2499624705666918
0.9191932554732802   0.8932054458696573   0.1798869849468221
0.5789353428378301   0.3077492463796058   0.5931917299379448
0.4186371652293935   0.5898119468875218   0.8082642808041916
0.0879240187512044   0.1956065919131818   0.4129273106668431
0.7474877459196428   0.3692890771025964   0.4997872943709828
0.2483774099123168   0.8654807432187794  -0.0000093760269546
0.7485161452853947   0.6286595600981588   0.0001792237092890
0.4125950870011521   0.1884538906055980   0.5903807661722061
0.3281128588219268   0.5752633950591624   0.6254664404922248
0.8252498048624348   0.0691720844818814   0.1399579682007606
0.6664257405727328   0.9177927904142389   0.6309612181037919
0.1670228582460961   0.5765623360269295   0.3681396535384572
0.4985640524617692   0.7484374217569389   0.3773515716332018
0.8343726681710570   0.1659676146649831   0.9474308366942948
0.0570980853846030   0.8296621092046955   0.6570795275870369
0.5543031419968413   0.3310867191634699   0.1577503554741782
0.1708451720836875   0.0918707196703817   0.6000986638143675
0.4197549886666298   0.4098364815305149   0.6912650791950798
```

```
    0.0784063755020692  0.4080417400825160  0.3063094801308422
    0.5784518225514872  0.9047866521086009  0.8099264093502753
    0.9183647819084145  0.0990076094774369  0.3170835520971051
    0.5762096245250707  0.0864353064295624  0.6918629493799549
    0.0759049979473330  0.5923223592497483  0.1905629018668556
    0.3301444941609855  0.6651028544120965  0.4431576680147068
    0.3377699791335348  0.3273330792543385  0.0543926993911857
    0.6680315491969240  0.8304523691520153  0.4448181067194182
    0.1591375819637427  0.3265088019676675  0.9468907333344293
    0.1685094318095420  0.4201119932468625  0.1268822554641300
    0.6675295068346433  0.0762390836132145  0.8733998085180528
    0.8297803178795573  0.9240342598359402  0.3525822347511851
    0.0548866757196265  0.1636868666560261  0.8407370203830427
    0.4392994927540249  0.1609320493884623  0.1584425563491838
    0.9373226046993841  0.6636303492887312  0.6549081702412184
    0.9390369211975677  0.3303592743208805  0.8463022018575223
    0.0809832918190479  0.7999032459003877  0.0879020184860281
    0.2873696231899088  0.0986782059035099  0.4336550609746134
    0.5805660033754277  0.6942557042346559  0.9104696380745853
    0.9172361127309778  0.6925151881495129  0.0888361803657946
    0.4171333287188895  0.7990364870856911  0.9113079148372031
    0.9179709363202359  0.3038752593899270  0.4081978815349466
LCaWO
 1.00000000000000000
    12.5455044516427101   0.0101420726370253  -0.0082737532981579
     0.0105829622424056  12.4922514106693807   0.0036112491548863
    -0.0096545830586942   0.0048358901322103  12.4613661656681014
     W    O   Ca   Li
    16   96   24   48
Direct
   0.7495451019138192  0.2478143157878838  0.7500406834104573
   0.2477048514166164  0.7523645468616155  0.2513907581122221
   0.7489030719205929  0.7465880551217896  0.2520497769124703
   0.2479798203742052  0.2514220722152712  0.7527697801421745
   0.2451381421809375  0.7464648791347568  0.7498468994411112
   0.7433567682821332  0.2473828850323387  0.2501122062973675
   0.2499526695076993  0.2493053643487760  0.2464329526293573
   0.7507331449898794  0.7485343944582672  0.7483711974440259
   0.4980213413196152  0.5005792703518385 -0.0007477047410212
  -0.0002306170073078  0.0018947304337361  0.4975504473197209
   0.4932898153921201  0.9970401233138402  0.5017900680471762
   0.9974938425701076  0.4991327348847274  0.0010205583051599
   0.9990335499487585  0.4943209592405280  0.4986737946846319
   0.4998595223656889  0.9957042584814090  0.9975431065742054
  -0.0037147558696211  0.9964310321303674  0.0003420478964039
   0.4989472908200250  0.4972972657515862  0.4971105769298907
   0.5491122393108817  0.3566730042541559  0.0229289563326267
   0.0455111903046803  0.8567581163877284  0.5216222446580441
   0.9489662788694929  0.6377967598949179  0.5223662647179950
   0.4462823821500616  0.1381728960346938  0.0222033469751261
   0.4470927943970147  0.8544694006990873  0.4754997542200966
   0.9462922388543611  0.3557585408432928  0.9783608241815648
   0.0489682593637720  0.1385122071302083  0.9753131109800858
   0.5458770304538177  0.6394323986356213  0.4713107289335356
   0.3486089771963780  0.0346110805331899  0.5537505018505825
   0.8582560614044431  0.5388631807584505  0.0552190978552780
   0.6371966225075937  0.5394737588426829  0.9446527359901521
   0.1456405199485326  0.0384901049939746  0.4505292739800610
   0.8600197356472333  0.4530177383512918  0.4467777408492759
   0.3589007935608677  0.9558231062146183  0.9441491223715470
   0.1374638696484714  0.9573512993457881  0.0526037698764462
   0.6374508967050040  0.4587664387355205  0.5530500994406643
   0.0298955690091013  0.5429340461760632  0.3505660847922443
   0.5309237119219704  0.0449884035486598  0.8497915322444782
   0.5267572352192685  0.9508143225014328  0.6493455113051723
   0.0314133488735010  0.4500340111734803  0.1483237885906381
   0.4644813589386151  0.4509322045624324  0.8513119235624137
   0.9701110467562846  0.9594308516189007  0.3507845893632538
   0.9642233282975045  0.0442896785008349  0.1467101703642711
   0.4648898753061423  0.5466482383652721  0.6469682489209364
   0.9530482962852396  0.1492039730501880  0.4642480408180433
   0.4464823786999716  0.6467740659404800  0.9628273434142337
   0.5505583914186994  0.8495874966363731  0.9603044180829128
   0.0529803021232799  0.3497647465553776  0.4607812269987030
   0.0483016778720600  0.6461762563385357  0.0363833479983378
   0.5405203152642150  0.1451810128527858  0.5395618965059253
   0.4443177203328842  0.3515268640457584  0.5389716996349128
   0.9469209152980929  0.8487845297983571  0.0360659065059708
   0.1439051826202195  0.4701869792138856  0.9577972710124066
   0.6384871178579059  0.9741521231951195  0.4566491721704459
   0.8567358569681524  0.9755531083820270  0.5476597570436177
   0.3513442612980425  0.4704278706885956  0.0420712486642304
   0.6446341655653959  0.0263280450087940  0.0396687498120283
   0.1443433749263427  0.5240723616356897  0.5450402892900074
   0.3515766641496394  0.5243973772954509  0.4530824026526288
   0.8518042329111647  0.0253974723579845  0.9541924996499777
   0.4747783802128778  0.9483981811395898  0.1411397284234495
   0.9710801606886040  0.4470898658009960  0.6428807287193220
   0.9697651164967206  0.5445412645795770  0.8569170693614772
   0.4579145172335759  0.0502086848285567  0.3604234511706512
   0.0377789167467036  0.0474328215783640  0.6429148961400192
   0.5248559726550955  0.5448980695027317  0.1428492673605766
   0.5245690736103125  0.4419710018775266  0.3570201273653779
   0.0228087253611442  0.9455607971276107  0.8568216425916588
   0.2961527737774262  0.7791913556723353  0.1090474455819379
```

```
0.7971453965413718  0.2745236884852122   0.6070830001581872
0.6962019500184697  0.2725877763250329   0.3938929593853097
0.1980738831862548  0.7753430012835711   0.8924755156201438
0.1973461356161364  0.2256015512622881   0.6085493730214985
0.6985770197523670  0.7199223656049349   0.1096974615018357
0.7966342615580829  0.7215964324137397   0.8921534742794397
0.2969943048393815  0.2245599130150116   0.3924570741188405
0.1003290929982484  0.3003352030656333   0.7906546940582232
0.6094181534268605  0.8012832076769465   0.2925509048193366
0.3880195859383589  0.6990570136274932   0.2903672600773143
0.8961782732149295  0.1955384804292235   0.7871140029103473
0.5987939541121566  0.1947152182735070   0.2127765334230044
0.0986243721155000  0.6961775436627451   0.7121820920653054
0.8967648768635930  0.7992464623112029   0.7110202267862926
0.3953607406250936  0.2978215357106075   0.2092576309063035
0.7763692105020646  0.0994671492934976   0.2972384008795779
0.2749008900355698  0.6006993051199613   0.7944308038408562
0.2764400010297288  0.3965134328270760   0.7079183363120481
0.7784041765445413  0.8965792218278459   0.2019649172702467
0.2172967885685578  0.6038733254804378   0.2045387865250381
0.7182063402252692  0.1012031107574495   0.7046575422735820
0.7208244855896694  0.8958422503465276   0.7910326100271227
0.2207502031361457  0.3958123668048478   0.2903518717244373
0.7008556576812109  0.2136265046419376   0.8974700329090544
0.1948776304325354  0.7123884556832650   0.3981702655657245
0.2915443581819308  0.7116683073688261   0.6012023787093767
0.7919373349386959  0.2125230379012474   0.1030567055739371
0.8037329900064808  0.7891754018751440   0.3965281702812037
0.2945219713657221  0.2815368490566405   0.8994107159148631
0.2015500790853536  0.2794159982911665   0.0999863719433049
0.7031691828683382  0.7837948953040386   0.5998183132325111
0.8945020386198372  0.7038520877732426   0.2243606567458331
0.3891867692194365  0.2001757291321857   0.7259895350441488
0.6069336634848749  0.2986263410107294   0.7267560190688545
0.1022393226308874  0.7984628187320224   0.2250571204615144
0.3883901123626278  0.7946639950053557   0.7725924337369392
0.8887304296368395  0.2929255273868682   0.2730732234768995
0.1068099854147802  0.1992800033804424   0.2736635667537606
0.6073863072780082  0.7000909926383859   0.7716825063956669
0.2077722490841618  0.8876175902175094   0.6938272243901152
0.7064038417498654  0.3881169729085641   0.1997679792926722
0.7234216349739594  0.6071507352074106   0.3074462934587539
0.2090471602387746  0.1070803354169222   0.7991545486356363
0.7883427855678102  0.3870997004441651   0.8047312672555638
0.2737815542790478  0.8925158156255170   0.3090529440541036
0.2892425793974140  0.1049813183244201   0.2027260096389955967
0.7887338814266763  0.6086035135158218   0.6943498094788857
0.2456443684670634  0.8725001095157822   0.4989643729501935
0.7468682165019188  0.3853948567275331   0.0003495544113871
0.7494502875578243  0.6105166181767542   0.4986352592079945
0.2489730071048619  0.1089358688670844  -0.0004823888953269
0.8638818902340580  0.4954028812404089   0.2503245857705508
0.3644808794705766  0.9947476199881011   0.7531410296239781
0.6340698826788417  0.4983091899082933   0.7489148162928108
0.1295061007536198  0.9974632345201918   0.2453041992646148
0.2475498312188916  0.6253413764918327   0.0035075599838813
0.7468880997463654  0.1244256353966728   0.5000874553086512
0.7482298476632454  0.8716169279157849   0.0027176354336692
0.2493881457398649  0.3724163277848382   0.4990082292043166
0.6176869802193661  0.9940830883951953   0.2553555288219233
0.1197406475168054  0.4974411791744139   0.7507925521008624
0.8768903701775740  0.9992951460559445   0.7485113388523955
0.3762348608882271  0.5035123866206054   0.2523141329760958
0.4993633569726384  0.2489963681031944   0.8751657056488098
0.9984388057414372  0.7460639646722456   0.3746192282873318
0.4976022383828622  0.7444280375835886   0.6240607732131908
0.9969482036893406  0.2472986043127722   0.1239506454425643
0.9981138753155340  0.2520083164367964   0.6253165358819442
0.4998648873536168  0.7478976512150527   0.1385597429309310
0.9965206678289839  0.7482141178081578   0.8744346230307541
0.4964767356903996  0.2496277594101635   0.3762960273060226
0.6463021037506484  0.1781181120046617   0.0522395127727932
0.1421637389700962  0.6749239286097225   0.5527933718888489
0.8527282114624047  0.8213843795980698   0.5506446571251994
0.3098769346523497  0.4386214531321682   0.8769569642443698
0.1258698411180520  0.9903783442131123   0.7626813551098691
0.6229107460393010  0.5085417614326400   0.2575695037433069
0.8707790271974414  0.4967956031360029   0.7487295574952195
0.3743418780937979  0.9805739908067896   0.2457283811310819
0.9318320157774400  0.8846487732537252   0.1924979460186931
0.5794613471294011  0.2951355557152790   0.5815313989348568
0.4277249505677512  0.5991405554103110   0.8017243019315879
0.1071871341977840  0.1952267074456311   0.4269305222935624
0.7466912000999967  0.3727736587285671   0.4994059757897263
0.2480947678435827  0.8748168756592388  -0.0010844981685342
0.7472582400507508  0.6216282422954632   0.0005092394741937
0.3911126707466523  0.1937011805702550   0.5738806622937092
0.3097498230068580  0.5570675984603288   0.6280220229918190
0.8093414344827033  0.0552289598134797   0.1353011343554739
0.6826565157476440  0.9393537699477761   0.6336033904218898
0.1827476035373348  0.5583326219056397   0.3651471465659052
0.4985083629265606  0.7495545067241294   0.3752131739374261
0.8452592994597566  0.1772364855837488   0.9496926389672754
0.0485112820826127  0.8378761680522300   0.6688576869625759
0.5484512562105607  0.3410103084465016   0.1719396871136492
```

```
  0.1937913766527400   0.0767251921521918   0.5949806678123818
  0.4295189664983358   0.3977106855492898   0.6964572200594347
  0.0684554742407641   0.3938944137765346   0.3015012535178333
  0.5685338945052397   0.8941896949210585   0.8025592933740623
  0.9321294777718918   0.1126084556181053   0.3046977908918176
  0.5637761457368926   0.1025111033812369   0.6964671304013864
  0.0650116893569647   0.6034671554199064   0.1962761602476492
  0.3339955047771029   0.6713201164568650   0.4503091211608025
  0.3506067681490487   0.3175487374959415   0.0518172790877032
  0.6663744587174608   0.8265661160870715   0.4515661996205297
  0.1466499363100042   0.3173840931027275   0.9487618675878821
  0.1863072675217136   0.4379142924931237   0.1266310351961582
  0.6858801840382999   0.0584990792680184   0.8720792063671252
  0.8181776667199698   0.9437154225959941   0.3532191070687823
  0.0496203016075710   0.1588550862702704   0.8280687865304611
  0.4460152789129543   0.1546063800305393   0.1712008472958131
  0.9466638003414962   0.6582584595790065   0.6689537055803791
  0.9465195644552293   0.3363426863701007   0.8315909641590229
  0.0875399409552860   0.7983655322277532   0.0780014278897466
  0.2993218495646787   0.0756603155646908   0.4068869084864814
  0.5862858673046093   0.6982352774522106   0.9202876222244364
  0.9083920933128579   0.6964005232530940   0.0784068633110130
  0.4104247515520900   0.7972858938598281   0.9213295719561575
  0.9144550274237788   0.2975310282036896   0.4187893303065306
LLNO
   1.00000000000000
    12.6993438462807777     0.0137700502095675     0.0161402112289772
     0.0134679780507001    12.7320020714613786     0.0129577195452331
     0.0162469850431498     0.0130545683460799    12.7475712221416106
   Nb   O   La   Li
   16   96   24   40
Direct
  0.7546398513718192   0.2428829159699482   0.7606177148333831
  0.2384452557638367   0.7425758647414512   0.2461076479539995
  0.7438931970820453   0.7388351940840768   0.2550206812191548
  0.2492980169407470   0.2471063492139666   0.7518888926292808
  0.2378426339292610   0.7483189233426050   0.7477869011565084
  0.7474326542149250   0.2472847600276906   0.2537085772500067
  0.2520392858471709   0.2433733099980476   0.2466801554624553
  0.7487766312167438   0.7466803344022481   0.7532621379662343
  0.4959865049184370   0.4886315810500887   0.0098553390686244
  0.0005073682814743  -0.0084320790120210   0.5048395798842946
  0.4986392024050122  -0.0015523185952332   0.5048438749405552
 -0.0013263952355026   0.4977752343558549   0.0071757934974835
 -0.0000703552001482   0.5050696999776988   0.5088966238986035
  0.5002920069532649  -0.0031609587261256   0.0066427258423997
 -0.0073499509514537  -0.0032893069513702   0.0036684520691834
  0.4973724449117153   0.4933896308962837   0.5084232067000601
  0.5568778059525256   0.3499370335927943   0.0307115830023389
  0.0528063624635130   0.8498215289627565   0.5286975772662339
  0.9426298210297419   0.6444182080550700   0.5321888946436284
  0.4471707358765715   0.1376516048251584   0.0385584623273114
  0.4538205327057638   0.8502004794112848   0.4617852927327204
  0.9449607654012675   0.3530974474414751   0.9757889754393100
  0.0503975685726885   0.1401596131216079   0.9729686275269749
  0.5527084995763311   0.6390088070073137   0.4808133068675894
  0.3473583635682448   0.0246794209251780   0.5473663349133062
  0.8548421836644067   0.5210588779791741   0.0653528898349038
  0.6402905822256697   0.5324295427250679   0.9551933753545162
  0.1490241198259939   0.0266401005179922   0.4491205828258538
  0.8510140722626037   0.4614190915544209   0.4576144590298001
  0.3511245134231453   0.9601609214242109   0.9579649195515803
  0.1411523173670377   0.9712157973874802   0.0611845757446315
  0.6472208344973077   0.4627664478453499   0.5581093088088225
  0.0234033714626478   0.5499966752747757   0.3627506066809530
  0.5300353807961753   0.0391464279915241   0.8575076846999496
  0.5171408746112722   0.9378798031708575   0.6460223604777038
  0.0331881023025698   0.4397812387699145   0.1563437149998167
  0.4713459699851554   0.4389534977806235   0.8631679021028250
  0.9676732805781504   0.9490524225659618   0.3589928430340306
  0.9634045669701182   0.0532039599938195   0.1491101298434227
  0.4642758862561693   0.5458802557093252   0.6553421943973257
  0.9523267262370845   0.1452120296805761   0.4755982979979174
  0.4436421758311833   0.6409919702741411   0.9720159883731009
  0.5431725037817429   0.8434424928615067   0.9777798705711953
  0.0507487568227872   0.3515049889718164   0.4738342750931634
  0.0513357200029264   0.6415632579973790   0.0340762622076342
  0.5493261481258614   0.1406054431316832   0.5343100648751027
  0.4488713809132607   0.3460717830193462   0.5335280486944141
  0.9419967185930532   0.8515486424238948   0.0343157056954889
  0.1472896998654517   0.4636867751319733   0.9622268986205247
  0.6450117344150168   0.9654508649988049   0.4537732480053840
  0.8593146434555063   0.9710392010494847   0.5634813885294631
  0.3504105816547251   0.4631945241525051   0.0620208956858637
  0.6463247510032182   0.0225602771352338   0.0477905251576403
  0.1438008950365591   0.5266323239954539   0.5625737594444492
  0.3518721947269914   0.5245981422588190   0.4603250607645275
  0.8484748087046036   0.0248571617874547   0.9473478831701411
  0.4704351038914765   0.9477436236291562   0.1562160739876537
  0.9612161223471044   0.4448356620945089   0.6502557117968860
  0.9723310412733840   0.5498007440029827   0.8642534539184809
  0.4634299309167066   0.0487359100542732   0.3566412509111835
  0.0400824046156657   0.0475337297650274   0.6466894740509941
  0.5210231327733829   0.5482632657649025   0.1510669008777206
  0.5327749684536310   0.4448257946995919   0.3636779725154594
```

```
 0.0267491351253857   0.9392206291279012   0.8627158420671864
 0.2999549633356688   0.7827585247645639   0.1028490326367610
 0.8013651372916215   0.2730669222112374   0.6074594266056512
 0.6940510005873527   0.2723153759427381   0.4075346101960046
 0.1986605786652142   0.7711851193635569   0.8946065990955534
 0.1975316445897341   0.2139709033499058   0.6013194264828405
 0.6877727974767431   0.7123819229200340   0.1117166343574152
 0.7905916542135241   0.7151496661764974   0.8999304206862415
 0.2963074129361487   0.2163866990043071   0.3995120727684314
 0.1059837270338259   0.2998092648780297   0.7814364423862180
 0.5938993922404365   0.7984425584641024   0.2887370957191491
 0.3855398091761763   0.6892536687300791   0.2797389790210645
 0.8977369977758417   0.1922357473868804   0.7819043071157210
 0.6056010687021337   0.1865974996934022   0.2268676892517764
 0.0996677026591616   0.6932718290680036   0.7260775893106339
 0.8891169568274686   0.7993825599379527   0.7274444416310587
 0.3968189217708768   0.2899295798388861   0.2255544555683757
 0.7868495368750038   0.1018241146299761   0.3105247757467084
 0.2837621028648363   0.5979763171234906   0.7921120519655767
 0.2743019165002471   0.3914688181957462   0.6926036445561783
 0.7713853137166974   0.8862574813587786   0.1998951847370372
 0.2190890613266189   0.5987314642757723   0.1911589561074692
 0.7227596229859887   0.0982290574380085   0.7097583535840133
 0.7078039448928720   0.8926808724077970   0.8077989194559414
 0.2120545015608141   0.3912586428312071   0.2934204541608983
 0.6968348806228786   0.2097161782400530   0.9007681555190109
 0.1874943517391265   0.7168864435478411   0.3904105976730596
 0.2989729395805500   0.7121523963534708   0.6025915449599235
 0.8008105663412978   0.2274304455881266   0.1121492843824671
 0.7857811109646040   0.7789052904804231   0.4014056834229145
 0.3094011851198216   0.2697252716336263   0.8905040903568647
 0.2000346445130465   0.2752084672197625   0.1033602472368340
 0.6896403454557645   0.7771522595604843   0.6082730862201218
 0.8881565105064717   0.6930974757673829   0.2282882455330609
 0.3916057071047073   0.1878378855927773   0.7129390500812189
 0.6072136018669742   0.2973553685902462   0.7209121443553462
 0.0979677411676745   0.7996390453016364   0.2221672899849337
 0.3936503138136713   0.7973905149053355   0.7846176795782063
 0.8980328073439636   0.2947778224024610   0.2893174621874280
 0.1013247124435966   0.1942267805971228   0.2833459018989042
 0.5934566881257436   0.6956069868744739   0.7852072924530999
 0.2202641614229733   0.8914728218020291   0.6921870649409007
 0.7170769921134916   0.3926239461918742   0.2210798209168971
 0.7179278914161178   0.5971649846951965   0.3135722796173329
 0.2190065566371113   0.0998290006654361   0.7999608102820261
 0.7763838830164442   0.3905624874759069   0.8076200786312999
 0.2784259635131590   0.8902095720959876   0.2983106816991708
 0.2771146903492273   0.0965406753416074   0.2003688027008497
 0.7726838564872550   0.6010037124492899   0.7010481664211798
 0.5006666793126129   0.2450769121967953   0.8758823932674277
-0.0075690877589717   0.7493727293249913   0.3769042126406784
 0.4958000782590242   0.7454998835484630   0.6250832857992247
-0.0042660259976778   0.2447205725791560   0.1190341295145403
 0.2473212067904607   0.8601716971935225   0.4967798627740306
 0.7508617694869493   0.3753158366751253   0.0018356728891625
 0.7484717062630089   0.6250590964924768   0.5079875298958689
 0.2405180049333271   0.1199364561665725  -0.0048588956636347
 0.8745622948596250   0.4991130232623247   0.2530240131450264
 0.3706566935128673  -0.0057943418303940   0.7551831797270470
 0.6226212970754363   0.4954171669738846   0.7584649679097921
 0.1164799952600276  -0.0073797173944479   0.2478537547546778
 0.0008313407753367   0.2444715194252120   0.6338998468258398
 0.4979932717587114   0.7402576248063314   0.1358051692825856
-0.0047593405676536   0.7446938917274702   0.8834352153584965
 0.5020678892392255   0.2488508737970978   0.3780186705787685
 0.2471419182139053   0.6198023282608338   0.0011836392826585
 0.7524776943632470   0.1156433683159614   0.5056247793653439
 0.7532591767836585   0.8656895282005070   0.0108331150965041
 0.2435260158885678   0.3731206547081954   0.5010980914909343
 0.6356217531573113  -0.0037439272044894   0.2546541217376342
 0.1179750559382046   0.4919791628028123   0.7588649175596429
 0.8806742478619018  -0.0061933953752424   0.7544307072653683
 0.3748286176338293   0.4985996745042495   0.2572816050252625
 0.7590209854788535   0.1236445619658474   0.0211623600471347
 0.2361202052439967   0.6219135006194669   0.4871122573083290
 0.7659901007689384   0.8718127866827420   0.5183379136526485
 0.2627641831662869   0.3665945780562357  -0.0141526184188956
 0.6204560397484364   0.5038844805253375   0.2480352899357984
 0.8722840458924982   0.5036313456288211   0.7675214684811443
 0.4254453198833500   0.0921900035783687   0.1993057679956904
-0.0067502306497627   0.7450008722941537   0.1286542825856449
 0.5000474489688095   0.2449061973657020   0.6350606883572505
 0.5599596902327235   0.8444451967698048   0.8309776257117981
 0.0570270223805426   0.3396815132562762   0.3241274137559584
 0.6936486903840623   0.4214724570734163   0.4034126280165850
 0.2232366587844637   0.8706117687641770  -0.0028253061881279
 0.3065886619600566   0.0654879580805515   0.3908721973982687
 0.3150123876338214   0.5659622814536431   0.6357594561596689
 0.6257591776021616  -0.0084261985170943   0.7229558249470941
 0.1204104010838033   0.5175006356780493   0.2626990196399251
 0.4839921046800369   0.7207551843757158   0.3792009848218282
-0.0040753269519845   0.7479535576544648   0.6294662685811121
 0.5084789321198394   0.2515646700580408   0.1265706794737304
 0.8146198836423625   0.4198055221839187   0.6019294871022197
 0.3205663305875094   0.9312284245975946   0.1478283364543982
```

```
   0.7566861838150062    0.6179353175334454    0.0136104763693120
   0.1877641280612992    0.0658746332186160    0.5964186827815379
   0.8720409811989042    0.9828341180616543    0.2473001750563742
   0.4317159780102822    0.6485956221001158    0.8238875853302003
   0.9387208195109142    0.1481934796970324    0.3285736936164412
  -0.0003146257509694    0.2484252281766807    0.8788832030148797
   0.6234736171582681    0.8195464189169593    0.4315837749148707
   0.1796944200845265    0.4230608163679259    0.1271213301685593
   0.6819256855258948    0.0566360237219393    0.8772585851461452
   0.3722504447013094    0.4696597813127921    0.7665180707055123
   0.1224371630387609    0.9896621843388431    0.7633577917162189
   0.4267597362823553    0.9016088228919766    0.3138355385886153
   0.6564057372552689    0.3163283164547294    0.5687597002266805
   0.1206699782395065    0.1765185645563284    0.4339655600253642
   0.5955465726609432    0.6796630902766642    0.9330229259343822
   0.3437637326006253    0.1725442286579683    0.5605640913413519
   0.3924624466888599    0.8091466833113560    0.9352840990759479
   0.8890862884193637    0.3132027248518049    0.4381819294302021
LLSnO
   1.00000000000000
     13.1290438137644507    0.0000000000000000    0.0000000000000000
      0.0000000000000000   13.1290438137644507    0.0000000000000000
      0.0000000000000000    0.0000000000000000   12.4818585963466386
   Sn   O   La   Li
    16   96   24   56
Direct
   0.2453270000000032    0.2570089999999965    0.2501960000000025
   0.7453270000000032    0.7570089999999965    0.7501960000000025
   0.7453270000000032    0.2570089999999965    0.7501960000000025
   0.2453270000000032    0.7570089999999965    0.2501960000000025
   0.9953270000000032    0.5070089999999965    0.0001960000000025
   0.4953270000000032    0.0070089999999965    0.5001960000000025
   0.9953270000000032    0.5070089999999965    0.5001960000000025
   0.4953270000000032    0.0070089999999965    0.0001960000000025
   0.2453270000000032    0.2570089999999965    0.7501960000000025
   0.7453270000000032    0.7570089999999965    0.2501960000000025
   0.7453270000000032    0.2570089999999965    0.2501960000000025
   0.2453270000000032    0.7570089999999965    0.7501960000000025
   0.9953270000000032    0.0070089999999965    0.5001960000000025
   0.4953270000000032    0.5070089999999965    0.0001960000000025
   0.9953270000000032    0.0070089999999965    0.0001960000000025
   0.4953270000000032    0.5070089999999965    0.5001960000000025
   0.3008722261300467    0.2909978099091138    0.4045722591937623
   0.8008722261300467    0.7909978099091136    0.9045722591937624
   0.6897827738699571    0.2909978099091138    0.9045722591937624
   0.1897827738699571    0.7909978099091136    0.4045722591937623
   0.9613381900908861    0.4514637738699530    0.1545722591937623
   0.4613381900908858    0.9514637738699530    0.6545722591937624
   0.0293158099091134    0.4514637738699530    0.6545722591937624
   0.5293158099091132    0.9514637738699530    0.1545722591937623
   0.6897827738699571    0.7230201900908865    0.5958187408062381
   0.1897827738699571    0.2230201900908863    0.0958187408062381
   0.3008722261300467    0.7230201900908865    0.0958187408062381
   0.8008722261300467    0.2230201900908863    0.5958187408062381
   0.0293158099091134    0.5625532261300425    0.8458187408062381
   0.5293158099091132    0.0625532261300425    0.3458187408062381
   0.9613381900908861    0.5625532261300425    0.3458187408062381
   0.4613381900908858    0.0625532261300425    0.8458187408062381
   0.3008722261300467    0.2230201900908863    0.9045722591937624
   0.8008722261300467    0.7230201900908865    0.4045722591937623
   0.6897827738699571    0.2230201900908863    0.4045722591937623
   0.1897827738699571    0.7230201900908865    0.9045722591937624
   0.9613381900908861    0.0625532261300425    0.6545722591937624
   0.4613381900908858    0.5625532261300425    0.1545722591937623
   0.0293158099091134    0.0625532261300425    0.1545722591937623
   0.5293158099091132    0.5625532261300425    0.6545722591937624
   0.6897827738699571    0.7909978099091136    0.0958187408062381
   0.1897827738699571    0.2909978099091138    0.5958187408062381
   0.3008722261300467    0.7909978099091136    0.5958187408062381
   0.8008722261300467    0.2909978099091138    0.0958187408062381
   0.0293158099091134    0.9514637738699530    0.3458187408062381
   0.5293158099091132    0.4514637738699530    0.8458187408062381
   0.9613381900908861    0.9514637738699530    0.8458187408062381
   0.4613381900908858    0.4514637738699530    0.3458187408062381
   0.0990822982861611    0.2038515092982003    0.7862402049430622
   0.5990822982861612    0.7038515092982001    0.2862402049430621
   0.8915727017138355    0.2038515092982003    0.2862402049430621
   0.3915727017138356    0.7038515092982001    0.7862402049430622
   0.0484844907018066    0.6532537017138385    0.5362402049430622
   0.5484844907017996    0.1532537017138385    0.0362402049430621
   0.9421695092981996    0.6532537017138385    0.0362402049430621
   0.4421695092981998    0.1532537017138385    0.5362402049430622
   0.8915727017138355    0.8101664907018000    0.2141517950569359
   0.3915727017138356    0.3101664907017999    0.7141517950569357
   0.0990822982861611    0.8101664907018000    0.7141517950569357
   0.5990822982861612    0.3101664907017999    0.2141517950569359
   0.9421695092981996    0.3607632982861640    0.4641517950569359
   0.4421695092981998    0.8607632982861642    0.9641517950569357
   0.0484844907018066    0.3607632982861640    0.9641517950569357
   0.5484844907017996    0.8607632982861642    0.4641517950569359
   0.0990822982861611    0.3101664907017999    0.2862402049430621
   0.5990822982861612    0.8101664907018000    0.7862402049430622
   0.8915727017138355    0.3101664907017999    0.7862402049430622
   0.3915727017138356    0.8101664907018000    0.2862402049430621
   0.0484844907018066    0.8607632982861642    0.0362402049430621
```

```
0.5484844907017996  0.3607632982861640  0.5362402049430622
0.9421695092981996  0.8607632982861642  0.5362402049430622
0.4421695092981998  0.3607632982861640  0.0362402049430621
0.8915727017138355  0.7038515092982001  0.7141517950569357
0.3915727017138356  0.2038515092982003  0.2141517950569359
0.0990822982861611  0.7038515092982003  0.2141517950569359
0.5990822982861612  0.2038515092982003  0.7141517950569357
0.9421695092981996  0.1532537017138385  0.9641517950569357
0.4421695092981998  0.6532537017138385  0.4641517950569359
0.0484844907018066  0.1532537017138385  0.4641517950569359
0.5484844907017996  0.6532537017138385  0.9641517950569357
0.2722372655625981  0.1063234447157162  0.6997408431252274
0.7722372655625981  0.6063234447157161  0.1997408431252274
0.7184177344373986  0.1063234447157162  0.1997408431252274
0.2184177344373985  0.6063234447157161  0.6997408431252274
0.1460125552842835  0.4800987344374015  0.4497408431252276
0.6460125552842836  0.9800987344374016  0.9497408431252274
0.8446424447157203  0.4800987344374015  0.9497408431252274
0.3446424447157204  0.9800987344374016  0.4497408431252276
0.7184177344373986  0.9076935552842794  0.3006501568747729
0.2184177344373985  0.4076935552842793  0.8006501568747730
0.2722372655625981  0.9076935552842794  0.8006501568747730
0.7722372655625981  0.4076935552842793  0.3006501568747729
0.8446424447157203  0.5339182655626010  0.5506501568747730
0.3446424447157204  0.0339182655626012  0.0506501568747730
0.1460125552842835  0.5339182655626010  0.0506501568747730
0.6460125552842836  0.0339182655626012  0.5506501568747730
0.2722372655625981  0.4076935552842793  0.1997408431252274
0.7722372655625981  0.9076935552842794  0.6997408431252274
0.7184177344373986  0.4076935552842793  0.6997408431252274
0.2184177344373985  0.9076935552842794  0.1997408431252274
0.1460125552842835  0.0339182655626012  0.9497408431252274
0.6460125552842836  0.5339182655626010  0.4497408431252276
0.8446424447157203  0.0339182655626012  0.4497408431252276
0.3446424447157204  0.5339182655626010  0.9497408431252274
0.7184177344373986  0.6063234447157161  0.8006501568747730
0.2184177344373985  0.1063234447157162  0.3006501568747729
0.2722372655625981  0.6063234447157161  0.3006501568747729
0.7722372655625981  0.1063234447157162  0.8006501568747730
0.8446424447157203  0.9800987344374016  0.0506501568747730
0.3446424447157204  0.4800987344374015  0.5506501568747730
0.1460125552842835  0.9800987344374016  0.5506501568747730
0.6460125552842836  0.4800987344374015  0.0506501568747730
0.4953270000000032  0.2570089999999965  0.3751960000000025
0.9953270000000032  0.7570089999999965  0.8751960000000025
0.4953270000000032  0.2570089999999965  0.8751960000000025
0.9953270000000032  0.7570089999999965  0.3751960000000025
0.9953270000000032  0.2570089999999965  0.1251960000000025
0.4953270000000032  0.7570089999999965  0.6251960000000025
0.9953270000000032  0.2570089999999965  0.6251960000000025
0.4953270000000032  0.7570089999999965  0.1251960000000025
0.2453270000000032  0.1302824849450194  0.5001960000000025
0.7453270000000032  0.6302824849450193  0.0001960000000025
0.7453270000000032  0.1302824849450194  0.0001960000000025
0.2453270000000032  0.6302824849450193  0.5001960000000025
0.1220535150549802  0.5070089999999965  0.2501960000000025
0.6220535150549804  0.0070089999999965  0.7501960000000025
0.8686014849450163  0.5070089999999965  0.7501960000000025
0.3686014849450165  0.0070089999999965  0.2501960000000025
0.7453270000000032  0.8837345150549833  0.5001960000000025
0.2453270000000032  0.3837345150549831  0.0001960000000025
0.2453270000000032  0.8837345150549833  0.0001960000000025
0.7453270000000032  0.3837345150549831  0.5001960000000025
0.1220535150549802  0.0070089999999965  0.7501960000000025
0.6220535150549804  0.5070089999999965  0.2501960000000025
0.8686014849450163  0.0070089999999965  0.2501960000000025
0.3686014849450165  0.5070089999999965  0.7501960000000025
0.4953270000000032  0.2570089999999965  0.6251960000000025
0.9953270000000032  0.7570089999999965  0.1251960000000025
0.4953270000000032  0.2570089999999965  0.1251960000000025
0.9953270000000032  0.7570089999999965  0.6251960000000025
0.9953270000000032  0.2570089999999965  0.3751960000000025
0.4953270000000032  0.7570089999999965  0.8751960000000025
0.9953270000000032  0.2570089999999965  0.8751960000000025
0.4953270000000032  0.7570089999999965  0.3751960000000025
0.6748316007920718  0.0775043992079278  0.3751960000000025
0.1748316007920719  0.5775043992079278  0.8751960000000025
0.3158233992079321  0.0775043992079278  0.8751960000000025
0.8158233992079320  0.5775043992079278  0.3751960000000025
0.1748316007920719  0.0775043992079278  0.1251960000000025
0.6748316007920718  0.5775043992079278  0.6251960000000025
0.8158233992079320  0.0775043992079278  0.6251960000000025
0.3158233992079321  0.5775043992079278  0.1251960000000025
0.3158233992079321  0.9365126007920677  0.6251960000000025
0.8158233992079320  0.4365126007920676  0.1251960000000025
0.6748316007920718  0.9365126007920677  0.1251960000000025
0.1748316007920719  0.4365126007920676  0.6251960000000025
0.8158233992079320  0.9365126007920677  0.8751960000000025
0.3158233992079321  0.4365126007920676  0.3751960000000025
0.1748316007920719  0.9365126007920677  0.3751960000000025
0.6748316007920718  0.4365126007920676  0.8751960000000025
0.3328466559062515  0.1761350433999795  0.0578411226296999
0.8328466559062516  0.6761350433999793  0.5578411226297001
0.6578073440937476  0.1761350433999795  0.5578411226297001
0.1578073440937477  0.6761350433999793  0.0578411226296999
```

```
   0.0762009566000202   0.4194893440937482   0.8078411226297001
   0.5762009566000204   0.9194893440937481   0.3078411226297000
   0.9144530433999789   0.4194893440937482   0.3078411226297000
   0.4144530433999791   0.9194893440937481   0.8078411226297001
   0.6578073440937476   0.8378829566000208   0.9425498773703004
   0.1578073440937477   0.3378829566000205   0.4425498773703005
   0.3328466559062515   0.8378829566000208   0.4425498773703005
   0.8328466559062516   0.3378829566000205   0.9425498773703004
   0.9144530433999789   0.5945286559062520   0.1925498773703006
   0.4144530433999791   0.0945286559062520   0.6925498773703004
   0.0762009566000202   0.5945286559062520   0.6925498773703004
   0.5762009566000204   0.0945286559062520   0.1925498773703006
   0.3328466559062515   0.3378829566000205   0.5578411226297001
   0.8328466559062516   0.8378829566000208   0.0578411226296999
   0.6578073440937476   0.3378829566000205   0.0578411226296999
   0.1578073440937477   0.8378829566000208   0.5578411226297001
   0.0762009566000202   0.0945286559062520   0.3078411226297000
   0.5762009566000204   0.5945286559062520   0.8078411226297001
   0.9144530433999789   0.0945286559062520   0.8078411226297001
   0.4144530433999791   0.5945286559062520   0.3078411226297000
   0.6578073440937476   0.6761350433999793   0.4425498773703005
   0.1578073440937477   0.1761350433999795   0.9425498773703004
   0.3328466559062515   0.6761350433999793   0.9425498773703004
   0.8328466559062516   0.1761350433999795   0.4425498773703005
   0.9144530433999789   0.9194893440937481   0.6925498773703004
   0.4144530433999791   0.4194893440937482   0.1925498773703006
   0.0762009566000202   0.9194893440937481   0.1925498773703006
   0.5762009566000204   0.4194893440937482   0.6925498773703004
LLTO
   1.00000000000000
    12.8416129762392366     0.0190612744611359     0.0044357059801743
     0.0189663037329721    12.8836236158031472    -0.0114991281917461
     0.0045818716734793    -0.0106243318165292    12.6387341701387168
   Ta   O    La   Li
   16   96   24   40
Direct
   0.2390440429175269   0.2642489992587140   0.2536982411988928
   0.7420614999648738   0.7614272043052275   0.7508225169474445
   0.7437331343686454   0.2641650110130213   0.7509806183234951
   0.2476094959173290   0.7527598671987279   0.2504386320990886
   0.9883193988318038   0.5074583014602211   0.0072754804394358
   0.4854376687740424   0.0090664332345608   0.5055663117939617
  -0.0113806522694159   0.5134656510609069   0.4988570577152155
   0.4936368119320194   0.0053398526298963   0.0068486614341959
   0.2408361315522594   0.2608611553887548   0.7550467234253174
   0.7474045980154668   0.7569072745542464   0.2489039645120864
   0.7449928592445505   0.2578956109888127   0.2463518905682841
   0.2440671214818710   0.7597047017406963   0.7464071607038285
  -0.0075687142895348   0.0064866965982066   0.5088846709347536
   0.4932317849125734   0.5008493388557274   0.0035535498751371
  -0.0093892382925769   0.0050666139194854  -0.0028198521318521
   0.4876616110010430   0.5031942317514547   0.4974766162438626
   0.2993940484553170   0.2853894532943482   0.3972577935491350
   0.7899666851273984   0.7923834633204958   0.8990468906862554
   0.6915280976471886   0.2851352832222286   0.8954519142094857
   0.1978082082383908   0.7850145144411382   0.3990755345369373
   0.9621807980576093   0.4530402769228813   0.1494750168693973
   0.4614958318364982   0.9450219508016177   0.6488171748944943
   0.0209422110817367   0.4559722267578030   0.6437299767915606
   0.5203402952470981   0.9458567336221964   0.1505320162877925
   0.6924136590687324   0.7274183004953332   0.5996946991396147
   0.1962552830224280   0.2308234997986240   0.1036014407766467
   0.3077177655178007   0.7330642607848936   0.1063345534642372
   0.7933944010923982   0.2292398718252948   0.5993065120993813
   0.0282977793472139   0.5640398701749073   0.8574504879874104
   0.5158293089052517   0.0485029641820344   0.3555321824163961
   0.9615765655884789   0.5618753562792175   0.3528775349425252
   0.4629818962980474   0.0483840524438021   0.8574640237419101
   0.3028415015031110   0.2221947085399325   0.8962200669022296
   0.8011148363272509   0.7387082026441366   0.3973818393816846
   0.6881779569158548   0.2186550146498408   0.3906151224390811
   0.1985510363145132   0.7279845867277640   0.8976468919128443
   0.9537380716034680   0.0567534680934360   0.6540732771882928
   0.4686949557786907   0.5585375590396098   0.1480841151647781
   0.0291978650266972   0.0636510565288339   0.1397826852860723
   0.5305216007427704   0.5479012728976503   0.6478297678234014
   0.6926016734442247   0.7935426826000792   0.1036563136777951
   0.1979608398781197   0.2897580179763636   0.6056088854706843
   0.3022176357147419   0.7893731252003428   0.6034515127185247
   0.8014555803675307   0.2824804268698410   0.1029772194774339
   0.0309184117820501   0.9547685505809103   0.3629014565724895
   0.5210247343114336   0.4478970047670476   0.8555292699683295
   0.9519213789294683   0.9559785325642313   0.8504424837211122
   0.4684884959767845   0.4515094831988268   0.3526587390537427
   0.0967955775478686   0.2077916987369381   0.7770096190768357
   0.6046348547089395   0.7124244626900794   0.2922052207927224
   0.8913275498850620   0.2035217029463591   0.2883241714593319
   0.3893464811446394   0.7114168034755761   0.7828566214146452
   0.0340903418564033   0.6583529043961051   0.5278286109899439
   0.5360517750265624   0.1505185079104401   0.0305648271779971
   0.9323419785283980   0.6531475570390698   0.0291691363627995
   0.4416634942560277   0.1548521480371312   0.5265633987741730
   0.8882159227764107   0.8127916250402073   0.2128328371576974
   0.3876541262681111   0.3186236135369127   0.7252175900084752
   0.1012103400111755   0.8165720490459257   0.7282426504091153
```

```
 0.6032728329502014   0.3112454964516367   0.2194931404120607
 0.9314967169199144   0.3702226412124094   0.4729887921569076
 0.4486868987078642   0.8589524227169327   0.9661271380559543
 0.0419695530537411   0.3647679539769005   0.9732179005711773
 0.5397766664203653   0.8629145438950310   0.4663328531864114
 0.0913544954401038   0.3082728344483066   0.2780267900918908
 0.5983363238292203   0.8121504831144504   0.7879629074513452
 0.8885470158405537   0.3103399819200049   0.7748846068237450
 0.3895164997491927   0.8128691564470076   0.2923183024540515
 0.0357160967849255   0.8594011742445142   0.0255538265310501
 0.5405964779696301   0.3546474961431798   0.5284170295192722
 0.9362127307559410   0.8650700556061910   0.5401439495757948
 0.4364043961803082   0.3603478218516269   0.0310531958395564
 0.8891736784704695   0.7296476563895945   0.7141431579473496
 0.3819477479882488   0.2035650678446580   0.2139302526936894
 0.1027526253642273   0.7062942332022355   0.2241781779884757
 0.5977972379207781   0.2165730827251075   0.7168259204153474
 0.9386222604977359   0.1477336269692910   0.9577469542758009
 0.4376042947978999   0.6486071774264648   0.4786455734793520
 0.0409807214478140   0.1512517250806515   0.4804497777320318
 0.5477512450363137   0.6481705723260671   0.9741405724056207
 0.2697858065163799   0.1137188588649161   0.7017687434326557
 0.7664871710003558   0.6154747274276001   0.1967953246490863
 0.7205653742791099   0.1093194573451820   0.1957013191676431
 0.2150515117095647   0.6161902776929669   0.6992217289011654
 0.1330798729218738   0.4852631470409138   0.4485728571449257
 0.6415777532659467   0.9813265807993435   0.9554159807033392
 0.8470262623751644   0.4795472863819416   0.9446131285921993
 0.3371441385724781   0.9786577811170254   0.4609323975970619
 0.7130749172432975   0.9107216635254145   0.2972757889926445
 0.2129178684469221   0.4038757486983470   0.8098068972679279
 0.2749314659242036   0.9076563839506658   0.7989162887688641
 0.7808050411879589   0.3974175977775835   0.3021803717600142
 0.8431089998874253   0.5420486817478132   0.5626380987785644
 0.3443690550110457   0.0329132514507172   0.0541190604069865
 0.1335415286511752   0.5367161569451139   0.0509465942564604
 0.6295228237522416   0.0418904041575752   0.5574397701252929
 0.2665580387923692   0.4042557120812075   0.2020926835171918
 0.7614670186548808   0.9095282714852772   0.6960872715746278
 0.7090173542967915   0.4067884748265375   0.6966305917486288
 0.2168326917485449   0.9019847135884810   0.1979417652988509
 0.1354902308738682   0.0278908277828287   0.9459155961876367
 0.6371879329510790   0.5332200276403559   0.4539963150526982
 0.8474698333782205   0.0341336366568858   0.4497504922663256
 0.3481366136749841   0.5338041345098837   0.9449512452367079
 0.7221301658635915   0.6159879878508782   0.7926497676356931
 0.2021611898118287   0.1161534845841937   0.3057006591933650
 0.2745065948609232   0.6129359620220499   0.3036177990530660
 0.7680902682068936   0.1134398331033723   0.7941156071520091
 0.8459003075209252   0.9793483076070149   0.0520792295687489
 0.3484536056572484   0.4705881009643427   0.5529836736577128
 0.1358168659852155   0.9817264750436753   0.5613121204308249
 0.6381095872120186   0.4743520768197664   0.0528648297867760
 0.4891958899361518   0.2590864641856831   0.3634306378302932
-0.0034177808622754   0.7512512549441170   0.8746218065243558
 0.4924503382942126   0.2594381080575293   0.8735834975697825
-0.0040692773557881   0.7586782244156633   0.3715730528920519
 0.9918457361690763   0.2600187193764804   0.1135165460251395
 0.4951571419440295   0.7526106510465873   0.6284441436510099
 0.0002438416805606   0.2701792512418686   0.6207391231399049
 0.4922813975703707   0.7505409997499235   0.1295302304082031
 0.2390668252896970   0.1392854924683451   0.5041773608858988
 0.7411539400182530  -0.6217190534039505  -0.0005002456786342
 0.7397083215284335   0.1392761644726687  -0.0012764390462586
 0.2386849321516012   0.6281340648727182   0.5027532708443766
 0.1189400682388148   0.5131684965476728   0.2487100043452180
 0.6101013296743101   0.0156454109233967   0.7489416503512879
 0.8762875027231559   0.5019011568998621   0.7546182176307163
 0.3616263683747428   0.0111002902578717   0.2494405284524512267
 0.7371045365670627   0.8917363909450002   0.4974770759610482
 0.2405048517795773   0.3852609333798414   0.0028935707627349
 0.2374838250521228   0.8782554200120909   0.0053444096508770
 0.7458456064833413   0.3865632876999477   0.4960899262625735
 0.1194369598725300   0.0118928103495275   0.7532140371380907
 0.6190192596613902   0.5022290253585121   0.2471105970088698
 0.8753531076719372   0.0085563069844593   0.2501623080065847
 0.3674193609161812   0.5063553005916004   0.7532242628380614
 0.0026045705787635   0.7489001253254881   0.1245890052457655
 0.5009572167406006   0.2625587379916708   0.1245884305618423
 0.0131338883767544   0.7638059489374737   0.6253355964595804
 0.4974459123396834   0.7417582702145905   0.8758769223685186
-0.0027665537107504   0.2676771257282803   0.8724110911936346
 0.4895331323206047   0.7410337201488063   0.3774887877320984
 0.6577239967117147   0.0742063588507371   0.3314492773198769
 0.1696282074641481   0.5865310552954249   0.9127702365039712
 0.3212496166986426   0.0758163394265142   0.8447360619923400
 0.8657015138437385   0.5157800219706347   0.2472835670474820
 0.1770184423707077   0.0946173624470380   0.0716233528359943
 0.6835324475202871   0.5757246001964240   0.6113902496445814
 0.8053411622281434   0.0833236004868430   0.6315134921141933
 0.3682293727105512   0.5062272706517641   0.2496988115019301
 0.3110239809746056   0.9428698548494990   0.6339304663739836
 0.6714823640293447   0.9489919325472707   0.1415186217254120
 0.2341872687952119   0.3861916323913634   0.4946028781250645
 0.8036563401717655   0.9416603870324033   0.8859768533884560
```

```
  0.1803195184908505  0.9231129035974281  0.4317178315124013
  0.3409793023439753  0.1821019698733597  0.0547435840957881
  0.8502237429332263  0.6893649045841850  0.5602225339981292
  0.5968893642220251  0.1943421619952560  0.5672247184137700
  0.1189767343637218  0.5038574551121949  0.7416270287179542
  0.5617208437272476  0.9046824257226405  0.3113271431656242
  0.9433437826648147  0.3429247470416888  0.3304722737829290
  0.4197831632318161  0.9047747093360002  0.8144259648636453
  0.6480781564288268  0.8359687352183924  0.9496618135214194
  0.3394027527733210  0.8320331993030677  0.4472730216569508
  0.4029062129539993  0.3015849977960455  0.5806850344719640
  0.8384461227734644  0.8323922968650733  0.0582136640641291
  0.7456781855078146  0.3810854981733215 -0.0004757708264059
  0.0478773169322151  0.1773381698812169  0.3372946502668972
  0.6196714131488410  0.5145477330471349  0.7748589489256390
  0.9120105490325190  0.1005090107799590  0.8055936624413551
  0.6772528169802721  0.6691797395459173  0.4410657958957808
  0.3208293958684387  0.6702604831166806  0.9696997613161854
  0.8244121898396310  0.1761104109769525  0.4470977455408234
  0.9082202190401105  0.9103927199459246  0.6972100610354042
  0.1177961469493761  0.0000258612331056  0.2473495633484562
  0.5596262246692874  0.3688855508355270  0.6767967332041791
LNdSbO
 1.00000000000000000
    12.6389450320814749    0.0038949815655358   -0.0005210920919075
     0.0038975370543952   12.6357289917442710    0.0044988024764395
    -0.0005013705759272    0.0044524047673732   12.6886983447725417
     Sb   O   Nd   Li
     16   96   24   40
Direct
  0.7437295013490102  0.2429154234794457  0.7487299052905164
  0.2444472822811940  0.7512851771097719  0.2491356880721070
  0.7512032348846978  0.7486360251602038  0.2487293877428219
  0.2451457026872961  0.2457770614453413  0.7468349459641229
  0.2510276432012037  0.7430428297981856  0.7468254478778602
  0.7454551924434866  0.2502845229521058  0.2485693452215753
  0.2463323626472656  0.2488804052321863  0.2493558706053354
  0.7498113905585991  0.7468164999445148  0.7516460191432992
  0.4900083384405909  0.4993917909238182 -0.0023120009000388
  0.0002815714331685 -0.0013518059575942  0.5051514469123665
  0.5018091684678371 -0.0007652984477970  0.5085995104543708
 -0.0075562337082870  0.4977318132047581 -0.0047536449807852
  0.0005031813436362  0.4966827368692166  0.4965971242073677
  0.4923528377486671 -0.0000574617723960  0.0055533668314126
 -0.0045888980054969 -0.0056232318013125  0.0050395798857484
  0.5006369859591583  0.4990904680493530  0.4987624800226608
  0.5357167043899822  0.3509069197262066  0.0365047670800266251
  0.0552340088305497  0.8532367994204063  0.5323708964802966
  0.9459154896863299  0.6415504062327669  0.5318240969346200
  0.4330742141828184  0.1476075496183983  0.0269558370867516
  0.4521954870866697  0.8487863300427451  0.4713683962092595
  0.9443902048390321  0.3459082523913733  0.9751580412761532
  0.0425613426858852  0.1458153257331312  0.9734007677638365
  0.5502664065262544  0.6480768156064899  0.4654008284118932
  0.3569565911183072  0.0230295406268395  0.5610165008472549
  0.8461740370015440  0.5309215216283319  0.0526277508318188
  0.6390722854049536  0.5187126484722214  0.9498284989576625
  0.1477262328679414  0.0326360831192425  0.4502629772730075
  0.8548733782024701  0.4729072987788944  0.4369227822464019
  0.3443832492208717  0.9656437209273051  0.9627832338216697
  0.1427712619275906  0.9679797086981523  0.0582894478134571
  0.6511045524947723  0.4633148199084019  0.5349261504370822
  0.0272339608001590  0.5590044532125612  0.3526374114002245
  0.5169328674453965  0.0530486786569620  0.8610857852300740
  0.5308922899081754  0.9455559621355991  0.6543478480747945
  0.0325294763913348  0.4505592659760539  0.1428962616985596
  0.4586079329617199  0.4427906570067285  0.8547184394751529
  0.9692259918542681  0.9434503951344277  0.3591163870394962
  0.9696480916096433  0.0453300586372337  0.1524667350808097
  0.4831907323116812  0.5500503212740813  0.6451361778024142
  0.9546071645677533  0.1474667357104357  0.4690984404385417
  0.4399934598308044  0.6459260969310315  0.9739970230004926
  0.5499177843487333  0.8546232095899369  0.9763329588621230
  0.0559161726122259  0.3494345427790908  0.4660670938870038
  0.0368888361788881  0.6465756450789111  0.0203411955809503
  0.5572811309744490  0.1484235335116603  0.5270695031636790
  0.4538606688595738  0.3499906170686510  0.5247064259884969
  0.9359184058447344  0.8495396023302011  0.0271137036726611
  0.1440724131059646  0.4675817068650160  0.9501109412634413
  0.6492118879985279  0.9581221831510333  0.4624903202265355
  0.8520616571676007  0.9744184231157712  0.5570894833516278
  0.3404501642655003  0.4696897163192122  0.0497016941234566
  0.6373694976918854  0.0270782015338169  0.0614756504932178
  0.1451662852111945  0.5226271754509422  0.5524824590931988
  0.3544082280636517  0.5352606536348070  0.4477427952595298
  0.8494819014077095  0.0345874700947561  0.9496867560275233
  0.4624550393594125  0.9453698615563962  0.1551137512232183
  0.9669924210179685  0.4388464053875687  0.6408549903686556
  0.9667793132709488  0.5324570006811729  0.8462485485398268
  0.4788401179797536  0.0452339263363070  0.3586907973664191
  0.0258121216071117  0.0556545240481538  0.6512317495973979
  0.5235107834580196  0.5477135022376128  0.1469811856230068
  0.5232295436130244  0.4543283977805749  0.3454183727345755
  0.0255263690214855  0.9504439773186321  0.8565032569013749
  0.2969382620568428  0.7807684612068124  0.1017354828365739
```

```
  0.7917211274314252   0.2710060175952660   0.6004807306843458
  0.6927157517446212   0.2886391600349645   0.3949664425247995
  0.1903692062639371   0.7733440948898581   0.8898483145241366
  0.1923125512449577   0.2238405383028006   0.5987437442888738
  0.7016261338757683   0.7136995566688672   0.1009799222532684
  0.8063552586975320   0.7235179406025942   0.8979117704709605
  0.2948306227830090   0.2197298049106695   0.3969122850646482
  0.0951348342188312   0.2944587514313688   0.7824119657137576
  0.6004887822832597   0.7964612884903686   0.2777354091516461
  0.3952384373422814   0.7000298221968746   0.2768208598506725
  0.8956485100214018   0.1998343321913708   0.7808396011260338
  0.5982629960032370   0.1995942590767039   0.2249691025021784
  0.1038501973835138   0.6911671354425020   0.7183362348301559
  0.8946886611887324   0.8021589926428823   0.7206398270701401
  0.3937058269924454   0.3024650822087079   0.2140844567068921
  0.7716188005185338   0.1059147508943885   0.3065363752041578
  0.2770419053096341   0.5959999792325853   0.8052022645254783
  0.2736607171966954   0.3946733220040428   0.7022141178186982
  0.7828159196188162   0.8941489995127313   0.1982876725375035
  0.2071932820654264   0.6067166013684968   0.1923276910712000
  0.7194581361387481   0.0955129485274825   0.6981721985205730
  0.7179666007303914   0.8960586556807546   0.7964674107692986
  0.2225170988973328   0.3974322032820001   0.2981291926198961
  0.6928622819396553   0.2199765898939759   0.8972229703948883
  0.2067533720888604   0.7264622129136866   0.3988704458640541
  0.3093582828960256   0.7167855398684060   0.6011487882088009
  0.7992655834792303   0.2212144621742428   0.1022744577789144
  0.7998974533520479   0.7663567731679737   0.3961951909082946
  0.2877644162007734   0.2789023877633510   0.8992808861481273
  0.1947265693445678   0.2788807113227885   0.1009940918744635
  0.6955050184180160   0.7801680653446050   0.6024915056761047
  0.9009070388047969   0.7004083496232785   0.2164274707478575
  0.3898083720842133   0.1896673940618379   0.7287619909000964
  0.5988744812984714   0.3012091282991058   0.7162455294385097
  0.0959837500707134   0.8028746305162204   0.2281493021764964
  0.3947859202407374   0.8000659326955887   0.7807082452666121
  0.8896910242993544   0.3083136266099239   0.2722520765242759
  0.0989143863809154   0.1984844555223386   0.2835556494194052
  0.6044235731328811   0.6930152519933365   0.7841924852599655
  0.2234123414938849   0.8900350637764344   0.6948184280136251
  0.7117693216969447   0.3999555690809086   0.1973207644206354
  0.7169847881442210   0.5980029081057338   0.2920007925296081
  0.2092820254946993   0.1008504319428507   0.8038705113675593
  0.7809234415793053   0.3894524282344728   0.8024595553424262
  0.2851471423627388   0.8994249175394613   0.2904149887468975
  0.2710589144639423   0.1026903194863680   0.1948669474915024
  0.7705039737515925   0.6003567844840807   0.6950896147476158
  0.4972389340567308   0.2442242803316134   0.8736960555802143
 -0.0002321257643494   0.7492234179110215   0.3873216153432283
  0.4992975524001607   0.7455159848486224   0.6374360943511581
 -0.0055119532971617   0.2436324513422999   0.1324973648789018
  0.2464246209326292   0.8765840217027513   0.5096035989231114
  0.7457853888239213   0.3709558393867446  -0.0009107960504189
  0.7571191932551686   0.6207179997336628   0.5094500417557136
  0.2422106103258709   0.1185739425512394  -0.0019775330681773
  0.8805490359012299   0.4983635706256181   0.2504284736203353
  0.3706143170290555  -0.0065106811698315   0.7552251023072355
  0.6239823708314305   0.4978629367283088   0.7564950845454629
  0.1209660097455885   0.0001864445597260   0.2572180074466791
 -0.0050338619420984   0.2506405675625574   0.6202265796232721
  0.4973451056293776   0.7490973422946654   0.1170633390257484
  0.0016151965989688   0.7553856253998794   0.8668817818095940
  0.4910837686738043   0.2408060365835434   0.3738956516929719
  0.2459088589468194   0.6331580454472722  -0.0061569603803846
  0.7526802655379099   0.1212963085257717   0.4945944809057759
  0.7422712937945030   0.8790489935450221  -0.0072992064934684
  0.2517167289617875   0.3781579089667525   0.5029249209610288
  0.6275575502990944   0.0002167701473928   0.2467275608430456
  0.1202391435733701   0.5046857950331903   0.7404740179474416
  0.8742178651045356  -0.0043436165330342   0.7494414191331668
  0.3647100894929655   0.5050900737683489   0.2522319078666739
  0.7335308849581238   0.1225222437407313   0.0079840726330984
  0.2383333817347849   0.6243201386088846   0.5061255801332886
  0.7737090989277114   0.8694526347951157   0.4929206598470591
  0.3013904870407120   0.4279113462444490  -0.1094648604836570
  0.6717592965928668   0.4414002354088583   0.3507964819177045
 -0.0006351187888197   0.7470548205831076   0.6265326863631873
  0.3733661319432055   0.0241569725293775   0.2627095175465475
  0.4811318231493086   0.2368978108456861   0.6281037679141898
  0.4982456116861810   0.7502644386262169   0.8773488185017215
 -0.0194150608688540   0.2608331371194305   0.3719866359709472
  0.7747218320287443   0.3717690975820167   0.4966636891806749
  0.2306713846130750   0.8715446384935732  -0.0159941161993144
  0.7373541478065962   0.6230672542847980  -0.0230125901720855
  0.2643662845776691   0.1215998342339534   0.5128350621730667
  0.8719435838443792   0.0072356683796298   0.2614984626321873
  0.6188546490577594   0.0084981742810464   0.7712279740162259
  0.1218517476860223   0.4981660926833202   0.2628176152509633
  0.5982270434877383   0.8027874020025261   0.4298244335913686
  0.0579392531027112   0.1452380614415694   0.8231679650488529
  0.4995950944605010   0.2290327637684194   0.1246267231458940
  0.8692438465251827   0.5021997415059892   0.7375055260244742
  0.3142799578700600   0.9282092566891003   0.1260761574436674
  0.6748966815855677   0.5576200604393514   0.1484313371852831
  0.9454715432378594   0.8470054472771817   0.1746017281103294
```

```
  0.3761697719079694  0.5032793171811062  0.7359310833900676
  0.0555804984730759  0.6693613113003953  0.1645867459520001
  0.8955726296197138  0.1835121609776637  0.9289554427718181
  0.3979027527035436  0.6916740615177923  0.4293540769496718
  0.3497802660934681  0.3188664322298430  0.0595285699661274
  0.1282752996095681  0.3164076373602761  0.9316329186575489
  0.1840980045815173  0.4272017118863173  0.0935908498077034
  0.6781245981588907  0.9281132049447475  0.6297663983763691
  0.5667051337852105  0.6229066506022963  0.3153040139412109
  0.1238915785614543 -0.0148431549669115  0.7437648642970134
  0.5613785310789366  0.3967504074748404  0.1884673580476925
  0.9386530672499300  0.3499640831706027  0.8246980547856251
  0.4367047873161362  0.8552019902013114  0.3203976216588225
  0.6235227979760264  0.3150625817321154  0.5648093024057563
  0.1223663723240263  0.1825709782256397  0.4326990921747408
  0.8866938534268723  0.6901190681223830  0.0630863595862249
LNdSbTeO
 1.00000000000000000
    12.6375352370432381   -0.0174794114834886    0.0150075607089967
    -0.0175349305020835   12.6295231858591190   -0.0086722473773333
     0.0150612492313625   -0.0083320521393533   12.6078692452981169
   Sb   Te    O   Nd   Li
    8    8   96   24   32
Direct
  0.7528198360656839  0.2479090278157983  0.7519261535566223
  0.7486956273579991  0.7439511436234790  0.2504334566175016
  0.2520243199059324  0.7483328768094640  0.7517176867249251
  0.2508486698467706  0.2452665465957835  0.2495515722711918
  0.0027525621156054 -0.0009167659429923  0.4961263454293828
  0.0010911288226594  0.4985963968555578 -0.0006346024239816
  0.5020560813407381  0.0010033922218860 -0.0013142307001553
  0.5011472431715736  0.5006825465553458  0.4990428575022061
  0.2550570688699886  0.7477073684514769  0.2493877680063818
  0.2502410417602387  0.2490668116513780  0.7475395623572609
  0.7531386062590064  0.2459117718542856  0.2519877317280111
  0.7542254403387835  0.7482812403081276  0.7532427559876137
  0.5037654238407564  0.5006541428832919 -0.0028520398571086
  0.4979896643012496  0.0014749707202408  0.4979831329082174
  0.0013280325110557  0.4984956715406749  0.5014042669663822
  0.0048446501910567  0.0016374979310279 -0.0000463666426137
  0.5533527498662826  0.3596979254745525  0.0294360065864818
  0.0369612561999991  0.8469887053360839  0.5261421532794104
  0.9494172713428670  0.6427290735128339  0.5282188954708338
  0.4487652656754385  0.1478002347265293  0.0269114976650164
  0.4430614397070005  0.8590600275064906  0.4754580002705938
  0.9487740369069338  0.3516673572824812  0.9716260976517882
  0.0552160450105854  0.1427481507590966  0.9690173415031269
  0.5506374933734224  0.6500050534205888  0.4709359992888544
  0.3585245499860860  0.0301288239948542  0.5543419344459081
  0.8546766787120736  0.5255284806972786  0.0565730906091820
  0.6469989323235557  0.5265554853821949  0.9469568688113630
  0.1507345512884465  0.0171237287510195  0.4469738438515271
  0.8587759880935795  0.4702815511333928  0.4461424579387227
  0.3561646183896812  0.9715003583234714  0.9441991576225371
  0.1472233194116202  0.9727115680309623  0.0501431522791702
  0.6525154181195576  0.4726361703473884  0.5383562834909051
  0.0322726587507437  0.5496758371162251  0.3587105476504335
  0.5292616862635442  0.0539730468001897  0.8502341749053726
  0.5326836579688577  0.9474377677458993  0.6392984737557305
  0.0315727351837937  0.4452250217476122  0.1464840345695928
  0.4735652812558583  0.4529938023986012  0.8504860646186855
  0.9725431019110005  0.9505644928831882  0.3423508832360972
  0.9762592371963055  0.0527675747194558  0.1449577350223830
  0.4752355712369684  0.5446239778815106  0.6514199098569837
  0.9503015963071674  0.1437857943644091  0.4625523150664246
  0.4548205643732400  0.6425584774849501  0.9587872407113492
  0.5594542246331478  0.8554551622723938  0.9716491186937084
  0.0528824822911580  0.3564544490656814  0.4748435304213822
  0.0555488211593523  0.6457332363813562  0.0266972634693252
  0.5548809028453351  0.1414968719165961  0.5244023533879409
  0.4536146302905675  0.3526367844540420  0.5307721157435527
  0.9503619769066308  0.8612776405988741  0.0424958308318174
  0.1496436546328818  0.4722855315706178  0.9429516450743231
  0.6383535523592596  0.9704774062071575  0.4408660151773807
  0.8478705292398372  0.9662790581439937  0.5317757716510135
  0.3571658454497275  0.4722433433714744  0.0381937961245084
  0.6490797033533381  0.0279424533619280  0.0539882377100603
  0.1439162018769738  0.5287851580090007  0.5533299032872346
  0.3520756891582230  0.5224787323745786  0.4537743747735609
  0.8605216403902413  0.0269367114906031  0.9501260024295788
  0.4751202556994966  0.9439419102076515  0.1454199334852150
  0.9684242522696840  0.4481751335382285  0.6427874011891067
  0.9713141764106890  0.5545781990047313  0.8543418576316214
  0.4686131361937365  0.0522104134043139  0.3555781973831275
  0.0217409111958800  0.0492259091446137  0.6445686941747674
  0.5231412686211714  0.5563259478805604  0.1394379258878281
  0.5328458156311366  0.4609534212929533  0.3462187349923359
  0.0301011570002028  0.9447860004316103  0.8587588609727701
  0.3056654380896882  0.7758600283490986  0.1067190070818342
  0.8095179199567138  0.2722193830713918  0.6054997401603006
  0.7027111278173930  0.2830960577345126  0.3959144968547845
  0.2016474203654778  0.7744848127777139  0.8985413717420607
  0.2020553535545357  0.2209172867879915  0.6045386801350625
  0.7007449903157206  0.7235376975768018  0.1020835057222888
  0.8061242680934636  0.7224526285404074  0.8943504375071343
```

```
 0.3054176851940025   0.2200267048098105   0.3966558754552854
 0.1076007357605352   0.2995371472467668   0.7737517098129408
 0.5981294566496558   0.7817392805336932   0.2790035206495527
 0.3960612076442276   0.6923620076409642   0.2793264656907632
 0.8991944781644876   0.1939852033265052   0.7790392078105199
 0.6084138870947936   0.1988713951327725   0.2259703473192660
 0.1064955980086575   0.6917173258258865   0.7300393713205111
 0.8968281474400200   0.7990381495066934   0.7159168839706198
 0.3975897266269167   0.3044401910287621   0.2232413269368358
 0.7784232053194544   0.1031614370495962   0.3063633438908461
 0.2836760494725284   0.5951907552362683   0.7980419958477660
 0.2817983053884115   0.3975471345617784   0.7011967638862366
 0.7738669269864347   0.8956400842235869   0.2126159547686852
 0.2230453898440233   0.6045817019008062   0.1991889680898721
 0.7247839844450369   0.1012607525233449   0.6963066546700382
 0.7251801908716483   0.8914391949725223   0.8013123138827278
 0.2214458614093479   0.3927624741551126   0.3035889057772915
 0.6977845206365367   0.2215906908555200   0.8977918021890656
 0.1999006411974825   0.7228741084878852   0.3924274838376660
 0.3047567874738327   0.7095495896352421   0.6043293420943491
 0.7997160813462580   0.2224566790108148   0.1081228875203273
 0.8019952945135755   0.7802104807217679   0.4009352884648448
 0.2996390405100343   0.2868037144036989   0.8902114028158401
 0.1973163675660301   0.2728997449099201   0.1036378190929628
 0.7086757332585252   0.7769625688940330   0.6077573082416292
 0.8980489769945880   0.6962195449687313   0.2268079758145272
 0.3981563951436550   0.2123496217243665   0.7209748083339832
 0.6054954190302568   0.3061755807543368   0.7206145465325018
 0.1122527084377616   0.7994929411028722   0.2199530099157289
 0.4032894214933012   0.7922483300309626   0.7751017500046233
 0.8990500541615043   0.2867364123629045   0.2786621059566129
 0.1038701479397280   0.1891520903407954   0.2786199267329267
 0.6099821606386914   0.6965333558888700   0.7757601837193056
 0.2271347701756627   0.8932178752993750   0.6951207097578336
 0.7287945487462026   0.3924769186757092   0.2119666246154822
 0.7183752486960133   0.5943009900012992   0.3043938278939022
 0.2284867851529859   0.1064351228737007   0.7982422045621105
 0.7818018047537937   0.3953442795693010   0.8045939775133548
 0.2865680170042397   0.8885956443615554   0.3015069475810533
 0.2829607057535501   0.1004475123069812   0.1942342339770373
 0.7823731893798886   0.6060730579688016   0.6986773117531517
 0.5092073193871413   0.2429083661407386   0.8777575551941688
 0.0084008940159599   0.7436925133641867   0.3712402044642712
 0.5049130609745645   0.7539918865965486   0.6221242125522987
 0.0078145806040882   0.2522404665031190   0.1210885522608928
 0.2512029132957032   0.8717387356730000   0.5044835379344139
 0.7579724225528458   0.3731409108368251  -0.0035102509806808
 0.7558135495567223   0.6243130705120347   0.5020156360798659
 0.2563416219313452   0.1246696113275640   0.0000901240485919
 0.8841414577947612   0.5012966561181867   0.2448832035973217
 0.3790377501907637  -0.0035242154556070   0.7523959054103919
 0.6377898715249637   0.4954023348543866   0.7490927601120656
 0.1382845709362563  -0.0018918199027739   0.2482659693213240
-0.0015289010692377   0.2435936730005437   0.6248595934456833
 0.5009468497791550   0.7547963427286798   0.1179237499075553
 0.0046865496853978   0.7485508744872272   0.8783286804783601
 0.4939710525274567   0.2476390113709499   0.3741744268850516
 0.2439258800893221   0.6265145092909087   0.0018947309439602
 0.7487998743538707   0.1261788982268693   0.5059716111222854
 0.7475620736340434   0.8764447366218650   0.0013126696254184
 0.2469190092659488   0.3697670363544874   0.4923315491727369
 0.6212769258793538   0.0026675033903192   0.2434351707462508
 0.1163302952784896   0.4999596321676222   0.7530659490979264
 0.8762118845022671   0.0055139696011045   0.7558596017691402
 0.3702012655368763   0.4925195633318430   0.2474250545104789
 0.7518441699833748   0.1245236742042651   0.0016999682483985
 0.2460307383197630   0.6233132827500452   0.4881401638053393
 0.7221673324911011   0.8735093805313957   0.5017563661219590
 0.2413581035608930   0.3725619909047496   0.0083876153628664
 0.1266903496084383  -0.0014285350371021   0.7472343918579384
 0.6277745815503827   0.4934396311668234   0.2143396145887633
 0.8770280520606897   0.5000660872245398   0.7532793856694369
 0.3781712942878023  -0.0032184541029811   0.2462085420622703
 0.0060705472067721   0.7423530401622277   0.1255243034394603
 0.2521109558690834   0.1215192669394772   0.4980550210728312
 0.5132387258169115   0.7619037578309691   0.8731598736643024
 0.7656482077929465   0.3722264017605442   0.5151515204599590
 0.2553927012725068   0.8750477379573591  -0.0010240919692090
 0.7542825846299052   0.6243952575892916   0.0012959798038391
 0.4318805820648238   0.6041501387454401   0.8075834510569264
 0.9296422594513403   0.1033134834891811   0.3029230903638048
 0.6276812651478574  -0.0006194442461298   0.7487266407175944
 0.1271000437129947   0.4959891777850501   0.2510217017330689
 0.4847260309857079   0.7674248755723396   0.3745070365094592
 0.0175972987767281   0.7353366710376521   0.6281403184928434
 0.6811784364845921   0.4330812653916534   0.3714503646494017
 0.9157309309790579   0.9228795999703941   0.1908609387081435
 0.4326164053297906   0.3874445610129242   0.6860261357013704
 0.8799052789907282   0.8245089166037827   0.5662672431400571
 0.5000576483353394   0.2509245841446487   0.1254933461564368
 0.3243027657679188   0.5586820371563035   0.6478328769834412
 0.8242369472467285   0.9325191699671207   0.3613908783823485
 0.3216693634421491   0.4414828555027175   0.8535606187970232
 0.5698407048308834   0.6267570123501132   0.3227778426418548
 0.0014325799772109   0.2487341377518441   0.8738404239102660
```

```
  0.5122489982514965  0.2389796100514402  0.6222800031569615
  0.0104660577535135  0.2565473397000192  0.3761601338557473
LNdTeO
   1.00000000000000
    12.5586675039427789    0.0000000000000000    0.0000000000000000
     0.0000000000000000   12.5586675039427789    0.0000000000000000
     0.0000000000000000    0.0000000000000000   12.5586675039427789
   Te   O    Nd   Li
   16   96   24   24
Direct
  0.7500000000000000  0.2500000000000000  0.7500000000000000
  0.2500000000000000  0.7500000000000000  0.2500000000000000
  0.7500000000000000  0.7500000000000000  0.2500000000000000
  0.2500000000000000  0.2500000000000000  0.7500000000000000
  0.2500000000000000  0.7500000000000000  0.7500000000000000
  0.7500000000000000  0.2500000000000000  0.2500000000000000
  0.2500000000000000  0.2500000000000000  0.2500000000000000
  0.7500000000000000  0.7500000000000000  0.7500000000000000
  0.5000000000000000  0.5000000000000000  0.0000000000000000
  0.0000000000000000  0.0000000000000000  0.5000000000000000
  0.5000000000000000  0.0000000000000000  0.5000000000000000
  0.0000000000000000  0.5000000000000000  0.0000000000000000
  0.0000000000000000  0.5000000000000000  0.5000000000000000
  0.5000000000000000  0.0000000000000000  0.0000000000000000
  0.0000000000000000  0.0000000000000000  0.0000000000000000
  0.5000000000000000  0.5000000000000000  0.5000000000000000
  0.5508748058318484  0.3560583870829312  0.0279226696456827
  0.0508748058318483  0.8560583870829312  0.5279226696456828
  0.9491251941681516  0.6439416129170688  0.5279226696456828
  0.4491251941681518  0.1439416129170687  0.0279226696456827
  0.4491251941681518  0.8560583870829312  0.4720773303543173
  0.9491251941681516  0.3560583870829312  0.9720773303543172
  0.0508748058318483  0.1439416129170687  0.9720773303543172
  0.5508748058318484  0.6439416129170688  0.4720773303543173
  0.3560583870829312  0.0279226696456827  0.5508748058318484
  0.8560583870829312  0.5279226696456828  0.0508748058318483
  0.6439416129170688  0.5279226696456828  0.9491251941681516
  0.1439416129170687  0.0279226696456827  0.4491251941681518
  0.8560583870829312  0.4720773303543173  0.4491251941681518
  0.3560583870829312  0.9720773303543172  0.9491251941681516
  0.1439416129170687  0.9720773303543172  0.0508748058318483
  0.6439416129170688  0.4720773303543173  0.5508748058318484
  0.0279226696456827  0.5508748058318484  0.3560583870829312
  0.5279226696456828  0.0508748058318483  0.8560583870829312
  0.5279226696456828  0.9491251941681516  0.6439416129170688
  0.0279226696456827  0.4491251941681518  0.1439416129170687
  0.4720773303543173  0.4491251941681518  0.8560583870829312
  0.9720773303543172  0.9491251941681516  0.3560583870829312
  0.9720773303543172  0.0508748058318483  0.1439416129170687
  0.4720773303543173  0.5508748058318484  0.6439416129170688
  0.9491251941681516  0.1439416129170687  0.4720773303543173
  0.4491251941681518  0.6439416129170688  0.9720773303543172
  0.5508748058318484  0.8560583870829312  0.9720773303543172
  0.0508748058318483  0.3560583870829312  0.4720773303543173
  0.0508748058318483  0.6439416129170688  0.0279226696456827
  0.5508748058318484  0.1439416129170687  0.5279226696456828
  0.4491251941681518  0.3560583870829312  0.5279226696456828
  0.9491251941681516  0.8560583870829312  0.0279226696456827
  0.1439416129170687  0.4720773303543173  0.9491251941681516
  0.6439416129170688  0.9720773303543172  0.4491251941681518
  0.8560583870829312  0.9720773303543172  0.5508748058318484
  0.3560583870829312  0.4720773303543173  0.0508748058318483
  0.6439416129170688  0.0279226696456827  0.0508748058318483
  0.1439416129170687  0.5279226696456828  0.5508748058318484
  0.3560583870829312  0.5279226696456828  0.4491251941681518
  0.8560583870829312  0.0279226696456827  0.9491251941681516
  0.4720773303543173  0.9491251941681516  0.1439416129170687
  0.9720773303543172  0.4491251941681518  0.6439416129170688
  0.9720773303543172  0.5508748058318484  0.8560583870829312
  0.4720773303543173  0.0508748058318483  0.3560583870829312
  0.0279226696456827  0.0508748058318483  0.6439416129170688
  0.5279226696456828  0.5508748058318484  0.1439416129170687
  0.5279226696456828  0.4491251941681518  0.3560583870829312
  0.0279226696456827  0.9491251941681516  0.8560583870829312
  0.3008748058318482  0.7779226696456828  0.1060583870829313
  0.8008748058318484  0.2779226696456827  0.6060583870829312
  0.6991251941681516  0.2779226696456827  0.3939416129170688
  0.1991251941681517  0.7779226696456828  0.8939416129170688
  0.1991251941681517  0.2220773303543173  0.6060583870829312
  0.6991251941681516  0.7220773303543172  0.1060583870829313
  0.8008748058318484  0.7220773303543172  0.8939416129170688
  0.3008748058318482  0.2220773303543173  0.3939416129170688
  0.1060583870829313  0.3008748058318482  0.7779226696456828
  0.6060583870829312  0.8008748058318484  0.2779226696456827
  0.3939416129170688  0.6991251941681516  0.2779226696456827
  0.8939416129170688  0.1991251941681517  0.7779226696456828
  0.6060583870829312  0.1991251941681517  0.2220773303543173
  0.1060583870829313  0.6991251941681516  0.7220773303543172
  0.8939416129170688  0.8008748058318484  0.7220773303543172
  0.3939416129170688  0.3008748058318482  0.2220773303543173
  0.7779226696456828  0.1060583870829313  0.3008748058318482
  0.2779226696456827  0.6060583870829312  0.8008748058318484
  0.2779226696456827  0.3939416129170688  0.6991251941681516
  0.7779226696456828  0.8939416129170688  0.1991251941681517
  0.2220773303543173  0.6060583870829312  0.1991251941681517
```

```
     0.7220773303543172    0.1060583870829313    0.6991251941681516
     0.7220773303543172    0.8939416129170688    0.8008748058318484
     0.2220773303543173    0.3939416129170688    0.3008748058318482
     0.6991251941681516    0.2220773303543173    0.8939416129170688
     0.1991251941681517    0.7220773303543172    0.3939416129170688
     0.3008748058318482    0.7220773303543172    0.6060583870829312
     0.8008748058318484    0.2220773303543173    0.1060583870829313
     0.8008748058318484    0.7779226696456828    0.3939416129170688
     0.3008748058318482    0.2779226696456827    0.8939416129170688
     0.1991251941681517    0.2779226696456827    0.1060583870829313
     0.6991251941681516    0.7779226696456828    0.6060583870829312
     0.8939416129170688    0.6991251941681516    0.2220773303543173
     0.3939416129170688    0.1991251941681517    0.7220773303543172
     0.6060583870829312    0.3008748058318482    0.7220773303543172
     0.1060583870829313    0.8008748058318484    0.2220773303543173
     0.3939416129170688    0.8008748058318484    0.7779226696456828
     0.8939416129170688    0.3008748058318482    0.2779226696456827
     0.1060583870829313    0.1991251941681517    0.2779226696456827
     0.6060583870829312    0.6991251941681516    0.7779226696456828
     0.2220773303543173    0.8939416129170688    0.6991251941681516
     0.7220773303543172    0.3939416129170688    0.1991251941681517
     0.7220773303543172    0.6060583870829312    0.3008748058318482
     0.2220773303543173    0.1060583870829313    0.8008748058318484
     0.7779226696456828    0.3939416129170688    0.8008748058318484
     0.2779226696456827    0.8939416129170688    0.3008748058318482
     0.2779226696456827    0.1060583870829313    0.1991251941681517
     0.7779226696456828    0.6060583870829312    0.6991251941681516
     0.5000000000000000    0.2500000000000000    0.8750000000000000
     0.0000000000000000    0.7500000000000000    0.3750000000000000
     0.5000000000000000    0.7500000000000000    0.6250000000000000
     0.0000000000000000    0.2500000000000000    0.1250000000000000
     0.2500000000000000    0.8750000000000000    0.5000000000000000
     0.7500000000000000    0.3750000000000000    0.0000000000000000
     0.7500000000000000    0.6250000000000000    0.5000000000000000
     0.2500000000000000    0.1250000000000000    0.0000000000000000
     0.8750000000000000    0.5000000000000000    0.2500000000000000
     0.3750000000000000    0.0000000000000000    0.7500000000000000
     0.6250000000000000    0.5000000000000000    0.7500000000000000
     0.1250000000000000    0.0000000000000000    0.2500000000000000
     0.0000000000000000    0.2500000000000000    0.6250000000000000
     0.5000000000000000    0.7500000000000000    0.1250000000000000
     0.0000000000000000    0.7500000000000000    0.8750000000000000
     0.5000000000000000    0.2500000000000000    0.3750000000000000
     0.2500000000000000    0.6250000000000000    0.0000000000000000
     0.7500000000000000    0.1250000000000000    0.5000000000000000
     0.7500000000000000    0.8750000000000000    0.0000000000000000
     0.2500000000000000    0.3750000000000000    0.5000000000000000
     0.6250000000000000    0.0000000000000000    0.2500000000000000
     0.1250000000000000    0.5000000000000000    0.7500000000000000
     0.8750000000000000    0.0000000000000000    0.7500000000000000
     0.3750000000000000    0.5000000000000000    0.2500000000000000
     0.7500000000000000    0.1250000000000000    0.0000000000000000
     0.2500000000000000    0.6250000000000000    0.5000000000000000
     0.7500000000000000    0.8750000000000000    0.5000000000000000
     0.2500000000000000    0.3750000000000000    0.0000000000000000
     0.1250000000000000    0.0000000000000000    0.7500000000000000
     0.6250000000000000    0.5000000000000000    0.2500000000000000
     0.8750000000000000    0.5000000000000000    0.7500000000000000
     0.3750000000000000    0.0000000000000000    0.2500000000000000
     0.0000000000000000    0.7500000000000000    0.1250000000000000
     0.5000000000000000    0.2500000000000000    0.6250000000000000
     0.5000000000000000    0.7500000000000000    0.8750000000000000
     0.0000000000000000    0.2500000000000000    0.3750000000000000
     0.7500000000000000    0.3750000000000000    0.5000000000000000
     0.2500000000000000    0.8750000000000000    0.0000000000000000
     0.7500000000000000    0.6250000000000000    0.0000000000000000
     0.2500000000000000    0.1250000000000000    0.5000000000000000
     0.3750000000000000    0.5000000000000000    0.7500000000000000
     0.8750000000000000    0.0000000000000000    0.2500000000000000
     0.6250000000000000    0.0000000000000000    0.7500000000000000
     0.1250000000000000    0.5000000000000000    0.2500000000000000
     0.5000000000000000    0.7500000000000000    0.3750000000000000
     0.0000000000000000    0.2500000000000000    0.8750000000000000
     0.0000000000000000    0.7500000000000000    0.6250000000000000
     0.5000000000000000    0.2500000000000000    0.1250000000000000
LNdWO
   1.00000000000000
    12.4563751420770163     0.0000000000000000     0.0000000000000000
     0.0000000000000000    12.4563733805716961     0.0000000000000000
     0.0000000000000000     0.0000000000000000    12.4563828098189848
     W    O    Nd   Li
    16   96   24   24
Direct
  0.7500000000000000   0.2500000000000000   0.7500000000000000
  0.2500000000000000   0.7500000000000000   0.2500000000000000
  0.7500000000000000   0.7500000000000000   0.2500000000000000
  0.2500000000000000   0.2500000000000000   0.7500000000000000
  0.2500000000000000   0.7500000000000000   0.7500000000000000
  0.7500000000000000   0.2500000000000000   0.2500000000000000
  0.2500000000000000   0.2500000000000000   0.2500000000000000
  0.7500000000000000   0.7500000000000000   0.7500000000000000
  0.5000000000000000   0.5000000000000000   0.0000000000000000
  0.0000000000000000   0.0000000000000000   0.5000000000000000
  0.5000000000000000   0.0000000000000000   0.5000000000000000
  0.0000000000000000   0.5000000000000000   0.0000000000000000
```

```
0.0000000000000000   0.5000000000000000   0.5000000000000000
0.5000000000000000   0.0000000000000000   0.0000000000000000
0.0000000000000000   0.0000000000000000   0.0000000000000000
0.5000000000000000   0.5000000000000000   0.5000000000000000
0.5528101098357530   0.3563465455440783   0.0294739368029272
0.0528101098357529   0.8563465455440784   0.5294739368029272
0.9471898901642470   0.6436534544559216   0.5294739368029272
0.4471898901642471   0.1436534544559217   0.0294739368029272
0.4471898901642471   0.8563465455440784   0.4705260631970728
0.9471898901642470   0.3563465455440783   0.9705260631970728
0.0528101098357529   0.1436534544559217   0.9705260631970728
0.5528101098357530   0.6436534544559216   0.4705260631970728
0.3563458939836133   0.0294741356821882   0.5528099065523369
0.8563458939836133   0.5294741356821882   0.0528099065523369
0.6436541060163867   0.5294741356821882   0.9471900934476631
0.1436541060163867   0.0294741356821882   0.4471900934476631
0.8563458939836133   0.4705258643178118   0.4471900934476631
0.3563458939836133   0.9705258643178118   0.9471900934476631
0.1436541060163867   0.9705258643178118   0.0528099065523369
0.6436541060163867   0.4705258643178118   0.5528099065523369
0.0294743068034176   0.5528098510791158   0.3563457163902271
0.5294743068034178   0.0528098510791158   0.8563457163902272
0.5294743068034178   0.9471901489208842   0.6436542836097728
0.0294743068034176   0.4471901489208841   0.1436542836097729
0.4705256931965823   0.4471901489208841   0.8563457163902272
0.9705256931965822   0.9471901489208842   0.3563457163902271
0.9705256931965822   0.0528098510791158   0.1436542836097729
0.4705256931965823   0.5528098510791158   0.6436542836097728
0.9471898901642470   0.1436534544559217   0.4705260631970728
0.4471898901642471   0.6436534544559216   0.9705260631970728
0.5528101098357530   0.8563465455440784   0.9705260631970728
0.0528101098357529   0.3563465455440783   0.4705260631970728
0.0528101098357529   0.6436534544559216   0.0294739368029272
0.5528101098357530   0.1436534544559217   0.5294739368029272
0.4471898901642471   0.3563465455440783   0.5294739368029272
0.9471898901642470   0.8563465455440784   0.0294739368029272
0.1436541060163867   0.4705258643178118   0.9471900934476631
0.6436541060163867   0.9705258643178118   0.4471900934476631
0.8563458939836133   0.9705258643178118   0.5528099065523369
0.3563458939836133   0.4705258643178118   0.0528099065523369
0.6436541060163867   0.0294741356821882   0.0528099065523369
0.1436541060163867   0.5294741356821882   0.5528099065523369
0.3563458939836133   0.5294741356821882   0.4471900934476631
0.8563458939836133   0.0294741356821882   0.9471900934476631
0.4705256931965823   0.9471901489208842   0.1436542836097729
0.9705256931965822   0.4471901489208841   0.6436542836097728
0.9705256931965822   0.5528098510791158   0.8563457163902272
0.4705256931965823   0.0528098510791158   0.3563457163902271
0.0294743068034176   0.0528098510791158   0.6436542836097728
0.5294743068034178   0.5528098510791158   0.1436542836097729
0.5294743068034178   0.4471901489208841   0.3563457163902271
0.0294743068034176   0.9471901489208842   0.8563457163902272
0.3028097762230786   0.7794741675173997   0.1063456456471443
0.8028097762230786   0.2794741675173998   0.6063456456471442
0.6971902237769214   0.2794741675173998   0.3936543543528557
0.1971902237769214   0.7794741675173997   0.8936543543528558
0.1971902237769214   0.2205258324826002   0.6063456456471442
0.6971902237769214   0.7205258324826003   0.1063456456471443
0.8028097762230786   0.7205258324826003   0.8936543543528558
0.3028097762230786   0.2205258324826002   0.3936543543528557
0.1063461036540274   0.3028100785352793   0.7794739478479753
0.6063461036540274   0.8028100785352793   0.2794739478479754
0.3936538963459726   0.6971899214647207   0.2794739478479754
0.8936538963459726   0.1971899214647206   0.7794739478479753
0.6063461036540274   0.1971899214647206   0.2205260521520245
0.1063461036540274   0.6971899214647207   0.7205260521520247
0.8936538963459726   0.8028100785352793   0.7205260521520247
0.3936538963459726   0.3028100785352793   0.2205260521520245
0.7794741202481269   0.1063465559385891   0.3028099300019390
0.2794741202481270   0.6063465559385892   0.8028099300019390
0.2794741202481270   0.3936534440614108   0.6971900699980610
0.7794741202481269   0.8936534440614108   0.1971900699980611
0.2205258797518730   0.6063465559385892   0.1971900699980611
0.7205258797518731   0.1063465559385891   0.6971900699980610
0.7205258797518731   0.8936534440614108   0.8028099300019390
0.2205258797518730   0.3936534440614108   0.3028099300019390
0.6971902237769214   0.2205258324826002   0.8936543543528558
0.1971902237769214   0.7205258324826003   0.3936543543528557
0.3028097762230786   0.7205258324826003   0.6063456456471442
0.8028097762230786   0.2205258324826002   0.1063456456471443
0.8028097762230786   0.7794741675173997   0.3936543543528557
0.3028097762230786   0.2794741675173998   0.8936543543528558
0.1971902237769214   0.2794741675173998   0.1063456456471443
0.6971902237769214   0.7794741675173997   0.6063456456471442
0.8936538963459726   0.6971899214647207   0.2205260521520245
0.3936538963459726   0.1971899214647206   0.7205260521520247
0.6063461036540274   0.3028100785352793   0.7205260521520247
0.1063461036540274   0.8028100785352793   0.2205260521520245
0.3936538963459726   0.8028100785352793   0.7794739478479753
0.8936538963459726   0.3028100785352793   0.2794739478479754
0.1063461036540274   0.1971899214647206   0.2794739478479754
0.6063461036540274   0.6971899214647207   0.7794739478479753
0.2205258797518730   0.8936534440614108   0.6971900699980610
0.7205258797518731   0.3936534440614108   0.1971900699980611
0.7205258797518731   0.6063465559385892   0.3028099300019390
```

```
     0.2205258797518730    0.1063465559385891    0.8028099300019390
     0.7794741202481269    0.3936534440614108    0.8028099300019390
     0.2794741202481270    0.8936534440614108    0.3028099300019390
     0.2794741202481270    0.1063465559385891    0.1971900699980611
     0.7794741202481269    0.6063465559385892    0.6971900699980610
     0.5000000000000000    0.2500000000000000    0.8750003158211267
     0.0000000000000000    0.7500000000000000    0.3750003158211268
     0.5000000000000000    0.7500000000000000    0.6249996841788733
     0.0000000000000000    0.2500000000000000    0.1249996841788733
     0.2500000000000000    0.8749999195710516    0.5000000000000000
     0.7500000000000000    0.3749999195710517    0.0000000000000000
     0.7500000000000000    0.6250000804289484    0.5000000000000000
     0.2500000000000000    0.1250000804289483    0.0000000000000000
     0.8749999912049573    0.5000000000000000    0.2500000000000000
     0.3749999912049573    0.0000000000000000    0.7500000000000000
     0.6250000087950427    0.5000000000000000    0.7500000000000000
     0.1250000087950428    0.0000000000000000    0.2500000000000000
     0.0000000000000000    0.2500000000000000    0.6249996841788733
     0.5000000000000000    0.7500000000000000    0.1249996841788733
     0.0000000000000000    0.7500000000000000    0.8750003158211267
     0.5000000000000000    0.2500000000000000    0.3750003158211268
     0.2500000000000000    0.6250000804289484    0.0000000000000000
     0.7500000000000000    0.1250000804289483    0.5000000000000000
     0.7500000000000000    0.8749999195710516    0.0000000000000000
     0.2500000000000000    0.3749999195710517    0.5000000000000000
     0.6250000087950427    0.0000000000000000    0.2500000000000000
     0.1250000087950428    0.5000000000000000    0.7500000000000000
     0.8749999912049573    0.0000000000000000    0.7500000000000000
     0.3749999912049573    0.5000000000000000    0.2500000000000000
     0.7500000000000000    0.1249999369170279    0.0000000000000000
     0.2500000000000000    0.6249999369170278    0.5000000000000000
     0.7500000000000000    0.8750000630829722    0.5000000000000000
     0.2500000000000000    0.3750000630829721    0.0000000000000000
     0.1250000109320543    0.0000000000000000    0.7500000000000000
     0.6250000109320544    0.5000000000000000    0.2500000000000000
     0.8749999890679456    0.5000000000000000    0.7500000000000000
     0.3749999890679457    0.0000000000000000    0.2500000000000000
     0.0000000000000000    0.7500000000000000    0.1250000458543647
     0.5000000000000000    0.2500000000000000    0.6250000458543647
     0.5000000000000000    0.7500000000000000    0.8749999541456353
     0.0000000000000000    0.2500000000000000    0.3749999541456353
     0.7500000000000000    0.3750000630829721    0.5000000000000000
     0.2500000000000000    0.8750000630829722    0.0000000000000000
     0.7500000000000000    0.6249999369170278    0.0000000000000000
     0.2500000000000000    0.1249999369170279    0.5000000000000000
     0.3749999890679457    0.5000000000000000    0.7500000000000000
     0.8749999890679456    0.0000000000000000    0.2500000000000000
     0.6250000109320544    0.0000000000000000    0.7500000000000000
     0.1250000109320543    0.5000000000000000    0.2500000000000000
     0.5000000000000000    0.7500000000000000    0.3749999541456353
     0.0000000000000000    0.2500000000000000    0.8749999541456353
     0.0000000000000000    0.7500000000000000    0.6250000458543647
     0.5000000000000000    0.2500000000000000    0.1250000458543647
t-LLZO
    1.00000000000000
     13.1000852327858865    0.0000000000000000    0.0000000000000000
      0.0000000000000000   13.1000852327858865    0.0000000000000000
      0.0000000000000000    0.0000000000000000   12.5669485226605389
    Zr    O    La   Li
    16    96    24   56
Direct
  0.2453270000000032  0.2570089999999965  0.2501960000000025
  0.7453270000000032  0.7570089999999965  0.7501960000000025
  0.7453270000000032  0.2570089999999965  0.7501960000000025
  0.2453270000000032  0.7570089999999965  0.2501960000000025
  0.9953270000000032  0.5070089999999965  0.0001960000000025
  0.4953270000000032  0.0070089999999965  0.5001960000000025
  0.9953270000000032  0.5070089999999965  0.5001960000000025
  0.4953270000000032  0.0070089999999965  0.0001960000000025
  0.2453270000000032  0.2570089999999965  0.7501960000000025
  0.7453270000000032  0.7570089999999965  0.2501960000000025
  0.7453270000000032  0.2570089999999965  0.2501960000000025
  0.2453270000000032  0.7570089999999965  0.7501960000000025
  0.9953270000000032  0.0070089999999965  0.5001960000000025
  0.4953270000000032  0.5070089999999965  0.0001960000000025
  0.9953270000000032  0.0070089999999965  0.0001960000000025
  0.4953270000000032  0.5070089999999965  0.5001960000000025
  0.3015512544919790  0.2915074041693172  0.4044451962048098
  0.8015512544919791  0.7915074041693173  0.9044451962048098
  0.6891037455080248  0.2915074041693172  0.9044451962048098
  0.1891037455080247  0.7915074041693173  0.4044451962048098
  0.9608285958306824  0.4507847455080207  0.1544451962048098
  0.4608285958306825  0.9507847455080206  0.6544451962048098
  0.0298254041693169  0.4507847455080207  0.6544451962048098
  0.5298254041693169  0.9507847455080206  0.1544451962048098
  0.6891037455080248  0.7225105958306828  0.5959458037951907
  0.1891037455080247  0.2225105958306827  0.0959458037951906
  0.3015512544919790  0.7225105958306828  0.0959458037951906
  0.8015512544919791  0.2225105958306827  0.5959458037951907
  0.0298254041693169  0.5632322544919749  0.8459458037951907
  0.5298254041693169  0.0632322544919749  0.3459458037951906
  0.9608285958306824  0.5632322544919749  0.3459458037951906
  0.4608285958306825  0.0632322544919749  0.8459458037951907
  0.3015512544919790  0.2225105958306827  0.9044451962048098
  0.8015512544919791  0.7225105958306828  0.4044451962048098
```

```
0.6891037455080248  0.2225105958306827  0.4044451962048098
0.1891037455080247  0.7225105958306828  0.9044451962048098
0.9608285958306824  0.0632322544919749  0.6544451962048098
0.4608285958306825  0.5632322544919749  0.1544451962048098
0.0298254041693169  0.0632322544919749  0.1544451962048098
0.5298254041693169  0.5632322544919749  0.6544451962048098
0.6891037455080248  0.7915074041693173  0.0959458037951906
0.1891037455080247  0.2915074041693172  0.5959458037951907
0.3015512544919790  0.7915074041693173  0.5959458037951907
0.8015512544919791  0.2915074041693172  0.0959458037951906
0.0298254041693169  0.9507847455080206  0.3459458037951906
0.5298254041693169  0.4507847455080207  0.8459458037951907
0.9608285958306824  0.9507847455080206  0.8459458037951907
0.4608285958306825  0.4507847455080207  0.3459458037951906
0.0974247145309676  0.2014872615684118  0.7855243488359716
0.5974247145309678  0.7014872615684118  0.2855243488359716
0.8932302854690291  0.2014872615684118  0.2855243488359716
0.3932302854690292  0.7014872615684118  0.7855243488359716
0.0508487384315949  0.6549112854690320  0.5355243488359716
0.5508487384315879  0.1549112854690321  0.0355243488359716
0.9398052615684114  0.6549112854690320  0.0355243488359716
0.4398052615684116  0.1549112854690321  0.5355243488359716
0.8932302854690289  0.8125307384315883  0.2148676511640263
0.3932302854690292  0.3125307384315881  0.7148676511640264
0.0974247145309676  0.8125307384315883  0.7148676511640264
0.5974247145309677  0.3125307384315881  0.2148676511640263
0.9398052615684114  0.3591057145309705  0.4648676511640264
0.4398052615684116  0.8591057145309707  0.9648676511640264
0.0508487384315949  0.3591057145309705  0.9648676511640264
0.5508487384315879  0.8591057145309706  0.4648676511640264
0.0974247145309676  0.3125307384315881  0.2855243488359716
0.5974247145309677  0.8125307384315883  0.7855243488359716
0.8932302854690291  0.3125307384315881  0.7855243488359716
0.3932302854690292  0.8125307384315883  0.2855243488359716
0.0508487384315949  0.8591057145309706  0.0355243488359716
0.5508487384315879  0.3591057145309705  0.5355243488359716
0.9398052615684114  0.8591057145309706  0.5355243488359716
0.4398052615684116  0.3591057145309705  0.0355243488359716
0.8932302854690291  0.7014872615684118  0.7148676511640264
0.3932302854690292  0.2014872615684118  0.2148676511640263
0.0974247145309676  0.7014872615684118  0.2148676511640263
0.5974247145309678  0.2014872615684118  0.7148676511640264
0.9398052615684114  0.1549112854690321  0.9648676511640264
0.4398052615684116  0.6549112854690320  0.4648676511640264
0.0508487384315949  0.1549112854690321  0.4648676511640264
0.5508487384315879  0.6549112854690320  0.9648676511640264
0.2730916119263374  0.1056402383059923  0.6964564079936489
0.7730916119263374  0.6056402383059923  0.1964564079936491
0.7175633880736594  0.1056402383059923  0.1964564079936491
0.2175633880736592  0.6056402383059923  0.6964564079936489
0.1466957616940074  0.4792443880736623  0.4464564079936491
0.6466957616940073  0.9792443880736623  0.9464564079936489
0.8439592383059965  0.4792443880736623  0.9464564079936489
0.3439592383059965  0.9792443880736623  0.4464564079936491
0.7175633880736594  0.9083767616940032  0.3039345920063514
0.2175633880736592  0.4083767616940032  0.8039345920063515
0.2730916119263374  0.9083767616940032  0.8039345920063515
0.7730916119263374  0.4083767616940032  0.3039345920063514
0.8439592383059965  0.5347726119263403  0.5539345920063515
0.3439592383059965  0.0347726119263404  0.0539345920063514
0.1466957616940074  0.5347726119263403  0.0539345920063514
0.6466957616940073  0.0347726119263404  0.5539345920063515
0.2730916119263374  0.4083767616940032  0.1964564079936491
0.7730916119263374  0.9083767616940032  0.6964564079936489
0.7175633880736594  0.4083767616940032  0.6964564079936489
0.2175633880736592  0.9083767616940032  0.1964564079936491
0.1466957616940074  0.0347726119263404  0.9464564079936489
0.6466957616940073  0.5347726119263403  0.4464564079936491
0.8439592383059965  0.0347726119263404  0.4464564079936491
0.3439592383059965  0.5347726119263403  0.9464564079936489
0.7175633880736594  0.6056402383059923  0.8039345920063515
0.2175633880736592  0.1056402383059923  0.3039345920063514
0.2730916119263374  0.6056402383059923  0.3039345920063514
0.7730916119263374  0.1056402383059923  0.8039345920063515
0.8439592383059965  0.9792443880736623  0.0539345920063514
0.3439592383059965  0.4792443880736623  0.5539345920063515
0.1466957616940074  0.9792443880736623  0.5539345920063515
0.6466957616940073  0.4792443880736623  0.0539345920063514
0.4953270000000032  0.2570089999999965  0.3751960000000025
0.9953270000000032  0.7570089999999965  0.8751960000000025
0.4953270000000032  0.2570089999999965  0.8751960000000025
0.9953270000000032  0.7570089999999965  0.3751960000000025
0.9953270000000032  0.2570089999999965  0.1251960000000025
0.4953270000000032  0.7570089999999965  0.6251960000000025
0.9953270000000032  0.2570089999999965  0.6251960000000025
0.4953270000000032  0.7570089999999965  0.1251960000000025
0.2453270000000032  0.1294149240388492  0.5001960000000025
0.7453270000000032  0.6294149240388492  0.0001960000000025
0.7453270000000032  0.1294149240388492  0.0001960000000025
0.2453270000000032  0.6294149240388492  0.5001960000000025
0.1229210759611505  0.5070089999999965  0.2501960000000025
0.6229210759611504  0.0070089999999965  0.7501960000000025
0.8677339240388463  0.5070089999999965  0.7501960000000025
0.3677339240388463  0.0070089999999965  0.2501960000000025
0.7453270000000032  0.8846020759611534  0.5001960000000025
```

```
  0.2453270000000032  0.3846020759611534  0.0001960000000025
  0.2453270000000032  0.8846020759611534  0.0001960000000025
  0.7453270000000032  0.3846020759611534  0.5001960000000025
  0.1229210759611505  0.0070089999999965  0.7501960000000025
  0.6229210759611504  0.5070089999999965  0.2501960000000025
  0.8677339240388463  0.0070089999999965  0.2501960000000025
  0.3677339240388463  0.5070089999999965  0.7501960000000025
  0.4953270000000032  0.2570089999999965  0.6251960000000025
  0.9953270000000032  0.7570089999999965  0.1251960000000025
  0.4953270000000032  0.2570089999999965  0.1251960000000025
  0.9953270000000032  0.7570089999999965  0.6251960000000025
  0.9953270000000032  0.2570089999999965  0.3751960000000025
  0.4953270000000032  0.7570089999999965  0.8751960000000025
  0.9953270000000032  0.2570089999999965  0.8751960000000025
  0.4953270000000032  0.7570089999999965  0.3751960000000025
  0.6767461416556670  0.0755898583443325  0.3751960000000025
  0.1767461416556672  0.5755898583443326  0.8751960000000025
  0.3139088583443367  0.0755898583443325  0.8751960000000025
  0.8139088583443368  0.5755898583443326  0.3751960000000025
  0.1767461416556672  0.0755898583443325  0.1251960000000025
  0.6767461416556670  0.5755898583443326  0.6251960000000025
  0.8139088583443368  0.0755898583443325  0.6251960000000025
  0.3139088583443367  0.5755898583443326  0.1251960000000025
  0.3139088583443367  0.9384271416556629  0.6251960000000025
  0.8139088583443368  0.4384271416556629  0.1251960000000025
  0.6767461416556670  0.9384271416556629  0.1251960000000025
  0.1767461416556672  0.4384271416556629  0.6251960000000025
  0.8139088583443368  0.9384271416556629  0.8751960000000025
  0.3139088583443367  0.4384271416556629  0.3751960000000025
  0.1767461416556672  0.9384271416556629  0.3751960000000025
  0.6767461416556670  0.4384271416556629  0.8751960000000025
  0.3333653239137426  0.1775012261746515  0.0567517218504549
  0.8333653239137427  0.6775012261746516  0.5567517218504550
  0.6572886760862565  0.1775012261746515  0.5567517218504550
  0.1572886760862566  0.6775012261746516  0.0567517218504549
  0.0748347738253482  0.4189706760862570  0.8067517218504550
  0.5748347738253481  0.9189706760862569  0.3067517218504549
  0.9158192261746512  0.4189706760862570  0.3067517218504549
  0.4158192261746511  0.9189706760862569  0.8067517218504550
  0.6572886760862565  0.8365167738253485  0.9436392781495455
  0.1572886760862566  0.3365167738253486  0.4436392781495456
  0.3333653239137426  0.8365167738253485  0.4436392781495456
  0.8333653239137427  0.3365167738253486  0.9436392781495455
  0.9158192261746512  0.5950473239137432  0.1936392781495456
  0.4158192261746511  0.0950473239137431  0.6936392781495455
  0.0748347738253482  0.5950473239137432  0.6936392781495455
  0.5748347738253481  0.0950473239137431  0.1936392781495456
  0.3333653239137426  0.3365167738253486  0.5567517218504550
  0.8333653239137427  0.8365167738253485  0.0567517218504549
  0.6572886760862565  0.3365167738253486  0.0567517218504549
  0.1572886760862566  0.8365167738253485  0.5567517218504550
  0.0748347738253482  0.0950473239137431  0.3067517218504549
  0.5748347738253481  0.5950473239137432  0.8067517218504550
  0.9158192261746512  0.0950473239137431  0.8067517218504550
  0.4158192261746511  0.5950473239137432  0.3067517218504549
  0.6572886760862565  0.6775012261746516  0.4436392781495456
  0.1572886760862566  0.1775012261746515  0.9436392781495455
  0.3333653239137426  0.6775012261746516  0.9436392781495455
  0.8333653239137427  0.1775012261746515  0.4436392781495456
  0.9158192261746512  0.9189706760862569  0.6936392781495455
  0.4158192261746511  0.4189706760862570  0.1936392781495456
  0.0748347738253482  0.9189706760862569  0.1936392781495456
  0.5748347738253481  0.4189706760862570  0.6936392781495455
t-LYZO
   1.00000000000000
    12.7218750532553848    0.0000000000000000    0.0000000000000000
     0.0000000000000000   12.7218750532553848    0.0000000000000000
     0.0000000000000000    0.0000000000000000   12.3605978511701888
   Zr   O   Y   Li
   16   96   24   56
Direct
  0.2453270000000032  0.2570089999999965  0.2501960000000025
  0.7453270000000032  0.7570089999999965  0.7501960000000025
  0.7453270000000032  0.2570089999999965  0.7501960000000025
  0.2453270000000032  0.7570089999999965  0.2501960000000025
  0.9953270000000032  0.5070089999999965  0.0001960000000025
  0.4953270000000032  0.0070089999999965  0.5001960000000025
  0.9953270000000032  0.5070089999999965  0.5001960000000025
  0.4953270000000032  0.0070089999999965  0.0001960000000025
  0.2453270000000032  0.2570089999999965  0.7501960000000025
  0.7453270000000032  0.7570089999999965  0.2501960000000025
  0.7453270000000032  0.2570089999999965  0.2501960000000025
  0.2453270000000032  0.7570089999999965  0.7501960000000025
  0.9953270000000032  0.0070089999999965  0.5001960000000025
  0.4953270000000032  0.5070089999999965  0.0001960000000025
  0.9953270000000032  0.0070089999999965  0.0001960000000025
  0.4953270000000032  0.5070089999999965  0.5001960000000025
  0.3039919146954438  0.2896604199909468  0.4059592668804096
  0.8039919146954435  0.7896604199909470  0.9059592668804093
  0.6866630853045603  0.2896604199909468  0.9059592668804093
  0.1866630853045602  0.7896604199909470  0.4059592668804093
  0.9626755800090526  0.4483440853045559  0.1559592668804095
  0.4626755800090528  0.9483440853045562  0.6559592668804093
  0.0279784199909465  0.4483440853045559  0.6559592668804093
  0.5279784199909466  0.9483440853045562  0.1559592668804095
```

```
0.6866630853045603   0.7243575800090530   0.5944317331195912
0.1866630853045602   0.2243575800090532   0.0944317331195910
0.3039919146954438   0.7243575800090530   0.0944317331195910
0.8039919146954435   0.2243575800090532   0.5944317331195912
0.0279784199909465   0.5656729146954393   0.8444317331195912
0.5279784199909466   0.0656729146954395   0.3444317331195909
0.9626755800090526   0.5656729146954393   0.3444317331195909
0.4626755800090528   0.0656729146954395   0.8444317331195912
0.3039919146954438   0.2243575800090532   0.9059592668804093
0.8039919146954435   0.7243575800090530   0.4059592668804096
0.6866630853045603   0.2243575800090532   0.4059592668804096
0.1866630853045602   0.7243575800090530   0.9059592668804093
0.9626755800090526   0.0656729146954395   0.6559592668804093
0.4626755800090528   0.5656729146954393   0.1559592668804095
0.0279784199909465   0.0656729146954395   0.1559592668804095
0.5279784199909466   0.5656729146954393   0.6559592668804093
0.6866630853045603   0.7896604199909470   0.0944317331195910
0.1866630853045602   0.2896604199909468   0.5944317331195912
0.3039919146954438   0.7896604199909470   0.5944317331195912
0.8039919146954435   0.2896604199909468   0.0944317331195910
0.0279784199909465   0.9483440853045562   0.3444317331195909
0.5279784199909466   0.4483440853045559   0.8444317331195912
0.9626755800090526   0.9483440853045562   0.8444317331195912
0.4626755800090528   0.4483440853045559   0.3444317331195909
0.0931938821977681   0.1963899129279167   0.7823012581650025
0.5931938821977679   0.6963899129279164   0.2823012581650027
0.8974611178022288   0.1963899129279167   0.2823012581650027
0.3974611178022286   0.6963899129279164   0.7823012581650025
0.0559460870720901   0.6591421178022318   0.5323012581650027
0.5559460870720833   0.1591421178022316   0.0323012581650027
0.9347079129279160   0.6591421178022318   0.0323012581650027
0.4347079129279164   0.1591421178022316   0.5323012581650025
0.8974611178022288   0.8176280870720837   0.2180907418349953
0.3974611178022286   0.3176280870720833   0.7180907418349954
0.0931938821977681   0.8176280870720837   0.7180907418349954
0.5931938821977679   0.3176280870720833   0.2180907418349953
0.9347079129279160   0.3548748821977711   0.4680907418349952
0.4347079129279164   0.8548748821977709   0.9680907418349954
0.0559460870720901   0.3548748821977711   0.9680907418349954
0.5559460870720833   0.8548748821977709   0.4680907418349952
0.0931938821977681   0.3176280870720833   0.2823012581650027
0.5931938821977679   0.8176280870720837   0.7823012581650025
0.8974611178022288   0.3176280870720833   0.7823012581650025
0.3974611178022286   0.8176280870720837   0.2823012581650027
0.0559460870720901   0.8548748821977709   0.0323012581650027
0.5559460870720833   0.3548748821977711   0.5323012581650025
0.9347079129279160   0.8548748821977709   0.5323012581650025
0.4347079129279164   0.3548748821977711   0.0323012581650027
0.8974611178022288   0.6963899129279164   0.7180907418349954
0.3974611178022286   0.1963899129279167   0.2180907418349953
0.0931938821977681   0.6963899129279164   0.2180907418349953
0.5931938821977679   0.1963899129279167   0.7180907418349954
0.9347079129279160   0.1591421178022316   0.9680907418349954
0.4347079129279164   0.6591421178022318   0.4680907418349952
0.0559460870720901   0.1591421178022316   0.4680907418349952
0.5559460870720833   0.6591421178022318   0.9680907418349954
0.2718616903733225   0.1022015694553175   0.6923011335629972
0.7718616903733226   0.6022015694553174   0.1923011335629972
0.7187933096266741   0.1022015694553175   0.1923011335629972
0.2187933096266741   0.6022015694553174   0.6923011335629972
0.1501344305446822   0.4804743096266771   0.4423011335629971
0.6501344305446822   0.9804743096266770   0.9423011335629972
0.8405205694553216   0.4804743096266771   0.9423011335629972
0.3405205694553217   0.9804743096266770   0.4423011335629971
0.7187933096266741   0.9118154305446781   0.3080898664370034
0.2187933096266741   0.4118154305446780   0.8080898664370033
0.2718616903733225   0.9118154305446781   0.8080898664370033
0.7718616903733226   0.4118154305446780   0.3080898664370034
0.8405205694553216   0.5335426903733256   0.5580898664370033
0.3405205694553217   0.0335426903733255   0.0580898664370033
0.1501344305446822   0.5335426903733256   0.0580898664370033
0.6501344305446822   0.0335426903733255   0.5580898664370033
0.2718616903733225   0.4118154305446780   0.1923011335629972
0.7718616903733226   0.9118154305446781   0.6923011335629972
0.7187933096266741   0.4118154305446780   0.6923011335629972
0.2187933096266741   0.9118154305446781   0.1923011335629972
0.1501344305446822   0.0335426903733255   0.9423011335629972
0.6501344305446822   0.5335426903733256   0.4423011335629972
0.8405205694553216   0.0335426903733255   0.4423011335629971
0.3405205694553217   0.5335426903733256   0.9423011335629972
0.7187933096266741   0.6022015694553174   0.8080898664370033
0.2187933096266741   0.1022015694553175   0.3080898664370034
0.2718616903733225   0.6022015694553174   0.3080898664370034
0.7718616903733226   0.1022015694553175   0.8080898664370033
0.8405205694553216   0.9804743096266770   0.0580898664370033
0.3405205694553217   0.4804743096266771   0.5580898664370033
0.1501344305446822   0.9804743096266770   0.5580898664370033
0.6501344305446822   0.4804743096266771   0.0580898664370033
0.4953270000000032   0.2570089999999965   0.3751960000000025
0.9953270000000032   0.7570089999999965   0.8751960000000025
0.4953270000000032   0.2570089999999965   0.8751960000000025
0.9953270000000032   0.7570089999999965   0.3751960000000025
0.9953270000000032   0.2570089999999965   0.1251960000000025
0.4953270000000032   0.7570089999999965   0.6251960000000025
0.9953270000000032   0.2570089999999965   0.6251960000000025
```

```
0.4953270000000032  0.7570089999999965  0.1251960000000025
0.2453270000000032  0.1291050973393051  0.5001960000000025
0.7453270000000032  0.6291050973393051  0.0001960000000025
0.7453270000000032  0.1291050973393051  0.0001960000000025
0.2453270000000032  0.6291050973393051  0.5001960000000025
0.1232309026606946  0.5070089999999965  0.2501960000000025
0.6232309026606946  0.0070089999999965  0.7501960000000025
0.8674240973393021  0.5070089999999965  0.7501960000000025
0.3674240973393022  0.0070089999999965  0.2501960000000025
0.7453270000000032  0.8849119026606975  0.5001960000000025
0.2453270000000032  0.3849119026606975  0.0001960000000025
0.2453270000000032  0.8849119026606975  0.0001960000000025
0.7453270000000032  0.3849119026606975  0.5001960000000025
0.1232309026606946  0.0070089999999965  0.7501960000000025
0.6232309026606946  0.5070089999999965  0.2501960000000025
0.8674240973393021  0.0070089999999965  0.2501960000000025
0.3674240973393022  0.5070089999999965  0.7501960000000025
0.4953270000000032  0.2570089999999965  0.6251960000000025
0.9953270000000032  0.7570089999999965  0.1251960000000025
0.4953270000000032  0.2570089999999965  0.1251960000000025
0.9953270000000032  0.7570089999999965  0.6251960000000025
0.9953270000000032  0.2570089999999965  0.3751960000000025
0.4953270000000032  0.7570089999999965  0.8751960000000025
0.9953270000000032  0.2570089999999965  0.8751960000000025
0.4953270000000032  0.7570089999999965  0.3751960000000025
0.6816059719374800  0.0707300280625199  0.3751960000000025
0.1816059719374798  0.5707300280625197  0.8751960000000025
0.3090490280625241  0.0707300280625199  0.8751960000000025
0.8090490280625239  0.5707300280625197  0.3751960000000025
0.1816059719374798  0.0707300280625199  0.1251960000000025
0.6816059719374800  0.5707300280625197  0.6251960000000025
0.8090490280625239  0.0707300280625199  0.6251960000000025
0.3090490280625241  0.5707300280625197  0.1251960000000025
0.3090490280625241  0.9432869719374758  0.6251960000000025
0.8090490280625239  0.4432869719374756  0.1251960000000025
0.6816059719374800  0.9432869719374758  0.1251960000000025
0.1816059719374798  0.4432869719374756  0.6251960000000025
0.8090490280625239  0.9432869719374758  0.8751960000000025
0.3090490280625241  0.4432869719374756  0.3751960000000025
0.1816059719374798  0.9432869719374758  0.3751960000000025
0.6816059719374800  0.4432869719374756  0.8751960000000025
0.3353845755680624  0.1804860394624024  0.0552269424641541
0.8353845755680621  0.6804860394624023  0.5552269424641540
0.6552694244319371  0.1804860394624024  0.5552269424641540
0.1552694244319369  0.6804860394624023  0.0552269424641541
0.0718499605375974  0.4169514244319373  0.8052269424641540
0.5718499605375974  0.9169514244319376  0.3052269424641540
0.9188040394624019  0.4169514244319373  0.3052269424641540
0.4188040394624019  0.9169514244319376  0.8052269424641540
0.6552694244319371  0.8335319605375978  0.9451640575358464
0.1552694244319369  0.3335319605375978  0.4451640575358465
0.3353845755680624  0.8335319605375978  0.4451640575358465
0.8353845755680621  0.3335319605375978  0.9451640575358464
0.9188040394624019  0.5970665755680625  0.1951640575358464
0.4188040394624019  0.0970665755680628  0.6951640575358464
0.0718499605375974  0.5970665755680625  0.6951640575358464
0.5718499605375974  0.0970665755680628  0.1951640575358464
0.3353845755680624  0.3335319605375978  0.5552269424641540
0.8353845755680621  0.8335319605375978  0.0552269424641541
0.6552694244319371  0.3335319605375978  0.0552269424641541
0.1552694244319369  0.8335319605375978  0.5552269424641540
0.0718499605375974  0.0970665755680628  0.3052269424641540
0.5718499605375974  0.5970665755680625  0.8052269424641540
0.9188040394624019  0.0970665755680628  0.8052269424641540
0.4188040394624019  0.5970665755680625  0.3052269424641540
0.6552694244319371  0.6804860394624023  0.4451640575358465
0.1552694244319369  0.1804860394624024  0.9451640575358464
0.3353845755680624  0.6804860394624023  0.9451640575358464
0.8353845755680621  0.1804860394624024  0.4451640575358465
0.9188040394624019  0.9169514244319376  0.6951640575358464
0.4188040394624019  0.4169514244319373  0.1951640575358464
0.0718499605375974  0.9169514244319376  0.1951640575358464
0.5718499605375974  0.4169514244319373  0.6951640575358464
```